\documentclass[reprint,showpacs,twocolumn,superscriptaddress,aps]{revtex4-2}

\usepackage{amsmath,amsthm,amssymb}
\usepackage{graphicx}  
\usepackage{xcolor}
\usepackage{mathtools}
\usepackage{bbm}
\usepackage{subfigure}
\usepackage{dcolumn}
\usepackage{braket}
\usepackage{lipsum}
\usepackage{bm}
\usepackage{comment}
\usepackage{soul}

\usepackage[colorlinks,citecolor=blue,linkcolor=blue,urlcolor=blue,hyperindex]{hyperref}

\definecolor{purple}{rgb}{0.7,0,0.7}
\definecolor{dkgreen}{rgb}{0,0.6,0}
\definecolor{brown}{rgb}{0.8,0.4,0}
\definecolor{midnightblue}{rgb}{0.39,0.58,0.93}

\usepackage[normalem]{ulem}

\newcommand{\avg}[1]{\left\langle #1 \right\rangle}

\usepackage{dsfont}
\usepackage{physics}

\begin{document}
\title{Quantum criticality under imperfect teleportation}

\author{Pablo Sala}
\thanks{These authors all contributed comparably to this work.}
\affiliation{Department of Physics and Institute for Quantum Information and Matter, California Institute of Technology, Pasadena, CA 91125, USA}
\affiliation{Walter Burke Institute for Theoretical Physics, California Institute of Technology, Pasadena, CA 91125, USA}

\author{Sara Murciano}
\thanks{These authors all contributed comparably to this work.}
\affiliation{Department of Physics and Institute for Quantum Information and Matter, California Institute of Technology, Pasadena, CA 91125, USA}
\affiliation{Walter Burke Institute for Theoretical Physics, California Institute of Technology, Pasadena, CA 91125, USA}

\author{Yue Liu}
\thanks{These authors all contributed comparably to this work.}
\affiliation{Department of Physics and Institute for Quantum Information and Matter, California Institute of Technology, Pasadena, CA 91125, USA}

\author{Jason Alicea}
 \affiliation{Department of Physics and Institute for Quantum Information and Matter, California Institute of Technology, Pasadena, CA 91125, USA}
 \affiliation{Walter Burke Institute for Theoretical Physics, California Institute of Technology, Pasadena, CA 91125, USA}

\date{\today}

\begin{abstract}
 Entanglement, measurement, and classical communication together enable teleportation of quantum states between distant parties, in principle with perfect fidelity. To what extent do correlations and entanglement of a many-body wavefunction transfer under \emph{imperfect} teleportation protocols?  We address this question for the case of an imperfectly teleported quantum critical wavefunction, focusing on the ground state of a critical Ising chain.  We demonstrate that imperfections, e.g., in the entangling gate adopted for a given protocol, effectively manifest as weak measurements acting on the otherwise pristinely teleported critical state.   
 Armed with this perspective, we leverage and further develop the theory of measurement-altered quantum criticality to quantify the resilience of critical-state teleportation. 
 We identify classes of teleportation protocols for which imperfection $(i)$ preserves both the universal long-range entanglement and correlations of the original quantum critical state, $(ii)$ weakly modifies these quantities away from their universal values, and $(iii)$ obliterates long-range entanglement altogether while preserving power-law correlations, albeit with a new set of exponents. We also show that mixed states describing the average over a series of sequential imperfect teleportation events retain pristine power-law correlations due to a `built-in' decoding algorithm, though their entanglement structure measured by the negativity depends on errors similarly to individual protocol runs.  
 These results may allow one to design teleportation protocols that optimize against errors---highlighting a potential practical application of measurement-altered criticality.

 \end{abstract} 

\maketitle

\tableofcontents

\section{Introduction}

The groundbreaking 1993 work of Bennett et al.~\cite{bennett1993} elevated teleportation from science fiction to a key concept in modern quantum science.  Quantum teleportation refers to the transfer of an unknown wavefunction from a sender, Alice, to a receiver, Bob, at a remote location. 
The protocol proceeds roughly as follows.  
First, Alice and Bob maximally entangle their qubits, e.g., by applying an appropriate unitary gate.  Alice then performs a 
projective measurement that destroys the quantum state that she wishes to teleport, and communicates her measurement outcome to Bob.  Finally, Bob applies an outcome-dependent unitary to his qubits to recover the target wavefunction.  Aside from revealing a fundamental aspect of quantum mechanics---relevant for phenomena including black-hole evaporation~\cite{lloyd2013}---teleportation admits potential applications in quantum communication, quantum networks, and quantum computation~\cite{meas_based1};  
for reviews see Refs.~\onlinecite{TeleportationReview, Pirandola_2015, review2}.  To date, quantum teleportation has been demonstrated experimentally in numerous settings (e.g., Refs.~\onlinecite{exp_1997,exp_1998_2,exp_2004_2,exp_2014,Herbst2015}), including ground-to-satellite teleportation over distances of 1400 kilometers~\cite{SatelliteTeleportation}.  

Teleportation protocols are far from unique.  In fact there exists an infinite family of protocols---distinguished, for instance, by the choice of initialization for Bob's qubits, the particular entangling gate employed, and Alice's measurement basis---all of which yield perfect teleportation in the ideal case.  Imperfections invariably occur, however, in experimental implementations due to decoherence, measurement and gate errors, etc. The generic presence of such imperfections raises many practically relevant questions, particularly in the context of teleporting \emph{many-body} wavefunctions. To what extent do protocol imperfections modify universal aspects of correlations and entanglement in a teleported quantum state?  Is there necessarily a sense in which one can regard imperfections as benign when assessing universal features, or can arbitrarily weak deviations from an ideal protocol qualitatively alter the teleported state's character?  Among the infinite set of possible teleportation protocols, are some more resilient to certain types of imperfections than others?  Can one optimize against expected errors with a judiciously chosen protocol?  How do errors manifest after averaging over many imperfect teleportation runs?

We address these questions for the special case of imperfect teleportation of quantum critical wavefunctions.  In particular, we focus on the ground state of the one-dimensional transverse-field Ising model tuned to the phase transition between paramagnetic and ferromagnetic phases, described by an Ising conformal field theory (CFT) with central charge $c = 1/2$. This setting offers an appealing test bed for diagnosing the influence of teleportation-protocol imperfections: 
First, the critical state exhibits universal power-law correlations among local observables, as well as universal long-range entanglement; both features provide useful diagnostics for assessing the quality of the teleported wavefunction.  Second, the fact that the critical point is gapless implies sensitivity of physical properties to perturbations---in turn hinting that protocol imperfections can potentially drastically alter the character of the teleported quantum critical wavefunction.  

We identify and analyze a class of protocol imperfections for which the teleported wavefunction received by Bob corresponds to Alice's original quantum critical state modified by the application of a \emph{non-unitary} operator (that reduces to the identity in the ideal case). Such imperfections arise, for example, when Alice and Bob fall short of maximally entangling their qubits, or when Alice misaligns her measurement basis relative to the optimal protocol. One can profitably view the resulting non-unitary operator as effectively encoding the action of a weak measurement on the critical wavefunction.  In other words, imperfectly teleporting the quantum critical wavefunction is formally equivalent to perfectly teleporting a weakly measured counterpart of that state.
This perspective unites the problem of quantum criticality under imperfect teleportation with the developing theory of `measurement-altered quantum criticality'.

The latter area has its roots in earlier works~\cite{vicari,Li2018,Lu22,Chan2019,deluca,Li2021,Friedman,Lin2023probingsign} but came into sharp focus with Garratt et al.'s demonstration that even arbitrarily weak measurements can qualitatively alter long-distance correlations in a Luttinger liquid~\cite{AltmanMeasurementLL}.  In particular, these authors established that weak measurements perturb the Luttinger-liquid action with an `imaginary-time quantum impurity'---enabling a renormalization group analysis of their effects similar to the classic Kane-Fisher (real space) impurity problem \cite{KaneFisher}.  Soon after, Ref.~\onlinecite{sun2023} showed that measurement-induced modifications of correlations are accompanied by qualitative changes in the Luttinger liquid's entanglement; see also Ref.~\onlinecite{Ashida23}.
These results have since been extended to quantum Ising spin chains~\cite{JianMeasurementIsing,EhudMeasurementIsing,usmeasurementaltered,Ma23,Paviglianiti2023,sun23} and $(2 + 1)$-dimensional quantum critical points~\cite{Lee2023} (see also Ref.~\cite{Zou2023}).  Various strategies have also been proposed for experimentally detecting measurement-altered criticality~\cite{AltmanMeasurementLL,usmeasurementaltered,Garratt23,McGinley23,McGinley23_2,Li23}.  In passing, we note that many recent works have explored measurement-induced phenomena in related contexts including entanglement phase transitions in monitored systems~\cite{Li2018,skinner2019,Li2019,Doggen23,Qian23,Sang23,Popperl23,Feng23}, where unitary dynamics intertwines with local measurements, and for efficiently generating long-range entangled quantum states~\cite{Verresen,nat21,nat22,Lu22,bravyi_2022,leeji2022,Zhu22}.

\begin{figure*}
    \centering   \includegraphics[width=\linewidth]{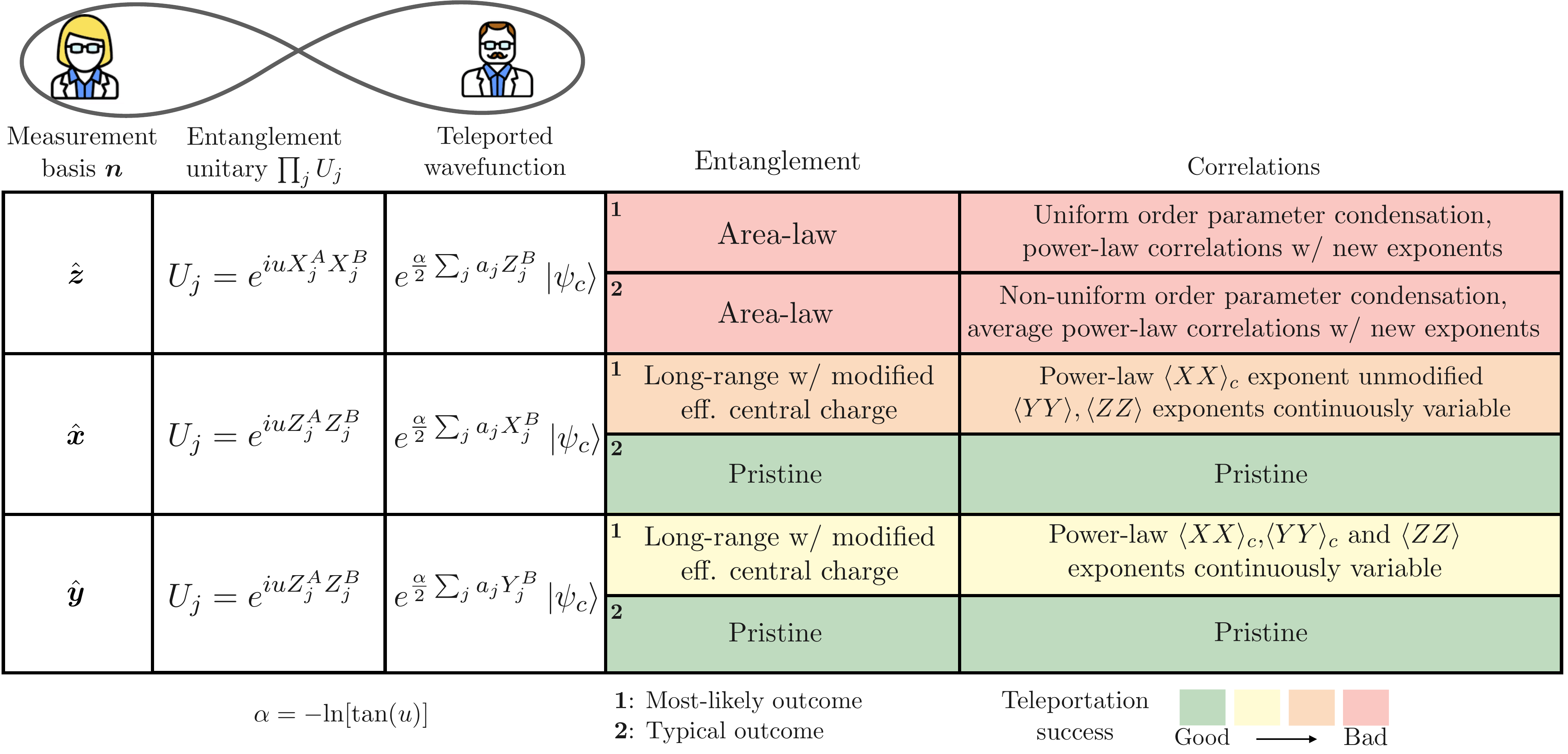}
    \caption{\textbf{Executive summary.}  The first two columns respectively indicate Alice's measurement basis and the imperfect entangling gate used in the protocol, leading to the teleported wavefunction in the third column.  Here $u$ denotes the entangling gate strength, with $u = \pi/4$ corresponding to the ideal protocol limit; $a_j$ denotes Alice's measurement outcome for qubit $j$; and $\ket{\psi_c}$ is Alice's original critical wavefunction.  The remaining columns summarize the entanglement and correlations in the imperfectly teleported state received by Bob.  Colors (from good/green to bad/red) represent imperfections that we classify in the main text as irrelevant, disguised marginal, marginal, and relevant.  Tools used to obtain these results include exact algebraic calculations of the resulting teleported states, analysis of an Ising CFT perturbed by an imperfection-induced `impurity', parent Hamiltonians for imperfectly teleported states, analytical calculations of correlations and entanglement based on Gaussian states, and tensor network techniques.} 
    \label{fig:main_fig}
\end{figure*}

By leveraging and advancing the theory of measurement-altered criticality, we derive detailed predictions for the fate of Ising quantum critical states under imperfect teleportation.  We explicitly obtain a family of imperfection-induced non-unitary operators---which indeed depend on the choice of protocol, as well as the measurement outcome that Alice communicates to Bob. These non-unitary operators apply to teleportation of generic many-body wavefunctions, as do some of the complementary tools that we employ to further analyze their impact on quantum critical states in both the limits of `weak' and `strong' imperfection.  The weak regime admits a renormalization group analysis of imperfection-induced imaginary-time quantum impurities similar to Refs.~\onlinecite{JianMeasurementIsing,EhudMeasurementIsing,usmeasurementaltered,Paviglianiti2023,sun2023}; in the strong regime we show that the character of the imperfectly teleported wavefunction relates deeply to `strange correlators'~\cite{You_2014}.  Our main results are summarized in Fig.~\ref{fig:main_fig}, and can be divided into three categories:

\emph{Relevant imperfection}.~Protocols yielding an imperfection-induced non-unitary operator that breaks $\mathcal{T}\times \mathbb{Z}_2$, where $\mathcal{T}$ denotes time reversal and $\mathbb{Z}_2$ is the spin-flip symmetry for the Ising spin system, are the least resilient to errors.  
Here, the associated imaginary-time quantum impurity comprises a relevant perturbation, and in the thermodynamic limit any amount of imperfection qualitatively alters both entanglement and correlations in the teleported wavefunction received by Bob.
Long-range entanglement present in the original quantum critical state becomes downgraded to area-law behavior~\cite{JianMeasurementIsing,sun2023}, independent of Alice's measurement outcome. 
Moreover, power-law correlations nevertheless persist---with apparently rigid decay exponents distinct from those in the pristine Ising theory.  We explain the coexistence of these properties through both the boundary CFT that incorporates the relevant perturbation~\cite{JianMeasurementIsing,AltmanMeasurementLL,EhudMeasurementIsing}, and by
deriving a long-range-interacting parent Hamiltonian for the imperfectly teleported wavefunction.  We further quantify how imperfection immediately and sharply modifies the distribution of order-parameter eigenvalues---i.e., full counting statistics---in that state.  (See also Ref.~\onlinecite{Tirrito} for a discussion of full counting statistics in a complementary setting with measurements.)

\emph{Marginal imperfection}.~In the weak-imperfection regime, teleportation protocols yielding non-unitary operators that preserve $\mathcal{T}\times \mathbb{Z}_2$ symmetry disrupt correlations and entanglement to a milder extent. 
For Alice's most likely measurement outcomes---which preserve translation symmetry, in some cases with an enlarged unit cell---the 
associated imaginary-time quantum impurity is generically marginal.  
Depending on the measurement basis used in the teleportation protocol, the marginal impurity can either arise at first order in the imperfection (in a sense made precise later on), or emerge as a higher-order effect under renormalization; we refer to the latter scenario as `disguised marginal imperfection'. In either case, Bob inherits an imperfectly teleported wavefunction featuring both power-law correlations and long-range entanglement, though the power-law exponents~\cite{EhudMeasurementIsing,JianMeasurementIsing,usmeasurementaltered,AltmanMeasurementLL} and effective central charge~\cite{entropyLuttinger,JianMeasurementIsing,Paviglianiti2023} characterizing the long-range entanglement vary continuously as one moves away from the ideal protocol limit.  The difference from relevant imperfection can be partly understood through the gentler impact of imperfection on full counting statistics.  In the extreme limit of `strong' imperfection---where the protocol transmits an asymptotically small amount of information from Alice's critical state---we recover the imperfectly teleported state's correlations and entanglement from a parent Hamiltonian (despite subtleties with convergence in the marginal case).

\emph{Irrelevant imperfection.}~Weak protocol imperfections yielding non-unitary operators that preserve $\mathcal{T}\times\mathbb{Z}_2$ symmetry generate irrelevant imaginary quantum impurities in two scenarios: $(i)$ for Alice's `typical' measurement outcomes (which we will define precisely)~\cite{AltmanMeasurementLL} and $(ii)$ for translation-invariant measurement outcomes provided one fine-tunes the measurement basis in a way that cancels off the marginal contribution highlighted above. Irrelevant imperfections are optimal in the sense that Bob's wavefunction inherits Alice's universal correlations and entanglement, albeit at sufficiently long length scales; such imperfections still inevitably corrupt short-distance properties.  In scenario $(ii)$ above, we show that a fine-tuned, finite level of imperfection actually \emph{revives} pristine universal properties in the imperfectly teleported state.  We further conjecture that Bob can, remarkably, continue to inherit these universal features even arbitrarily deep in the regime of strong imperfection, where the imperfection-induced non-unitary operators corrupting Alice's original quantum critical state are far from the identity. 

The hierarchy of imperfections that we identify---relevant, marginal, disguised marginal, and irrelevant---are associated with progressively greater resilience of the teleported quantum critical state to protocol errors.  These results provide concrete guidelines for choosing teleportation protocols that most faithfully transfer universal quantum critical properties between parties.  We stress that the imperfectly teleported wavefunction received by Bob can overlap negligibly with the original quantum critical wavefunction---while still (in the irrelevant case) displaying the same universal features. In essence, the metrics we study quantify the quality of teleporting within a family of quantum critical wavefunctions in the same universality class, rather than a specific many-body state.  Wavefunction overlaps, by contrast, in no way isolate long-distance characteristics, and at least in this context do not appear to be the `right' metric for assessing quality of the teleportation protocol. Finally, we address the fate of the teleported (mixed) state when a specific protocol is run many times, as would naturally occur when Bob wants to perform tomography of the teleported state with Alice obtaining different measurement outcomes sampled according to the Born rule. We find that power-law exponents encoded in the mixed state are faithfully teleported by virtue of the outcome-dependent unitary Bob applies in each teleportation run, which one can view as a `decoding step' (as in, e.g., Ref.~\onlinecite{lu2023}) baked into the protocol.  Imperfections can, nevertheless, nontrivially alter the entanglement structure of the mixed state as measured by the negativity.  More broadly, our work highlights many open questions, ranging from studying imperfect teleportation of other classes of entangled many-body wavefunctions to identifying concrete applications of this family of problems.  

We organize the remainder of the paper as follows.  As a warm-up, Sec.~\ref{sec:teleportation} briefly reviews imperfect teleportation at the single-qubit level.  Section~\ref{sec:wavefunction} then discusses the imperfect many-body teleportation problem from a mostly general viewpoint that sets the stage for our subsequent analysis of imperfectly teleported quantum critical wavefunctions.  Section~\ref{sec:Ising} reviews the correlations, entanglement structure, and full counting statistics for the ground state of a critical quantum Ising chain; these properties provide the baseline that we compare against when exploring the fate of Ising  criticality under imperfect teleportation in  Sec.~\ref{sec:ising_teleported}. Section~\ref{sec:general_main} generalizes our imperfect teleportation protocols by allowing an additional error source with similar consequences.  Section~\ref{sec:decoding} diagnoses errors in mixed states describing an ensemble of imperfectly teleported critical wavefunctions.  Conclusions and an outlook appear in Sec.~\ref{sec:outlook}, followed by numerous appendices that provide supplemental details of our analysis.

\section{Primer: Single-qubit teleportation}
\label{sec:teleportation}

\subsection{Canonical protocol} \label{sec:canonical}

Imperfect teleportation of a single qubit~\cite{bennett1993} already contains ingredients essential for the many-body case of our interest, and hence we begin with this simpler problem.  Following the standard protocol, consider a three-qubit Hilbert space $\mathcal{H}_{A_1}\otimes \mathcal{H}_{A_2}\otimes \mathcal{H}_{B}$; two qubits belong to Alice ($A_{1,2}$) and the third belongs to Bob ($B$).  In what follows $X_q, Y_q, Z_q$ denote Pauli operators acting on qubit $q$.  

Suppose that Alice wishes to transfer the state of qubit $A_1$ to Bob---without revealing her precise wavefunction.  In the computational ($Z$) basis, the state to be teleported reads $\ket{\psi_{A_1}}=c_1 \ket{\uparrow_{A_1}} + c_2 \ket{\downarrow_{A_1}}$ with $c_1, c_2 \in \mathbb{C}$.  Starting from a product state $\ket{\uparrow_{A_2}\uparrow_B}$ for the remaining two qubits, applying a Hadamard gate on qubit $A_2$ followed by a CNOT entangling gate~\cite{nielsen2002quantum} yields the maximally entangled state $\ket{\phi^+}=( \ket{\downarrow_{A_2}\downarrow_B}+\ket{\uparrow_{A_2}\uparrow_B})/\sqrt{2}$ 
that provides a resource for teleportation.  Next, Alice performs a Bell measurement on her two qubits.  Adopting similar logic used to generate maximal entanglement, Alice can accomplish this measurement by performing a CNOT gate, then a Hadamard gate on $A_2$, and finally measuring her individual qubits in the computational basis with outcomes $z_{A_1}$ and $z_{A_2}$.  (All measurement outcomes appear with the same probability of $1/4$.)  
Bob's resulting unnormalized single-qubit state reads \begin{equation}
\begin{aligned}
    \ket{\psi_{z_{A_1}z_{A_2}}}=&\bra{z_{A_1} z_{A_2}}U_{A_1A_2B}\ket{\psi_{A_1} \uparrow_{A_2}}\ket{\uparrow_{B}}.
\end{aligned}
\end{equation}
Here $U_{A_1A_2B}\equiv H_{A_2}\text{CNOT}_{A_2 A_1}\text{CNOT}_{A_2 B}H_{A_2}$ incorporates the Hadamard and CNOT gates used for entanglement generation and Bell measurement \footnote{In the standard teleportation protocol $U_{A_1A_2B}\equiv H_{A_1}\text{CNOT}_{A_1 A_2}\text{CNOT}_{A_2 B}H_{A_2}$. Nonetheless, our different choice of the controlled qubit (i.e., $A_1$ instead of $A_2$) only results in different measurement-dependent unitaries.}.  Using the fact that $H_{A_2}$ squares to the identity, one can equivalently express $U_{A_1A_2B}=\widetilde{\text{CNOT}}_{A_2A_1}\widetilde{\text{CNOT}}_{A_2B}$ as a product of two-qubit entangling unitaries 
\begin{equation}
     \widetilde{\text{CNOT}}_{ij}=e^{i\frac{\pi}{4}(1-X_i) (1-X_j)}.
     \label{CNOTtilde}
\end{equation}
After some algebra one finds that Bob's wavefunction becomes $\ket{\psi_{A_1}}$, up to a unitary $W_{z_{A_1}z_{A_2}}$ dependent on Alice's measurement outcomes $z_{A_{1,2}}$, i.e., $\ket{\psi_{z_{A_1}z_{A_2}}} = W_{z_{A_1}z_{A_2}} \ket{\psi_{A_1}}$.
After Alice classically communicates these outcomes to Bob, he can `undo' that unitary---completing the teleportation perfectly.

Notice that perfect teleportation can also arise upon replacing $\widetilde{\text{CNOT}}_{ij}$ by $U^{\frac{\pi}{4}}_{ij}=e^{i\frac{\pi}{4}X_i X_j}$, which mods out the single-qubit gates in Eq.~\eqref{CNOTtilde} yet similarly generates maximal entanglement when acting on $\ket{\uparrow_{A_2}\uparrow_{B}}$. 
As before Alice's measurement outcomes remain equally likely, and one finds that, after she performs her measurement, Bob's state reduces to $\ket{\psi_{A_1}}$ up to a (different) measurement-outcome-dependent unitary $V_{z_{A_1}z_{A_2}}$ 
that he can undo.  Below we consider the latter unitary entangling gate since it simplifies the analysis of imperfect protocols. 

As alluded to in the introduction, failure to perform the preceding operations exactly yields imperfect fidelity of the teleported state.  Such imperfections can arise from different error sources.  We primarily focus on imperfection in the two-qubit entangling gate $U^{\pi/4}_{ij}$; that is, suppose that the protocol unintentionally implemented $U_{ij} = e^{i u_{ij} X_i X_j}$, with $u_{ij} \in [0,\pi/2]$.  For $u_{ij} \neq \pi/4$ this unitary does \emph{not} maximally entangle computational-basis eigenstates, and moreover leads to unequal probabilities for Alice's measurement outcomes.  Hence, by repeatedly measuring, Alice can acquire some information about the unknown state $\ket{\psi_{A_1}}$.  Following the measurement phase of the protocol, Bob's unnormalized state correspondingly becomes
\begin{align}
    \ket{\psi_{z_{A_1}z_{A_2}}}&=\bra{z_{A_1} z_{A_2}}U_{A_1A_2}U_{A_2B}\ket{\psi_{A_1} \uparrow_{A_2}}\ket{\uparrow_{B}}
    \nonumber \\
    &= V_{z_{A_1}z_{A_2}} P_{z_{A_1}z_{A_2}} \ket{\psi_{A_1}}.
    \label{BobPsi}
\end{align}
On the second line, $V_{z_{A_1}z_{A_2}}$ is the same outcome-dependent unitary from the preceding paragraph, while crucially 
\begin{align}
    P_{z_{A_1}z_{A_2}} = e^{\frac{1}{2}z_{a_1}(\alpha_{A_1A_2}+z_{A_2}\alpha_{A_2B})\hat{Z}_B}
\end{align}
is a non-unitary operator dependent on the amount of imperfection through the parameters 
\begin{equation}
    \alpha_{ij} =-\ln[\tan(u_{ij})]. 
    \label{alphaij}
\end{equation}
For a perfect protocol with $u_{ij} = \pi/4$, $\alpha_{ij}$ vanishes, and hence $P_{z_{A_1}z_{A_2}}$ reduces to the identity.  For $u_{ij} \neq \pi/4$ the imperfect protocol generically teleports to Bob a non-unitarily modified cousin of Alice's $A_1$ qubit state \footnote{A loophole occurs when $u_{A_1 A_2} = u_{A_2 B}$. In this fine-tuned limit, even with a highly imperfect protocol for which the identical $u$'s are far from $\pi/4$, perfect teleportation still arises in the measurement outcome sector with $z_{A_2} = -1$.  The probability for accessing this  measurement outcome vanishes in the extreme limit where the protocol fails to entangle Alice's and Bob's qubits.  We neglect this non-generic limit in what follows.}. 
 In the extreme limit where $u_{ij}$ approaches 0, $\alpha_{ij}$ diverges; here the non-unitary operator completely obliterates the content of Alice's original state, and Eq.~\eqref{BobPsi} simply reduces to Bob's initial wavefunction.  Incidentally, this limit shows why entangling-gate imperfections non-unitarily (as opposed to unitarily) corrupt the teleported state: When the protocol altogether fails to entangle Alice's and Bob's qubits, Bob's wavefunction before and after Alice measures her qubits must be unchanged for any choice of $\ket{\psi_{A_1}}$, which can only occur by modifying the latter with a non-unitary operator.  

Imperfections in the initialization and measurement stages provide additional error sources that also non-unitarily corrupt the teleported state. Suppose, for instance, that Alice inadvertently measures along a quantization axis tilted away from the $Z$ direction and with a finite projection along $X$---thus no longer achieving a perfect Bell measurement.  The first line of Eq.~\eqref{BobPsi} would then be modified via $\bra{z_{A_1}z_{A_2}} \rightarrow \bra{z_{A_1}z_{A_2}}S_{A_1}S_{A_2}$, where $S_{A_j}$ is a spin-rotation operator acting on qubit $A_j$ that encodes the measurement misalignment.  In the extreme case where the quantization axis rotates all the way to the $X$ direction, $U_{A_1A_2}$ entirely fails to entangle Alice's $A_{1,2}$ qubits, and once again no information about $\ket{\psi_{A_1}}$ can be transmitted to Bob.  Following the above logic, $P_{z_{A_1} z_{A_2}}$ must again be a non-unitary operator that erases the content of Alice's $A_1$ qubit state.  Similar reasoning holds for misalignment in the initialization of the $A_1$ and $B$ qubits.  

Next we will explore a simplified version of the teleportation protocol reviewed here.  This simplification will turn out to greatly streamline explicit, closed-form derivation of imperfectly teleported quantum states both at the single- and many-qubit levels.

\subsection{Simplified protocol}
\label{sec:simplified}

Let us retain the key ingredients underlying quantum teleportation---entanglement, measurement, and classical communication---while distilling the setup by reducing the number of degrees of freedom.  Specifically, we replace the three-qubit Hilbert space ($\mathcal{H}_{A_1}\otimes \mathcal{H}_{A_2}\otimes \mathcal{H}_{B}$) with that of just two qubits ($\mathcal{H}_{A}\otimes \mathcal{H}_{B}$).  
Bob once again initializes his qubit into the state $\ket{\uparrow_B}$.  Alice's now sole qubit, $A$, realizes some general state $\ket{\psi_{A}}=c_1 \ket{\uparrow_{A}} + c_2 \ket{\downarrow_{A}}$ that she would like to transfer to Bob in a similar vein as in the previous subsection.  (Alice need not know her precise qubit state, e.g., she may have received it through a separate teleportation protocol.)  An ideal protocol proceeds by  
entangling $A$ and $B$ via the unitary gate $U_{AB}^{\pi/4} = e^{i \frac{\pi}{4} X_A X_B}$ and then projectively measuring $A$ in the computational basis. Importantly, as for the standard protocol, in this perfect implementation all measurement outcomes are equally likely.  
Upon Alice communicating her measurement outcome to Bob, he can perform an outcome-dependent unitary to perfectly recover Alice's qubit state.

Imagine, however, that the actual entangling gate employed had the form $U_{AB} = e^{i u X_A X_B}$ for some $u \in [0,\pi/2]$.  With $u = \pi/4$ we recover the ideal case, but otherwise the protocol is imperfect.  This unitary sends the initial state $\ket{\psi_A}\ket{\uparrow_B}$ to 
\begin{align}
  \ket{\psi_U}&=\ket{\uparrow_A}[\cos(u)c_1\ket{\uparrow_B}+i\sin(u)c_2\ket{\downarrow_B}]
  \nonumber \\
  &+\ket{\downarrow_A}[\cos(u)c_2\ket{\uparrow_B}+i\sin(u)c_1\ket{\downarrow_B}]. 
\end{align}
Also as for the standard protocol, Alice's measurement outcomes become biased by imperfection. 
After Alice measures her qubit with outcome $z_A = \pm 1$, Bob's unnormalized wavefunction becomes
\begin{equation}
  \begin{split}
  \ket{\psi_{z_A=+1}}=&\cos(u)c_1 \ket{\uparrow_B}+i\sin(u)c_2 \ket{\downarrow_B}\\
  \ket{\psi_{z_A=-1}}=&\cos(u)c_2 \ket{\uparrow_B}+i\sin(u)c_1 \ket{\downarrow_B}.
  \end{split}
  \label{zAcases}
\end{equation}
 One can equivalently recast Eqs.~\eqref{zAcases}, modulo an overall normalization, in the compact form 
 \begin{equation}
  \begin{split}
  \ket{\psi_{z_A}}=V_{z_A} P_{z_A} \ket{\psi_A}.
  \end{split}
  \label{psizA}
\end{equation}
Here
\begin{equation}
    V_{z_A} = e^{i\frac{\pi}{4}(1-z_A)X_B}e^{-i\frac{\pi}{4}z_AZ_B}
\end{equation}
is an outcome-dependent unitary operator present even in the optimal protocol, while imperfection is encoded through the non-unitary operator
\begin{equation}
    P_{z_A} = e^{\frac{\alpha}{2}z_AZ_B}
\end{equation}
dependent on $u$ through the parameter
\begin{equation}
    \alpha = -\ln[\tan(u)].
    \label{alpha_def}
\end{equation}
Once Bob learns Alice's measurement outcome and undoes the $W_{z_A}$ unitary, the final teleported wavefunction reads
\begin{equation}
    \ket{\psi_{z_A}^{\rm tele}} = P_{z_A} \ket{\psi_A}.
\end{equation}
The above analysis closely parallels the results from Eqs.~\eqref{BobPsi} through~\eqref{alphaij} for the canonical three-qubit teleportation protocol.  Similar to the latter case, imperfection in our simplified protocol, encoded through the single parameter $u \neq\pi/4$, nonunitarily corrupts the teleported state, completely erasing the content of Alice's qubit in the extreme case where $u \rightarrow 0$ and $\alpha \rightarrow \infty$.  This similarity is not accidental: In fact the upper line from Eq.~\eqref{BobPsi} maps to Eq.~\eqref{psizA} (modulo a unitary that does not depend on the state being teleported) in the limit where $u_{A_2B} = \pi/4$ and $u_{A_1 A_2} = u$.  Thus our simplified teleportation protocol captures the situation in which imperfection appears in only one of the two entangling gates employed in the canonical protocol.

\section{Many-body teleportation}
\label{sec:wavefunction}

\begin{figure}[ht]
    \centering
    \includegraphics[width=0.99\linewidth]{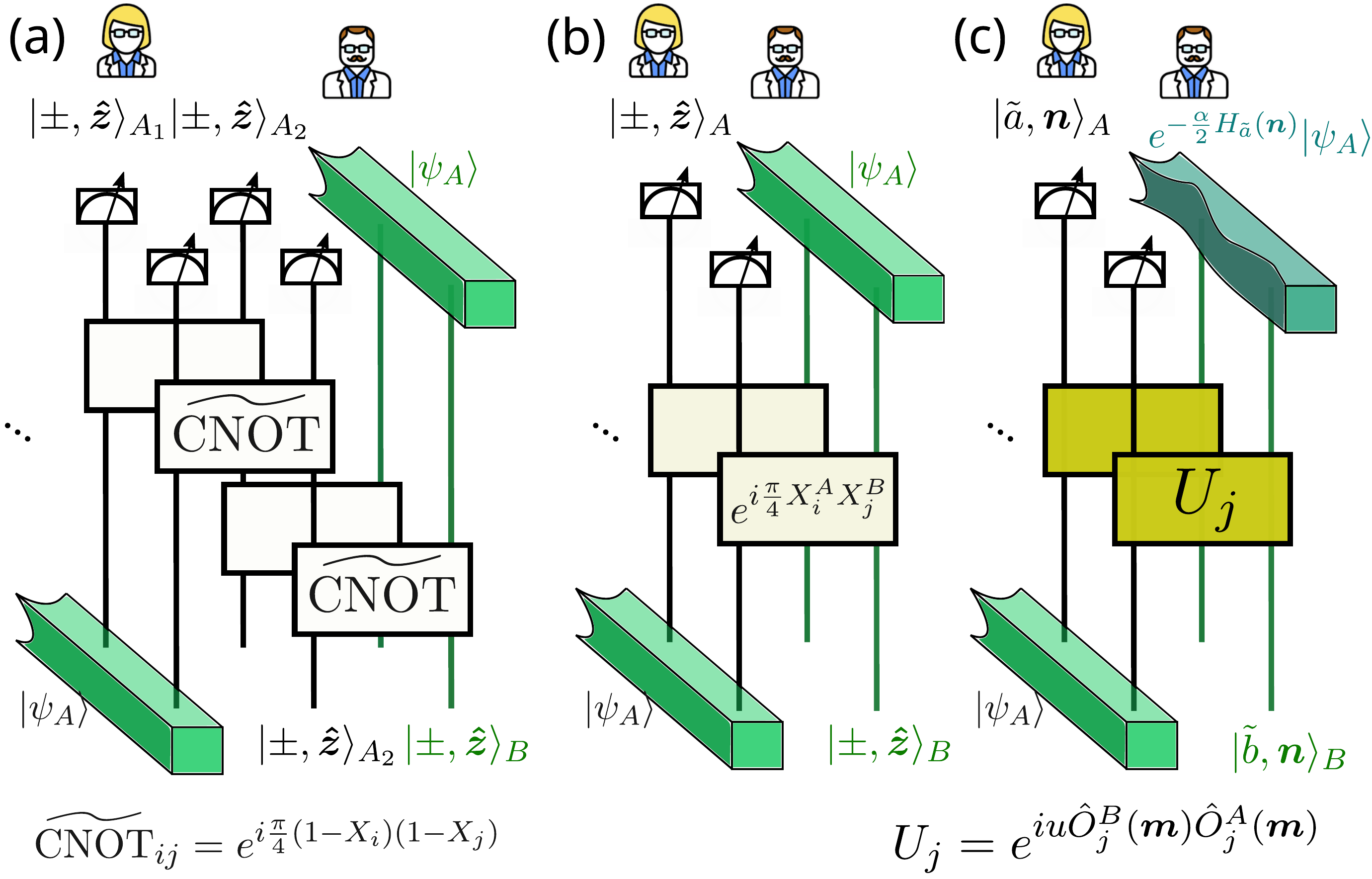}
    \caption{\textbf{Many-body quantum teleportation.} (a) Perfect quantum teleportation of an $L$-qubit wavefunction $\ket{\psi_A}$. After coupling Alice's qubits $A_1,\, A_2$ to Bob's $B$ (via two consecutive entangling unitaries $\widetilde{\text{CNOT}}$), performing properly aligned single-qubit measurements, and implementing an outcome-dependent unitary, $\ket{\psi_A}$ (light green prism) is perfectly teleported to Bob. (b) The same perfect setup can be simplified---see Sec.~\ref{sec:simplified}---by reducing Alice's number of degrees of freedom from $2L$ to $L$. (c) Errors in the `strength' of the entangling unitary (deviations from $u = \pi/4) $, or improper alignment of the measurement basis ($\bm{m}$ not parallel to $\bm{n}^\perp$), leads to imperfect teleportation where Bob receives a non-unitarily corrupted wavefunction $e^{-\frac{\alpha}{2} H_{\tilde{a}}}\ket{\psi_A}$ (deformed dark green solid) with $\alpha=-\ln[\tan(u)]$ and $H_{\tilde a}$ a Hermitian operator.} 
    \label{fig:protocol}
\end{figure}

Next we turn to the richer problem of imperfectly teleporting a many-body wavefunction from Alice to Bob.  Although we are primarily interested in teleporting quantum critical states, in much of this section we keep our discussion general.  Many-body teleportation of an $L$-qubit state can proceed by running the protocol reviewed in Sec.~\ref{sec:teleportation} separately on each constituent qubit---respectively requiring a total of $3L$ and $2L$ qubits in the canonical and simplified teleportation protocols.  Figures~\ref{fig:protocol}(a) and (b) sketch the corresponding protocols in the ideal case of perfect teleportation.  

In the following we focus on the simplified protocol with imperfectly applied entangling gates and establish the formalism that we will use to assess the nature of the resulting teleported state.  Once again, this scenario accounts for imperfection in one of the two entangling gates used in the canonical protocol (see Sec.~\ref{sec:canonical}). For a many-body state that breaks spin-rotation symmetry (always the case for the present paper), the impact of such protocol imperfections depends on the specific entangling gate used as we will emphatically see later in the context of teleporting Ising criticality.  We therefore generalize the teleportation protocol from Sec.~\ref{sec:simplified} in a way that incorporates freedom in the intialization of Bob's qubits and the gate that entangles with Alice's qubits.  

Suppose that Bob's initial wavefunction is a product state $\ket{\psi_B}= 
|\tilde b,\bm{n}\rangle$. 
On the right side we introduced short-hand notation $\tilde{b}=\{b_j\}$, where $b_k = \pm 1$ is the spin eigenvalue for qubit $k$ along the quantization axis set by the unit vector $\bm{n}$; i.e., $b_k$ is the eigenvalue for the Pauli operator $\hat{O}^B_k(\bm{n}) = \bm{n} \cdot \bm{\sigma}_k^B$. 
Alice's initial wavefunction---which we leave unspecified for now---can be expanded in the analogous basis as $\ket{\psi_A} = \sum_{\tilde a}c_{\tilde a} \ket{\tilde a_,\bm{n}}$; as above $\tilde a = \{a_j \}$, and the $c_{\tilde a}$ coefficients encode Alice's entanglement and correlations.
En route to teleportation, Alice and Bob employ the unitary 
\begin{equation}\label{eq:unitary} 
U 
=\prod_j e^{iu \hat{O}^A_j(\bm{n}^{\perp})\hat{O}^B_j(\bm{n}^{\perp})} 
\end{equation}
that pairwise entangles their respective qubits, transforming their combined state to $\ket{\psi_U}=U\ket{\psi_{A}}\ket{\psi_B}$.  Here $\bm{n}\cdot \bm{n}^{\perp}=0$ such that $\hat{O}_k(\bm{n}^{\perp})=\bm{n}^{\perp}\cdot \bm{\sigma}_k$ acting on $\ket{\tilde s,\bm{n}}$ sends $s_k \rightarrow -s_k$~\footnote{Notice that this choice is unique. Given $\bm{n}=(\sin(\theta)\cos(\phi), \sin(\theta)\sin(\phi), \cos(\theta))$, then the unique $\bm{n}^{\perp}$ satisfying these constraints is given by $\bm{n}^{\perp}=(-\cos(\theta)\cos(\phi), -\cos(\theta)\sin(\phi), \sin(\theta))$. Nonetheless, a similar result holds also when considering $U_{AB}
=\prod_j e^{iu \hat{O}^A_j(\bm{n}\times \bm{n}^{\perp})\hat{O}^B_j(\bm{n}\times\bm{n}^{\perp})}$}; the $A,B$ superscript designates whether the operator acts on Alice or Bob; and $u$ specifies the strength of the entangling gate.
 Next, Alice projectively measures all of her spins along the same quantization axis $\bm{n}$ used for Bob's initial state---obtaining particular measurement outcomes 
 $\tilde{a}$ with probability $p_{\tilde{a}}$.   All measurement outcomes are equally likely in the ideal $u = \pi/4$ protocol limit---similar to single-qubit teleportation---but imperfection biases the distribution as we discuss further in Sec.~\ref{sec:ising_teleported}. The wavefunction correspondingly evolves via $\ket{\psi_U}  \to \ket{\tilde a,\bm{n}}\ket{\psi_{\tilde{a}}}$, 
 where  
\begin{align}\label{eq:psis} 
  \ket{\psi_{\tilde{a}}} = \frac{1}{\sqrt{p_{\tilde{a}}}}\bra{\tilde a,\bm{n}}U\ket{\psi_A}|\tilde b,\bm{n}\rangle
\end{align}
is Bob's normalized wavefunction after the measurement stage of the protocol.

One can alternatively view Eq.~\eqref{eq:psis} as the wavefunction resulting from weakly measuring Bob's initial qubit state.  From this perspective, Alice's qubits constitute ancillary degrees of freedom that are entangled with Bob via the unitary $U$ and then projectively measured---thereby mediating the weak measurement. 
 Reference \onlinecite{usmeasurementaltered} indeed encountered expressions similar to 
Eq.~\eqref{eq:psis} in the context of measurement-altered Ising criticality.  There, both Alice's and Bob's initial wavefunctions exhibited nontrivial quantum correlations; analytical progress in treating the analogue of Eq.~\eqref{eq:psis} could then be carried out only perturbatively for entangling gates close to the identity.  
Here, by contrast, 
any entanglement in $\ket{\psi_{\tilde{a}}}$ descends exclusively from Alice's initial state $\ket{\psi_A}$---enabling further non-perturbative evaluation.  
As detailed in Appendix~\ref{sec:general}, using properties of Pauli operators for the unitary in Eq.~\eqref{eq:unitary}, and the fact that Alice's and Bob's Hilbert spaces are isomorphic, Bob's  wavefunction can be exactly recast as
\begin{equation} \label{eq:compact}
    \ket{\psi_{\tilde{a}}}= \frac{1}{\sqrt{\mathcal{N}}}e^{i\frac{\pi}{4}H_{\tilde{b}\leftarrow \tilde{a}}(\bm{n}^\perp)} e^{i\frac{\pi}{4}H_{\tilde{a}}(\bm{n})}e^{-\frac{\alpha}{2}H_{\tilde{a}}(\bm{n})}\ket{\psi_A}.
\end{equation}
Above $\alpha$ parametrizes the degree of protocol imperfection and once again depends on $u$ through Eq.~\eqref{alpha_def}; we will often interchangeably refer to the entangling gate strength by quoting $\alpha$ and/or $u$ depending on context.  The exponentials acting on $\ket{\psi_A}$ involve Hermitian operators
\begin{align}
    H_{\tilde{b}\leftarrow \tilde{a}}(\bm{n}^\perp)&=\sum_{j}(1-a_jb_j)\hat{O}^B_j(\bm{n}^{\perp})
    \\
    H_{\tilde{a}}(\bm{n})&=-\sum_j a_j\hat{O}^B_j(\bm{n}).
    \label{Han_def}
\end{align}
Notice that $\{\hat{O}_j(\bm{n}^{\perp}),\hat{O}_j(\bm{n})\}= 0$.  The normalization factor $\mathcal{N}$ relates to the probability $p_{\tilde{a}}$ according to 
\begin{equation}
    \mathcal{N}=[\sin(u)\cos(u)]^{-L}p_{\tilde{a}}.
    \label{Nprelation}
\end{equation} 
For $u=\pi/4$ (perfect teleportation protocol), equal probability of all measurement outcomes implies $p_{\tilde{a}}=2^{-L}$ and hence $\mathcal{N}=1$, consistent with the fact that the non-unitary operator becomes trivial in that limit.

Equation~\eqref{eq:compact} closely resembles Eq.~\eqref{psizA} and similarly involves unitary and non-unitary operators acting on Alice's wavefunction $\ket{\psi_A}$.  As in the one-qubit limit, knowledge of Alice's measurement outcome allows Bob to undo the unitaries---yielding the final teleported state
\begin{equation} \label{eq:compact_2}
    \ket{{\psi}^{\rm tele}_{\tilde{a}}}= \frac{1}{\sqrt{\mathcal{N}}}e^{-\frac{\alpha}{2}H_{\tilde{a}}(\bm{n})}\ket{\psi_A}.
\end{equation}
[Later we will examine both Eqs.~\eqref{eq:compact} and~\eqref{eq:compact_2} to analyze the impact of teleportation imperfections, since each form offers advantages in different regimes.]  Also as in the one-qubit example, the measurement-outcome-dependent non-unitary factor $e^{-\frac{\alpha}{2}H_{\tilde{a}}(\bm{n})}$ becomes the identity at $u = \pi/4$ (aka $\alpha = 0$) but otherwise becomes nontrivial, spoiling perfect teleportation.  See Fig.~\ref{fig:protocol}(c) for a sketch of the imperfect teleportation protocol discussed here.

As highlighted already in the introduction, subtle questions arise when specifically addressing imperfect teleportation of a quantum critical wavefunction.  To what extent is Bob's final wavefunction $\ket{\psi^{\rm tele}_{\tilde a}}$ the ground state of a critical chain?  Under what conditions can one regard the imperfect teleportation as weak?  That is, does small non-zero $\alpha$ perturbatively modify the structure of $\ket{\psi^{\rm tele}_{\tilde a}}$ compared to the ideal $\alpha = 0$ case, or is the effect inherently non-perturbative?  What is the relation between changes in entanglement and changes in critical correlations resulting from imperfect teleportation?  In the thermodynamic limit where system size $L \rightarrow \infty$, does $\ket{\psi^{\rm tele}_{\tilde a}}$ necessarily tend to a simple product state as $\alpha$ increases?

\subsection{Imperfection-induced modification of full counting statistics}\label{subsec:FCS} 

At $\alpha \neq 0$, the imperfection-induced non-unitary operator in Eq.~\eqref{eq:compact_2} redistributes the weight on the elements of Alice's original wavefunction $\ket{\psi_A}$.  This redistribution becomes particularly transparent upon expanding $\ket{\psi_A}$ in terms of $H_{\tilde a}(\bm{n})$ eigenstates, since the non-unitary operator acts very simply in that basis.  Below we explore general properties of Alice's imperfectly teleported wavefunction from this vantage point---leading to additional metrics, complementary to correlations and entanglement, for diagnosing the quality of the teleported state.  Our discussion here will additionally allow us to anticipate various scenarios for the impact of both weak and strong protocol imperfections.  

Since $H_{\tilde a}(\bm{n})$ in Eq.~\eqref{Han_def} is just a sum of $L$ commuting Pauli operators, $H_{\tilde a}(\bm{n})/2$ admits eigenvalues $m = -L/2,-L/2 + 1,\ldots,L/2-1,L/2$ with degeneracy $D_m=\binom{L}{m+L/2}$~\footnote{Notice that $H_{\tilde a}(\bm{n})$ is unitarily equivalent to the uniform Hamiltonian $H_{\text{uni}}(\bm{n})=-\sum_j \hat{O}^B_j(\bm{n})$. Hence an eigenspace labeled by eigenvalue $m$ shares the same degeneracy. }.  The corresponding eigenvalue distribution in a many-body wavefunction is known as full counting statistics and, at least for quantum critical states, exhibits universal features~\cite{Eisler03,Lamacraft08} (for a review see Sec.~\ref{sec:Ising}). 
Alice's initial state can always be decomposed into different $m$ sectors via $\ket{\psi_A} = \sum_m c_m \ket{\psi_m}$.  Here $c_m$ denotes the amplitude for the normalized component $\ket{\psi_m}$ that has $H_{\tilde a}(\bm{n})/2$ eigenvalue $m$, and determines the full counting statistics through $|c_m|^2$.  
In terms of the intensive variable $f = m/L + 1/2$---which ranges from 0 to 1---the function $P(f,L) = L |c_{m = (f-1/2)L}|^2$ \footnote{The relation between $P(f,L)$ and $c_m$ follows upon assuming that $m$ and $f$ can be treated as continuous variables. 
 This assumption is expected to hold generally for sufficiently large system sizes, except for deep in the tails of the distribution near $f = 0,1$, e.g., for $f$ of order $1/L$.  } gives the probability density for finding Alice's initial state with quantum number $f$ in system size $L$.  We stress that a quantum-critical initial state $\ket{\psi_A}$ will generically superpose all possible $m$ sectors compatible with symmetries.  Furthermore, given that 
 $D_m$ grows  combinatorially as $|m|$ decreases from $L/2$ towards zero,  $P(f,L)$ is expected to invariably peak at some intermediate $\mathcal{O}(1)$ value(s) of $f$ and eventually decay to zero at both $f = 0$ and 1 in the thermodynamic limit. 
 
Applying the non-unitary operator $e^{-\frac{\alpha}{2}H_{\tilde a}(\bm{n})}$ to $\ket{\psi_A}$ restructures the wavefunction by sending $c_{m} \rightarrow e^{-\alpha m}c_m$ and hence yields a modified probability density
\begin{equation}
    P_{\rm modified}(\alpha,f,L) \propto e^{-2\alpha L f} P(f,L).
    \label{Pmodified}
\end{equation}
In particular, the new exponential heavily favors $f$'s in the interval 0 to $\sim (\alpha L)^{-1}$ (which for any fixed nonzero $\alpha$ becomes vanishingly small as $L \rightarrow \infty$) and exponentially suppresses weight for larger $f$'s.  That exponential suppression, however, competes with the growth in $P(f,L)$ as $f$ increases from 0. Upon turning on $\alpha$ to nominally small values, one can then ask whether the most probable $f$ values predicted by $P_{\rm modified}$ (w1) evolve smoothly with $\alpha$ or (w2) non-perturbatively shift, for any nonzero $\alpha$, toward the interval near zero favored by the exponential in Eq.~\eqref{Pmodified}.  
We will encounter examples of both types later in this paper.

In either case (w1) or (w2), increasing $\alpha$ to large values shifts the probability weight captured by $P_{\rm modified}$ deep into the left tail of the distribution near $f = 0$.  Taking $\alpha \rightarrow \infty$ at fixed $L$ generically yields a product state.  
 In particular, here we obtain $e^{-\frac{\alpha}{2}H_{\tilde a}(\bm{n})}\ket{\psi_A} \rightarrow \ket{\psi_{m = -L/2}}=\ket{\tilde{a},\bm{n}}$, corresponding to Alice's measured state.  Upon accounting also for the unitaries in Eq.~\eqref{eq:compact}, Bob's wavefunction $\ket{\psi_{\tilde a}}$ reduces to his initialized product-state wavefunction $|\tilde b,\bm{n}\rangle$.  Since $\alpha \rightarrow \infty$ implies $u \rightarrow 0$, corresponding to the limit where Alice and Bob's entangling unitary approaches the identity, this result appears trivial---yet need not apply in the order of limits of interest here wherein $L\rightarrow \infty$ first, followed by $\alpha \rightarrow \infty$. The naively expected product state with $m = -L/2$ is non-degenerate, whereas the next $m = -L/2 + 1$ sector exhibits degeneracy $L$ that diverges in the thermodynamic limit; hence we expect that corrections to the naively expected product state persist in the above order of limits.  
  Similar in spirit to (w1) and (w2) above, two possible cases also appear in the regime of strongly imperfect teleportation: With $L\rightarrow \infty$, the teleported wavefunction at large $\alpha$, aka small $u$, (s1) may be perturbatively accessible starting from the naively expected product state, and thus exhibit properties such as full counting statistics that evolve smoothly with $u$, or (s2) may undergo non-perturbative changes for any non-zero $u$.

The remainder of this section further explores the small-$\alpha$ and large-$\alpha$ regimes.  

\subsection{Small-\texorpdfstring{$\alpha$}{alpha} regime}
\label{sec:small_alpha}

To make analytical progress at small $\alpha$, corresponding to $u$ near $\pi/4$, we now specialize to the case where $\ket{\psi_A}$ is the ground state of a quantum critical chain governed by a CFT.  We can then formulate a continuum field-theoretic description of correlations in the teleported state received by Bob that highlights universal consequences of protocol imperfections---similar to techniques employed for measurement-altered quantum criticality \cite{AltmanMeasurementLL}.  We will attack the final teleported state $\ket{\psi_{\tilde a}^{\rm tele}}$ from Eq.~\eqref{eq:compact_2}; this form proves most illuminating in the weak-imperfection regime where comparison with Alice's initial state is natural.  

Long-distance correlations in Alice's original quantum critical state can be extracted from a Euclidean path integral with accompanying CFT action $\mathcal{S}_{\rm CFT}$.  For an infinite chain, $\mathcal{S}_{\rm CFT}$ describes fields living at spatial coordinate $x$ and imaginary time $\tau$ each spanning the range $(-\infty,\infty)$.  One can view imaginary-time evolution from $\tau = -\infty$ to $0$ as preparing Alice's initial ket $\ket{\psi_A}$, while imaginary time evolution from $\tau = 0$ to $+\infty$ prepares the dual bra $\bra{\psi_A}$.  The measurement-induced non-unitary factor $e^{-\frac{\alpha}{2}H_{\tilde a}(\bm{n})}$ in Eq.~\eqref{eq:compact} disrupts the pristine CFT action via a `defect-line' perturbation acting at all spatial positions $x$ but only at $\tau = 0$ \cite{AltmanMeasurementLL}.  \emph{To leading order in $\alpha$}, the defect-line action takes the form
\begin{equation}\label{eq:defect}
  \delta\mathcal{S} = \alpha\int_x \omega(x)\chi(x,\tau=0).
\end{equation}
Here $\omega(x)$ is a scalar function and $\chi(x)$ is a CFT field such that $\omega(x)\chi(x)$ represents the continuum limit of  $a_j \hat{O}^B_j({\bm{n}})$ appearing in $H_{\tilde a}(\bm{n})$ [Eq.~\eqref{Han_def}].  The detailed structure of $\omega$ and $\chi$ depends on Alice's measurement basis (i.e., the vector $\bm{n}$) and particular measurement outcome. 

Within this continuum formulation, we can utilize renormalization group techniques to efficiently assess the extent to which the final teleported state $\ket{\psi_{\tilde a}^{\rm tele}}$ obtained by Bob showcases the distinguishing features of Alice's critical behavior at $\alpha \ll 1$.  More precisely, weakly imperfect teleportation yields profoundly different consequences depending on whether the defect-line action is relevant, marginal, or irrelevant.  [We caution that, as we will encounter later, properly assessing relevance in some cases requires considering $\mathcal{O}(\alpha^2)$ corrections that are not displayed in Eq.~\eqref{eq:defect}.]  The relevance or otherwise of the defect-line perturbation influences critical correlations and entanglement, as well as full counting statistics captured by the probability density $P_{\rm modified}$, in a ways that we will later characterize in detail. 

\subsection{Large-\texorpdfstring{$\alpha$}{alpha} regime} \label{sec:largealpha}

At $\alpha \gg 1$ (equivalently $u\approx e^{-\alpha} \ll 1$), further analytical progress is possible without specifying Alice's initial wavefunction $\ket{\psi_A}$.  Here it is more natural to examine Bob's wavefunction $\ket{\psi_{\tilde a}}$ given in Eq.~\eqref{eq:compact}.  The reason is that at $u = 0$, for which Alice and Bob remain unentangled throughout, Eq.~\eqref{eq:compact} reduces precisely to Bob's initial state $|\tilde b,\bm{n}\rangle$---whereas $\ket{\psi_{\tilde a}^{\rm tele}}$ in Eq.~\eqref{eq:compact_2} contains additional outcome-dependent unitary factors that somewhat mask the triviality of the teleportation protocol in this extreme limit.  We will assume for the remainder of this subsection that at small nonzero $u$, Bob inherits only faint imprints of Alice's state that can be captured by applying a `weak' entanglement-generating operator to $|\tilde b,\bm{n}\rangle$.  In other words, we postulate that small $u$ represents either a marginal or irrelevant perturbation to the trivial protocol that transmits no information about Alice's state.

As detailed in Ref. \onlinecite{usmeasurementaltered}, assuming $ \bra{\tilde a,\bm{n}}\psi_A\rangle\neq 0$ we can exactly write $\ket{\psi_{\tilde{a}}}$ as 
\begin{equation}
\begin{aligned}
 \ket{\psi_{\tilde{a}}} &= \frac{1}{\sqrt{\mathcal{N}}}\avg{\tilde a,\bm{n}|\psi_A}\sum_{N_f=0}^L e^{-\alpha N_f}e^{i\frac{\pi}{2} N_f}  \\ & \times\sum_{i_1<\dots<i_{N_f}}q(i_1, \dots, i_{N_f})\left(\prod_{j=1}^{N_f}\hat{O}^B_{i_j}(\bm{n}^\perp)\right)\ket{\tilde{b},\bm{n}}.
\end{aligned}
\end{equation}
Here $N_f$ counts the number of spin flips on Bob's initial state generated by a given piece of the entangling gate $U$, and we have defined the generalized `strange correlator'~\cite{You_2014}
\begin{equation}
q(i_1, \dots, i_{N_f})\equiv  \frac{\avg{\tilde a,\bm{n}|\prod_{j=1}^{N_f}\hat{O}^A_{i_j}(\bm{n}^\perp)|\psi_A}}{\avg{\tilde{a},\bm{n}|\psi_A}}
\label{f_def}
\end{equation}
involving normalized matrix elements between Alice's (generally very different) initial and post-measurement wavefunctions.  Expanding about the $\alpha\to \infty$ ($u = 0$) limit 
allows one to recast the expression above as 
\begin{equation}
\begin{aligned}
    \ket{\psi_{\tilde{a}}} &\approx \frac{1}{\sqrt{\mathcal{N}}}\underbrace{\exp\left(iu\sum_j q(j) \hat{O}^B_j(\bm{n}^\perp)\right)}_{=U^\prime} \\ & \times \underbrace{\exp\left(-\frac{u^2}{2}\sum_{j\neq k}V_{jk}\hat{O}^B_j(\bm{n}^\perp)\hat{O}^B_k(\bm{n}^\perp)\right)}_{=e^{-\frac{u^2}{2} H'_{\tilde a}}}\ket{\tilde{b},\bm{n}}.
    \label{psi_m_goal}
\end{aligned}
\end{equation}  
In the first line we absorbed constants into the normalization, and in the second line
\begin{equation}
  V_{jk} = q(j,k)-q(j)q(k)     
  \label{Vjk}
\end{equation}
can be viewed as a connected strange correlator.  
Equation~\eqref{psi_m_goal} is consistent with Ref.~\onlinecite{usmeasurementaltered}, where we performed a similar analysis but starting from a critical state for Bob, i.e., replacing $|\tilde b,\bm{n}\rangle$ with a quantum critical wavefunction.   As we discussed there, the large-$\alpha$, small-$u$ expansion employed above is expected to hold provided $V_{jk}$ is sufficiently local.  In some cases Alice's measured state $|\tilde a,\bm{n}\rangle$  may yield $V_{jk}$ that does not decay to zero sufficiently rapidly with $|j-k|$, spoiling the validity of the expansion in Eq.~\eqref{psi_m_goal} (likely indicating that Bob's state $\ket{\psi_{\tilde a}}$ can not be obtained by perturbatively modifying his initial state, contrary to our assumption). 

Equations~\eqref{eq:compact} and~\eqref{psi_m_goal} encode quantum correlations for Bob's wavefunction $\ket{\psi_{\tilde{a}}}$ in a very different manner.  In Eq.~\eqref{eq:compact} correlations are encoded primarily through Alice's initial state $\ket{\psi_A}$, while in Eq.~\eqref{psi_m_goal} they are captured by the structure of $q(j)$ and $V_{jk}$---where information about Alice's state is embedded.  One advantage of the latter representation is that, when the large-$\alpha$ expansion is valid, it permits using a variant of Witten's conjugation method~\cite{Witten982} to find a parent Hamiltonian for which $\ket{\psi_{\tilde{a}}}$ is the unique ground state.  Let us specialize to the case where $q(j)$ and $V_{jk}$ are purely real, so that in Eq.~\eqref{psi_m_goal} $U'$ is a unitary operator and $H^{\tilde a}_m$ is Hermitian.  (The case with imaginary parts can be treated by re-partitioning the exponentials to factor out a purely unitary component.) First observe that $|\tilde b,\bm{n}\rangle$ is the unique ground state of the commuting-projector Hamiltonian $H_0 = \sum_j \Gamma^\dagger_j\Gamma_j = \sum_j \Gamma_j$ 
with projectors $\Gamma_j=\frac{1}{2}[1-b_j\hat{O}^B_j(\bm{n})]$.  Physically, $H_0$ represents a position-dependent Zeeman field of uniform strength $1/2$ pointing along or against $\bm{n}$ such that Bob's initial state is energetically optimal.  Using $e^{-\frac{u^2}{2} H'_{\tilde a}} (U')^\dagger\ket{\psi_{\tilde a}}\propto |\tilde b,\bm{n}\rangle$, the state $\ket{\psi_{\tilde a}}$ is annihilated by the new set of operators $\overline{\Gamma}_j = U^\prime e^{\frac{u^2}{2} H'_{\tilde a}} \Gamma_j  e^{-\frac{u^2}{2} H'_{\tilde a}}(U^\prime)^\dagger$, which satisfy $\overline{\Gamma}_j^2 = \overline{\Gamma}_j$ but are not Hermitian and 
do not commute at different sites.  It follows that $\ket{\psi_{\tilde a}}$ is the unique ground state of the frustration-free parent Hamiltonian 
$H_{\rm parent} = \sum_j \overline{\Gamma}_j^\dagger \overline{\Gamma}_j$, which explicitly reads (modulo a constant) 
\begin{equation}
\begin{aligned}
    H_{\rm parent}&= \frac{1}{4}U^\prime \left[\sum_je^{2u^2 \hat{O}_j^B(\bm{n}^\perp)\sum_{k\neq j} V_{jk} \hat{O}_k^B(\bm{n}^\perp)}\right. \\ & \left. \!\!\!\!\!\!\!\!\!\!\!\!\!\!\!\!\! -\sum_j b_j \left\{\hat{O}_j^B(\bm{n}),e^{-u^2 \hat{O}_j^B(\bm{n}^\perp)\sum_{k\neq j} V_{jk} \hat{O}_k^B(\bm{n}^\perp)} \right\} \right](U^\prime)^\dagger.
    \label{Hau_exact}
\end{aligned}
\end{equation}
Equation~\eqref{Hau_exact} in general is not a sum of local operators. Nonetheless, boldly expanding the terms in brackets to $\mathcal{O}(u^2)$ yields the illuminating form 
\begin{align}
    H_{\rm parent}  \approx U' \bigg[&-\frac{1}{2}\sum_j b_j\hat{O}^B_j(\bm{n}) \bigg] (U')^\dagger
    \nonumber \\
    &+  u^2 \sum_{j\neq k}V_{jk}\hat{O}^B_j(\bm{n}^\perp) \hat{O}^B_k(\bm{n}^\perp).
    \label{Hau}
\end{align}

Nonzero $u$ manifests in two ways at this order.  First, already at order $u$ the orientations of the local Zeeman fields in $H_{\rm parent}$ rotate slightly compared to their orientations in $H_0$---though their magnitudes remain fixed at $1/2$.  This contribution originates entirely from the $U'$ unitary in Eq.~\eqref{psi_m_goal}.  Teleportation-wise, these rotations imprint information about Alice's initial state only at the single-qubit level, i.e., with this effect alone, Bob's wavefunction $\ket{\psi_{\tilde a}}$ would remain a product state and thus not inherit any of Alice's correlations and entanglement.   Second, and much more importantly, 
the non-unitary piece $e^{-\frac{u^2}{2} H'_{\tilde a}}$ from Eq.~\eqref{psi_m_goal} generates Ising spin-spin interactions of strength $u^2 V_{jk}$ in $H_{\rm parent}$.  These Ising interactions couple spin components that are orthogonal to the spin components appearing in the Zeeman field terms from the top line of Eq.~\eqref{Hau}, even when accounting for the $U'$ rotations.  One can therefore always (for any measurement outcome $\tilde a$ and any initialization $\tilde b$) unitarily transform the parent Hamiltonian to recast the Zeeman terms in the top line as spatially uniform without altering the bottom-line interactions.  It follows that when the large-$\alpha$ expansion holds, Bob's wavefunction $\ket{\psi_{\tilde a}}$ is the ground state of a transverse-field Ising model with weak but long-range Ising interactions dependent on both Alice's initial state and her particular measurement outcome.  The Ising interactions, albeit weak, enrich the structure of $\ket{\psi_{\tilde a}}$ beyond a simple product-state form and thus do impart some information about Alice's correlations and entanglement.  We will later exploit such parent Hamiltonians to intuitively understand the character of teleported states in highly imperfect protocols.

Earlier we remarked that the large-$\alpha$ expansion is expected to be controlled provided $V_{jk}$ decays sufficiently rapidly with $|j-k|$.  The structure of Eq.~\eqref{Hau} more quantitatively suggests that the expansion is controlled provided $V_{jk}$ decays faster than $1/|j-k|$.  In that case the Ising interaction term in $H_{\rm parent}$ is an extensive operator---just like the Zeeman term---with eigenvalues generically scaling linearly with system size.  One can then always make the $V_{jk}$ contribution a small perturbation by taking $u$ sufficiently small (i.e., $\alpha$ sufficiently large).  Conversely, if $V_{jk}$ decays as or slower than $1/|j-k|$, then the Ising interactions are no longer extensive and may exhibit eigenvalues that grow \emph{faster} than extensively with system size, thus potentially dominating over the Zeeman contribution for any nonzero $u$.  The large-$\alpha$ expansion is expected to break down in the latter case.

\vspace{0.1in}

Hereafter we specialize to imperfect teleportation of Ising quantum criticality realized in the transverse-field Ising chain (see next section for a review).  Our specific goals include (1) characterizing the final teleported state received by Bob for different protocols, considering both Alice's most likely measurement outcomes as well as typical outcomes; (2) identifying optimal teleportation protocols in the sense of resilience to imperfections; and (3) characterizing the mixed state describing the average over a series of sequential runs of a specific teleportation protocol via correlations and the entanglement negativity.

\section{Ising Quantum criticality review}\label{sec:Ising}

Suppose now that Alice's initial wavefunction $\ket{\psi_A}$ realizes the ground state $\ket{\psi_c}$ of a quantum Ising spin chain tuned to criticality, described by the Hamiltonian
\begin{equation}\label{eq:Hamiltonian_Ising}
H_c=-J\sum_j(Z_jZ_{j+1}+X_j).
\end{equation}
We take $J>0$ and consider a chain of length $L$ with periodic boundary conditions; moreover, here and below we ease the notation by neglecting superscripts $A/B$ on Pauli operators whenever the qubits that they act on is clear from context. In this section we briefly review the phenomenology of Ising criticality needed to investigate imperfect teleportation of Alice's quantum critical state.  

Equation~\eqref{eq:Hamiltonian_Ising} preserves translation symmetry, global $\mathbb{Z}_2$ spin flip symmetry generated by $G\equiv \prod_j X_j$, and an antiunitary time reversal symmetry $\mathcal{T}$ that leaves $X_j$ and $Z_j$ invariant but sends $Y_j\to -Y_j$. By applying a Jordan-Wigner transformation to Majorana fermion operators
\begin{align}\label{eq:JW}
  \gamma_{Aj} = \left(\prod_{k<j}X_k\right) Z_j,~~~\gamma_{Bj} = \left(\prod_{k<j}X_j\right) i X_j Z_j,
\end{align}  
the Hamiltonian maps to a free-fermion problem:
\begin{equation}\label{eq:majorana}
  H = iJ \sum_j(\gamma_{Aj+1}-\gamma_{Aj})\gamma_{Bj}.
\end{equation}
One can correspondingly express all correlation functions in terms of two-point fermion correlators using Wick’s theorem. Taking into account the non-local strings present in the Jordan-Wigner transformation,  correlation functions of the physical spin operators $X$, $Z$ and $Y$ can be expressed in terms of Toeplitz matrices, whose entries only depend on the difference between the two indices of the matrix~\cite{barouch1970,barouch1971}. Long-distance ground-state spin-spin correlations follow from the asymptotic behavior of Toeplitz determinants and are given by 
\begin{equation}
\begin{split}\label{eq:correlators_critical}
\langle X_iX_j \rangle_c\sim &\frac{1}{|i-j|^2},\quad \langle Y_iY_j \rangle\sim \frac{1}{|i-j|^{9/4}},\\
& \langle Z_iZ_j \rangle\sim \frac{1}{|i-j|^{1/4}}.
\end{split}
\end{equation}
The subscript $c$ indicates a connected correlator and is only needed for $X$, given that one-point $Y$ and $Z$ expectation values vanish by symmetry. Note that $Z$ correlations exhibit by the far the slowest decay, followed by $X$ and then $Y$. 
For a general Pauli operator $\hat{O}_j(\bm{n})=\bm{n}\cdot \bm{\sigma}_j$, its two-point correlation function $\langle\hat{O}_i(\bm{n})\hat{O}_j(\bm{n})\rangle$ will be dominated at long distances by the slowest-decaying contribution.

The free-fermion representation in Eq.~\eqref{eq:majorana} also enables an efficient computation of the entanglement entropy. 
Bipartitioning the system between a subregion $R$ and its complement $\overline{R}$, the reduced density matrix in $R$ is given by $\rho_R=\mathrm{Tr}_{\overline{R}}(\ket{\psi_c}\bra{\psi_c})$. Then the entanglement between $R$ and its complement reads $S_R=-\mathrm{Tr}(\rho_R\ln \rho_R)$. A more general family of functions quantifying the entanglement---dubbed R\'enyi entropies---are given by $S_R^{(n)}=1/(1-n)\ln\mathrm{Tr}\rho_R^n$. We notice the relation $S_R=\lim_{n\to 1} S_R^{(n)}$ known as the replica limit~\cite{Calabrese_2004}.
For the critical Ising chain, the leading-order contributions in the subsystem size $|R|=\ell$ for a single interval on the infinite line ($L\to \infty$) read~\cite{Calabrese_2004} 
\begin{equation} \label{eq:pris_ent}
S^{(n)}_R=\frac{c}{6}\left(\frac{n+1}{n}\right)\ln(\ell/\epsilon),\quad S_R=\frac{c}{3}\ln(\ell/\epsilon),
\end{equation}
where $\epsilon$ is an ultraviolet cutoff.
The logarithm prefactors are universal and are related to the central charge $c = 1/2$ of the underlying Ising CFT governing long-distance, low-energy properties at criticality.

The CFT description follows upon changing variables to left-moving ($\gamma_L$) and right-moving ($\gamma_R$) Majorana fermion fields via $\gamma_A = \gamma_R+\gamma_L$ and $\gamma_B = \gamma_R - \gamma_L$, and then taking the continuum limit of Eq.~\eqref{eq:majorana}. This procedure yields the continuum Hamiltonian
\begin{equation}
  \mathcal{H} = -iv \int_x(\gamma_R \partial_x \gamma_R - \gamma_L\partial_x \gamma_L)
  \label{HCFT}
\end{equation}
with $v\propto J$; the associated Euclidean action describes precisely the $c = 1/2$ Ising CFT.  The Ising CFT admits three primary fields: the identity $\mathbbm{1}$, the spin field $\sigma$ (scaling dimension $\Delta_{\sigma}=1/8$), and the energy operator $\varepsilon=i\gamma_R \gamma_L$ (scaling dimension $\Delta_{\varepsilon}=1$).  Physically, $\sigma$ represents the continuum limit of the microscopic order parameter $Z_j$, while $\varepsilon$ represents the perturbation generated by moving off of criticality in a $\mathbb{Z}_2$-preserving manner (e.g., by increasing or decreasing the transverse-field strength).  Microscopic spin components relate to CFT fields according to the following dictionary:
\begin{equation}
\begin{split}\label{eq:dictionary}
 X_j-\langle X\rangle &\sim\varepsilon +\cdots,\\
 Y_j&\sim i\partial_{\tau}\sigma+\cdots
, \\
  Z_j &\sim \sigma+\cdots.
\end{split}
\end{equation}
On the left side, $\langle X\rangle = 2/\pi$ 
is the ground state expectation value of $X_j$; subtracting this factor merely removes a contribution proportional to the identity field $\mathbbm{1}$.  On the right side, the fields are evaluated on coarse-grained positions corresponding to lattice site $j$, $\tau$ denotes imaginary time, and the ellipses account for subleading terms with higher scaling dimension.  Equation~\eqref{eq:dictionary} is not only compatible with symmetries, but also immediately reproduces the scaling behavior in Eq.~\eqref{eq:correlators_critical} given the scaling dimensions for the CFT fields provided above.

Finally, we briefly review the full counting statistics for the critical ground state of the Ising spin chain studied previously in Refs.~\onlinecite{Eisler03,Lamacraft08}.  As anticipated in Sec.~\ref{subsec:FCS}, this property will yield insight about the resilience of Ising criticality under imperfect teleportation protocols.  We will first examine how eigenvalues for the global operator 
\begin{equation}
    M^{\tilde{a}}_X=\sum_j a_jX_j/2
    \label{MXdef}
\end{equation}
are distributed in the critical ground state.  
Here the $a_j$'s take on $\pm 1$ values, and hence $M_{X}^{\tilde a}$ admits eigenvalues $m = -L/2,-L/2+1,\ldots,L/2$; later we will associate $a_j$ with Alice's measurement outcomes.   In the special case where $a_j = +1$ for all $j$, $M_X^{\tilde a}$
reduces to the global transverse magnetization.  

The mean $\langle M_X^{\tilde a} \rangle$ depends non-universally on the choice of $a_j$ while the variance reads
\begin{equation}
    \langle (M^{\tilde a}_X)^2\rangle - \langle M^{\tilde a}_X\rangle^2 = \frac{1}{4}\sum_{i,j}a_i a_j\langle X_i X_j\rangle_c.
    \label{MXvar}
\end{equation}
Given that the connected two-point $X$ correlator decays faster than $1/|i-j|$, the right-hand side is dominated by the short-distance region with $i$ near $j$ (for any pattern of $a_j$ signs). 
One thus generically obtains $\langle (M^{\tilde a}_X)^2\rangle - \langle M^{\tilde a}_X\rangle^2 \propto L$, i.e., the variance grows linearly with system size.  In terms of the intensive variable $f = m/L + 1/2$ used in Sec.~\ref{subsec:FCS}, these properties are compatible with $M_X^{\tilde a}$ exhibiting full counting statistics of a Gaussian form 
\begin{equation}
     P(f,L)\propto e^{-\kappa L(f-\overline{f})^2},
     \label{PMX}
\end{equation}  
where $\kappa$ and $\overline{f}$ are $a_j$-dependent constants.  This conclusion is consistent with Ref.~\onlinecite{Eisler03} in the limit of uniform $a_j$.  
Due to the even faster decay of two-point $Y$ correlators, we expect $M^{\tilde{a}}_Y=\sum_j a_jY_j/2$ to similarly exhibit Gaussian-like full counting statistics.

Next we consider the global longitudinal magnetization 
\begin{equation}
    M^{+}_Z=\sum_j Z_j/2.
    \label{MZdef}
\end{equation}
[Contrary to Eq.~\eqref{MXdef}, we do not allow for unspecified $a_j$ signs above since drawing conclusions about full counting statistics in the general case is not straightforward here.]
The mean $\langle M^+_Z\rangle$ vanishes by symmetry while the variance is
\begin{equation}
    \langle(M^+_Z)^2\rangle = \frac{1}{4}\sum_{i,j}\langle Z_i Z_j\rangle.
    \label{MZvar}
\end{equation}
Slow decay of the two-point $Z$ correlator implies that the right side, unlike Eq.~\eqref{MXvar}, is dominated by the long-distance region with $|i-j| \gg 1$.  This contribution yields a variance $\langle(M^+_Z)^2\rangle \propto L^{7/4}$ that diverges with system size faster than the linear growth obtained for $M_X^{\tilde a}$.  Thus the longitudinal magnetization eigenvalues are distributed more broadly in the critical ground state relative to the transverse magnetization eigenvalues: the characteristic spread of the intensive variable $f$ is $\delta f \sim 1/L^{1/8}$ for the former versus $\delta f \sim 1/L^{1/2}$ for the latter.  Full counting statistics for the longitudinal magnetization is non-Gaussian, however, and exhibits a universal double-peak structure~\cite{Lamacraft08} discussed in more detail in Appendix~\ref{app:longitudinal}.   

\section{Imperfect teleportation of Ising criticality}\label{sec:ising_teleported}

We are now in position to examine Bob's final wavefunction in Eq.~\eqref{eq:compact_2} resulting from imperfect teleportation of Alice's Ising quantum critical state $\ket{\psi_A} = \ket{\psi_c}$.  The non-unitary operator $e^{-\frac{\alpha}{2}H_{\tilde{a}}(\bm{n})}$ encoding imperfection depends on the quantization axis $\bm{n}$ defining Bob's initial wavefunction and Alice's measurement basis, her particular measurement outcome $\tilde a$, and the imperfection strength quantified by $\alpha$.  We will specifically address the extent to which the universal features of Ising criticality reviewed in the previous section survive with protocols employing different $\bm{n}$ vectors, assuming for simplicity throughout this section that Bob initializes his qubits into the uniform state $\ket{\psi_B} = |\tilde {b},\bm{n}\rangle$ with all $b_j = +1$. Sections~\ref{sec:relevant} through~\ref{sec:irrelevant} analyze the impact of protocol imperfections upon post-selecting for Alice's highest probability measurement outcomes;  Sec.~\ref{sec:typical} then examines typical outcomes.   

One can anticipate the hierarchy of protocol imperfections that we uncover by viewing the problem through the lens of full counting statistics.  For ${\bm n} = \hat{\bm{x}}$, the non-unitary operator becomes $e^{-\frac{\alpha}{2}H_{\tilde{a}}(\bm{n})} = e^{\alpha M_X^{\tilde a}}$ and hence alters full counting statistics of the global operator $M_X^{\tilde a}$ defined in Eq.~\eqref{MXdef}.  In Sec.~\ref{sec:Ising} we argued that, for any $\tilde a$, the distribution of $M_X^{\tilde a}$ eigenvalues in Alice's original quantum critical wavefunction obeys a Gaussian distribution, Eq.~\eqref{PMX}.   Following Eq.~\eqref{Pmodified}, Bob's final wavefunction exhibits the modified eigenvalue distribution 
\begin{equation}
    P_{\rm modified} \propto e^{-2\alpha L f} e^{-\kappa L(f-\overline{f})^2}
\end{equation}
that remains Gaussian. 
Imperfection merely shifts the peak location via $\overline{f} \rightarrow \overline{f} -\alpha/\kappa$; $L$-independence of the shift descends from the fact that $\langle (M^{\tilde a}_X)^2\rangle - \langle M^{\tilde a}_X\rangle^2 \propto L$.  The gradual change of full counting statistics with $\alpha$ in the thermodynamic limit implies that with ${\bm n} = \hat{\bm{x}}$, imperfections smoothly restructure the weight in Alice's original wavefunction, regardless of which measurement outcome she obtains during the teleportation protocol.  This property, in turn, hints that correlations and entanglement in the imperfectly teleported state generically also evolve smoothly with $\alpha$ relative to the ideal case.  (We caution that full counting statistics by itself does not reveal whether long-distance properties, e.g., scaling dimensions, evolve nontrivially with $\alpha$, though short-distance properties certainly will.)  Analogous conclusions apply to protocols utilizing ${\bm n} = \hat{\bm{y}}$.

Protocols with ${\bm n} = \hat{\bm{z}}$, by contrast, can behave qualitatively differently in this regard.  Let us consider uniform $\tilde a$ with all $a_j = +1$---which, along with the all $a_j = -1$ state, corresponds to Alice's most probable measurement outcome (see below).  In this case, the imperfection-induced non-unitary operator instead becomes $e^{-\frac{\alpha}{2}H_{\tilde{a}}(\bm{n})} = e^{\alpha M_Z^{+}}$, now modifying full counting statistics of $M_Z^{+}$ from Eq.~\eqref{MZdef}.  Recall that in this case the full counting statistics for Alice's original critical wavefunction exhibits a universal double-peak structure whose variance is parametrically broader with system size compared to the distribution for $M_X^{\tilde a}$.  Full counting statistics for Bob's imperfectly teleported wavefunction is once again modified according to Eq.~\eqref{Pmodified}, though the broader variance now translates into a more effective reshaping of the distribution by the imperfection-induced exponential factor $e^{-2\alpha L f}$.  

In Appendix~\ref{app:longitudinal} we introduce an analytical function that approximates the pristine double-peak distribution obtained in Ref.~\onlinecite{Lamacraft08}; we then use that function to simply extract the imperfection-modified distribution that Bob inherits.  This analysis reveals that in the thermodynamic limit arbitrarily small $\alpha$ obliterates the double-peak structure---yielding a single peak located in the distribution's tail near $f = 0$.  (The precise peak location in this altered distribution nevertheless varies smoothly with $\alpha$.)  Consequently, such protocol imperfections non-perturbatively restructure the weights in Alice's original quantum critical state in a manner that, owing to the explicit breaking of $f\rightarrow 1-f$ symmetry, yields a net non-zero magnetization in Bob's final state.  The non-perturbative restructuring further strongly suggests that long-distance correlations and entanglement in Bob's state are no longer captured by the Ising CFT---even for arbitrarily weak imperfection.  


\subsection{Relevant imperfection}\label{sec:relevant}

\begin{figure*}[ht]
    \centering
    \includegraphics[width=\linewidth]{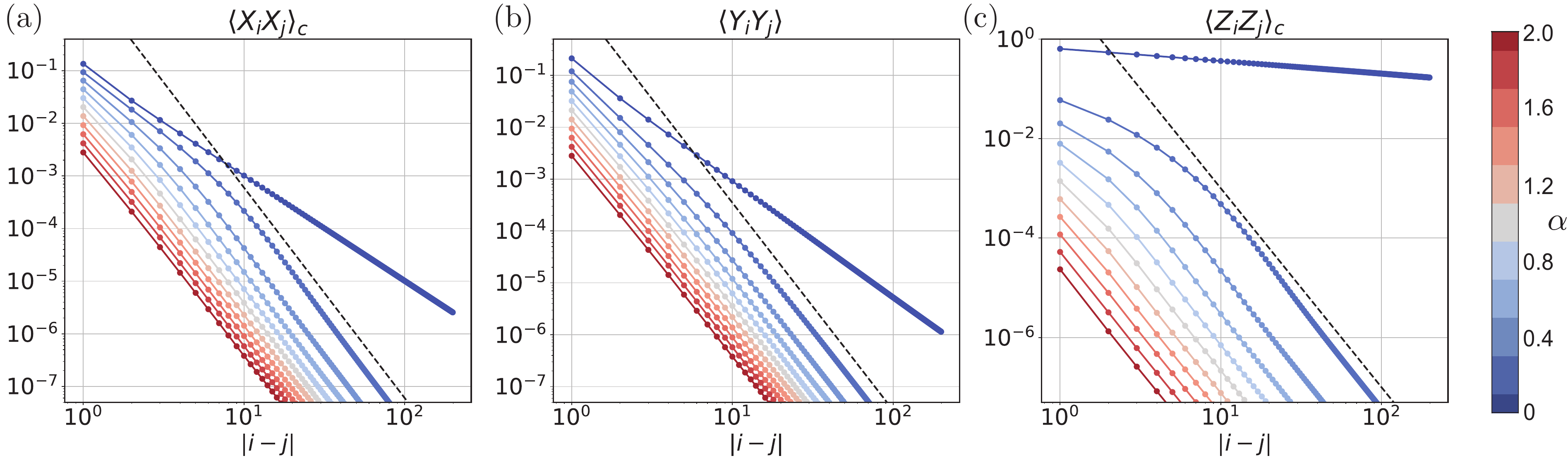}
    \caption{\textbf{Correlations with relevant imperfection.} Correlations in the teleported state received by Bob [Eq.~\eqref{eq:Z}] following an imperfect protocol with $\bm{n} = \hat{\bm{z}}$ and with Alice's most probable measurement outcome.  Data points were obtained using infinite DMRG for $\alpha \in [0,2]$ in steps of $0.2$.  For all $\alpha \neq 0$ shown, we obtain non-perturbatively altered power-law correlations with apparently rigid exponents that agree well with the analytical predictions in Eqs.~\eqref{eq:correlators_critical2_Z} and \eqref{eq:correlators_critical_Z} (dashed lines).}  
    
\label{fig:Z_correlations}
\end{figure*}

We proceed by exploring protocols with ${\bm n} = \hat{\bm{z}}$, for which full counting statistics suggests the most dramatic impact of imperfection.  Numerically, we find that whenever the entangling unitary $e^{i u X_j^A X_j^B}$ employed in the protocol is imperfect, i.e., with $u \neq \pi/4$, Alice's most probable measurement outcomes are uniform states $\ket{\tilde{a},\hat{\bm{z}}}$ with all $a_j = +1$ or all $a_j = -1$.  Focusing for concreteness on the $a_j =+1$ uniform outcome, Bob's final state explicitly reads
\begin{equation}\label{eq:Z}
\ket{\psi^{\mathrm{tele}}_{\tilde{a}}}=\frac{1}{\sqrt{\mathcal{N}}}e^{\alpha M_Z^+}\ket{{\psi}_c},
\end{equation}
where $M_Z^+$ is the global longitudinal magnetization defined in Eq.~\eqref{MZdef}.  Neither correlation functions nor entanglement scaling are amenable to exact analytical computations since $M_Z^+$ does not map to a local operator under a Jordan-Wigner transformation.  Nevertheless, below we pursue several strategies for diagnosing universal aspects of these quantities.

As a first approach we appeal to a parent Hamiltonian for the imperfectly teleported state $\ket{\psi^{\rm tele}_{\tilde a}}$.  Section~\ref{sec:largealpha} already previewed a parent Hamiltonian operative in the large-$\alpha$ limit---which we revisit shortly---but here we seek an alternate form valid at any $\alpha$.  While the existence of a general Hermitian parent Hamiltonian is not guaranteed, one can construct a \emph{strictly local} non-Hermitian parent $H_{\alpha}$ which has $\ket{\psi^{\mathrm{tele}}_{\tilde{a}}}$ as its unique (right) ground state:
\begin{equation}
\begin{aligned}
    H_\alpha&\equiv e^{\alpha M_Z^+}H e^{-\alpha M_Z^+}
    \\ &=-J\sum_j Z_j Z_{j+1} -\sum_j\left(h_xX_j -ih_yY_j \right)
\end{aligned}
\end{equation}
with $h_x=J\cosh(\alpha)$ and $h_y=J\sinh(\alpha)$.
Since $H$ and $H_\alpha$ are related by a similarity transformation, they share the same (purely real) gapless energy spectrum for any value of $\alpha$; their corresponding wavefunctions, however, are non-unitarily related to one another. 
This viewpoint suggests that imperfectly teleported critical states can exhibit atypical behavior inherited from non-Hermiticity of their parent Hamiltonians.  
In fact, Ref.~\onlinecite{Dora_2022} studied a non-unitarily perturbed free-fermion Hamiltonian that also features a gapless spectrum yet exhibits a finite `non-Hermitian coherence length' $\xi_{\text{NH}}$, beyond which two striking features emerge: 
First, correlations remain power-law, albeit with enhanced exponents encoding faster decay, and second, the spatial entanglement entropy 
saturates to a finite value $S\sim \frac{c}{3}\ln(\xi_{\text{NH}})$ with increasing subsystem size.    
We will show that the state in Eq.~\eqref{eq:Z} similarly exhibits anomalous power-law correlations coexisting with area-law entanglement.

Following Sec.~\ref{sec:small_alpha}, we can gain further quantitative insight at small $\alpha$ by translating the non-unitary operator in Eq.~\eqref{eq:Z} into a defect-line action $\delta S$ [Eq.~\eqref{eq:defect}] perturbing the pristine Ising CFT at imaginary time $\tau = 0$.  Using the continuum-limit expansion $M_Z^+ \sim \int_x \sigma$, one obtains 
\begin{equation}
  \delta S \sim \alpha \int_x \sigma(x,\tau = 0). 
  \label{deltaSZ}
\end{equation}
Equation~\eqref{deltaSZ} constitutes a strongly relevant perturbation that induces a flow to a new fixed point far from the Ising CFT~\cite{usmeasurementaltered,JianMeasurementIsing}.  That is, arbitrarily weak teleportation imperfection of the type studied here---which we dub relevant imperfection---qualitatively alters the universal long-distance properties encoded in Alice's original quantum critical wavefunction, consistent with the dramatic restructuring of full counting statistics arising for any $\alpha \neq 0$.  

The Ising CFT specifically flows to a fixed point that pins $\langle \sigma\rangle \neq 0$ along the defect line, yielding a boundary condition that chops the $1+1$-dimensional Euclidean spacetime into decoupled upper ($\tau>0$) and lower ($\tau<0$) halves. 
Scaling arguments~\cite{usmeasurementaltered} reveal that Bob's final wavefunction exhibits a net longitudinal magnetization that grows non-analytically with $\alpha$: $\langle Z_j \rangle \sim \langle \sigma(x_j,\tau = 0) \rangle \sim \alpha^{1/7}$.  Furthermore, using the dictionary in Eq.~\eqref{eq:dictionary}, connected two-point $X$ and $Z$ correlators in Bob's final state can be calculated by evaluating the  correlators $\langle \sigma(x)\sigma(x') \rangle$ and $\langle \varepsilon(x)\varepsilon(x') \rangle$ along the boundary at the new fixed point.  Appendix~\ref{app:boundary} details the boundary CFT calculation, which yields power-law correlations
\begin{equation}\label{eq:correlators_critical2_Z}
\langle X_iX_j \rangle_c  \sim \frac{1}{ |i-j|^4},\quad \langle Z_iZ_j \rangle_c \sim \frac{1}{|i-j|^{4}}
\end{equation}
with entirely different exponents from those of the unperturbed Ising CFT.  (As we show below, $Y$ correlators also admit power-law exponent 4, again in contrast to the pure Ising CFT.)

The modified power-law correlations captured above assuming small $\alpha$ hold for any $\alpha \neq 0$ in the thermodynamic limit.  In the opposite regime of large $\alpha$ we can exploit the asymptotic expansion developed in Sec.~\ref{sec:largealpha} to express the penultimate form of Bob's wavefunction (prior to applying the final outcome-dependent unitary) in the useful form [Eq.~\eqref{psi_m_goal}] 
\begin{equation}
\label{eq:XX_measZbasis}
\ket{\psi_{\tilde{a}}}\approx\frac{1}{\sqrt{\mathcal{N}}} e^{i u\sum_j q(j)X_j}e^{-\frac{u^2}{2}\sum_{j\neq k}V_{jk}X_jX_k}\ket{\psi_B}.
\end{equation}
Recall that $u \approx e^{-\alpha}$ 
at $\alpha \gg 1$, $q(j)$ is defined in Eq.~\eqref{f_def}, $V_{jk}$ is a connected strange correlator defined in Eq.~\eqref{Vjk}, and $\ket{\psi_B}$ is Bob's initial state with all $Z_j = +1$. 
 Due to translation symmetry $q(j)$ is just a constant; the strange correlator was evaluated in Ref.~\onlinecite{usmeasurementaltered} and scales as $V_{jk} \sim 1/|j-k|^4$.  As discussed in Appendix~\ref{app:perturbative}, one can equivalently express two-point $X$ correlators evaluated in the wavefunction~\eqref{eq:XX_measZbasis} as 
\begin{equation}\label{eq:large1}
    \bra{\psi_{\tilde a}} X_i X_j \ket{\psi_{\tilde a}}_c = \frac{1}{\mathcal{Z}} \sum_{\{x_k\} = \pm 1}x_i x_j e^{-\beta \mathcal{H}},
\end{equation}
where $\mathcal{Z}$ is a classical partition function corresponding to the classical power-law-interacting Hamiltonian $\mathcal{H} = \sum_{j \neq k} V_{jk} x_j x_k$ with inverse temperature $\beta = u^2$.  The leading contribution in a high-temperature series expansion gives $\langle X_i X_j \rangle_c 
\approx u^2V_{ij} \sim 1/|i-j|^4$ in harmony with Eq.~\eqref{eq:correlators_critical2_Z} obtained at small $\alpha$.
We further argue in Appendix~\ref{app:perturbative} that, given the fast decay of $V_{jk}$ with separation, $Z$ and $Y$ correlators also display the same power law,
\begin{equation}\label{eq:correlators_critical_Z}
\begin{split}
\langle Z_iZ_j \rangle_c \sim \frac{1}{|i-j|^{4}},\quad  \langle Y_iY_j \rangle \sim \frac{1}{|i-j|^{4}},
\end{split}
\end{equation}
which for $Z$ also agrees with Eq.~\eqref{eq:correlators_critical2_Z}.  

Our DMRG simulations presented in Fig.~\ref{fig:Z_correlations} support universality of the anomalous power-laws captured above at small- and large-$\alpha$.  Curves in each panel cover $\alpha \in [0,2]$ in steps of 0.2, while, as a guide, the dashed black line represents the slope for a power law with exponent 4.  (Details of our DMRG simulations presented here and below can be found in Appendix~\ref{app:dmrg}.) The simulations agree well with our predictions for $X$, $Y$, and $Z$ correlators.  To summarize, with any degree of imperfection in the protocol considered here, two-point correlators of an arbitrary Pauli operator display `fast' algebraic decay with a common exponent that does not depend on $\alpha$ and differs strongly relative to pristine Ising CFT expectations.  

\begin{figure}
    \centering
    \includegraphics[width=\linewidth]{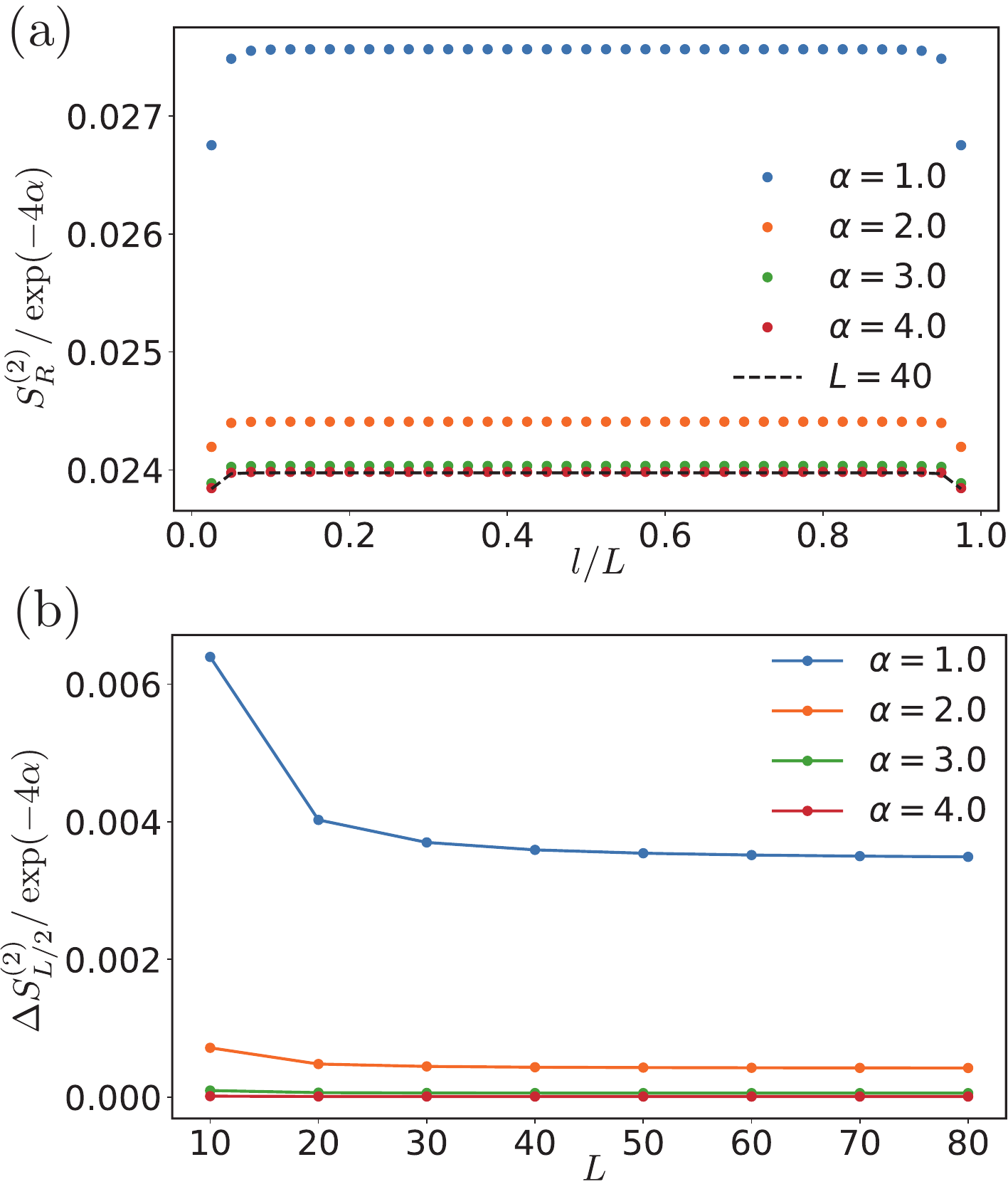}
    \caption{\textbf{Area-law entanglement for relevant imperfection.} 
    (a) Second R\'enyi entropy of the teleported state in Eq.~\eqref{eq:Z} for system size $L=40$ with periodic boundary conditions and different $\alpha$'s obtained using finite DMRG. The dashed line refers to the perturbative analytical prediction in Eq.~\eqref{eq:Renyi2}. (b) System-size dependence of the difference in the half-chain entanglement entropy between the perturbative analytical approach [as obtained by expanding Eq.~\eqref{eq:XX_measZbasis}] and numerical evaluation of Eq.~\eqref{eq:Z} using finite DMRG.  The data show excellent agreement with area-law entanglement emerging from relevant imperfection as calculated analytically in the large-$\alpha$ limit.  }
    \label{fig:Z_expansion}
\end{figure}

Does logarithmic growth of entanglement entropy survive relevant imperfect teleportation?   Reference~\onlinecite{JianMeasurementIsing} provided numerical results together with CFT arguments suggesting that even arbitrarily weak non-unitary operators of the form in Eq.~\eqref{eq:Z} destroy logarithmic growth in favor of area-law entanglement entropy that saturates to a value dependent on $\alpha$ but not subsystem size.  
We can further justify this conclusion at the lattice level in the large-$\alpha$ regime. 
Consider again the state in Eq.~\eqref{eq:XX_measZbasis}.  
Upon further expanding this perturbative expression to order $\mathcal{O}(u^2)$, which we expect is legitimate given the fast power-law decay of $V_{jk}$, we can explicitly obtain the reduced density matrix $\rho_R$ and evaluate the second R\'enyi entropy $S^{(2)}_{R}$ at $\mathcal{O}(u^4)$.  The result (derived in Appendix~\ref{app:EE}) reads $S^{(2)}_R\propto u^4 \sum_{j\in R, k \in \overline{R}}V_{jk}^2$ and indeed saturates to a constant that does not depend on the subsystem size $|R|$. Figure~\ref{fig:Z_expansion} shows that this expansion is well-behaved in the large-$\alpha$ regime (as expected), with the difference between the entanglement entropies numerically computed from the exact wavefunction 
$\ket{\psi^{\rm tele}_{\tilde a}}$ in Eq.~\eqref{eq:Z} using DMRG, and from the preceding perturbative treatment, decreasing with increasing system size and with increasing $\alpha$.

Therefore, the imperfectly teleported wavefunction $\ket{\psi^{\mathrm{tele}}_{\tilde{a}}}$ showcases power-law decaying correlations yet exhibits short-range area-law entanglement, as hinted at earlier from our non-Hermitian parent Hamiltonian discussion.  
Additional insight into this unusual coexistence emerges by examining the Hermitian parent Hamiltonian derived in the large-$\alpha$, small-$u$ limit in Sec.~\ref{sec:largealpha} [Eq.~\eqref{Hau}]. 
In the present context the parent Hamiltonian (after applying a unitary to undo the unimportant $U'$ rotation) explicitly reads 
\begin{equation} \label{eq:parent_Z}
    H_{\rm parent} \approx -\sum_j Z_j + u^2\sum_{j\neq k}V_{jk}X_j X_k.
\end{equation}
The dominant Zeeman field in the first term reflects the fact that at $u = 0$, the protocol trivially returns Bob's initial state with all $Z_j = +1$; the subdominant second term is an $O(u^2)$ anti-ferromagnetic Ising-like interaction with power-law decay dictated by the strange correlator $V_{jk}$ that encodes information about Alice's critical state.
References~\onlinecite{Tagliacozzo,Vodola_2016} analyzed the phase diagram of such long-range-interacting Hamiltonians for general power-law behavior of $V_{jk}$.   
In particular, for our Zeeman-field-dominated model with fast-decaying $V_{jk}\sim|j-k|^{-4}$, the ground state realizes an area-law-entangled paramagnetic gapped phase but with connected correlation functions displaying algebraic decay at long distances---precisely as found above. 
Given these atypical properties, it is interesting to ask whether the imperfectly teleported state received by Bob can be efficiently expressed as a matrix product state (MPS). Using the expression of the parent Hamiltonian in Eq.~\eqref{eq:parent_Z}, and the proof in Ref.~\onlinecite{Saito_19}, one finds that this state can indeed be efficiently approximated by an MPS (see Appendix~\ref{app:MPS_rel} for details).

\subsection{Marginal imperfection}\label{sec:marginal}

Here we examine protocols with $\bm{n} = \hat{\bm{x}}$.  
Similar to the previous subsection, we find numerically that Alice's most probable measurement outcome for any imperfect strength of the entangling unitary $e^{i u Z_j^A Z_j^B}$ is the uniform state $\ket{\tilde{a},\hat{\bm{x}}}$ with all $a_j = +1$.
Focusing on this measurement outcome, Bob's final state reads
\begin{equation}\label{eq:X}
\ket{\psi^{\mathrm{tele}}_{\tilde{a}}}=\frac{1}{\sqrt{\mathcal{N}}}e^{\alpha M_X^+}\ket{{\psi}_c}
\end{equation}
with $M_X^+=\sum_jX_j/2$ the global transverse magnetization.  In the weak-imperfection limit $\alpha \ll 1$ one can again map the non-unitary operator acting on Alice's pristine quantum critical state to a defect-line action perturbing the Ising CFT.  Expanding $M_X^+ \sim \int_x \varepsilon$ yields
\begin{equation}
    \delta S \sim \alpha \int_x \varepsilon(x,\tau = 0),
\end{equation}
which is marginal (hence the name marginal imperfection).  Marginality indicates that in this case imperfection in the teleportation protocol smoothly modifies universal aspects of correlations and entanglement as Refs.~\onlinecite{usmeasurementaltered,JianMeasurementIsing,EhudMeasurementIsing} established in the context of measurement-altered criticality.  Specifically, Refs.~\onlinecite{usmeasurementaltered,EhudMeasurementIsing} exploited CFT calculations~\cite{CABRA_1994,NT2011} to compute two-point $Z$ correlators in the presence of an $\varepsilon$ line defect.  Reference~\onlinecite{JianMeasurementIsing} evaluated $Z$ correlators as well as the von Neumann entanglement entropy of the state in Eq.~\eqref{eq:X} for arbitrary $\alpha$ by mapping the problem to a single-bond-defect Ising model and using  the results of Ref.~\onlinecite{eisler}.  We will pursue a complementary approach that allows explicit microscopic computation, for any $\alpha$, of two-point $X, Y,$ and $Z$ correlators together with R\'enyi entropies with any index $n$. 

 \begin{figure}[ht]
    \centering
    \includegraphics[width=\linewidth]{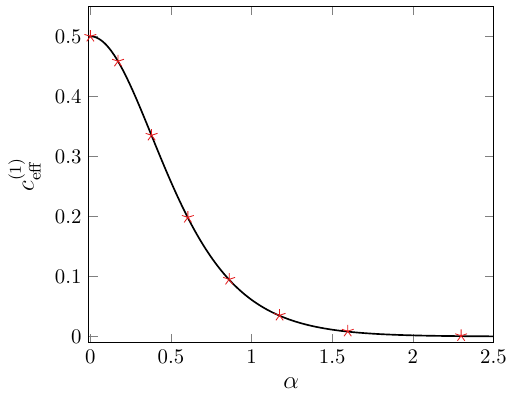}
    \caption{{\bf Effective central charge resulting from marginal imperfection.} 
 The solid line corresponds to the analytical prediction in Eq.~\eqref{eq:c_eff1} while the symbols are obtained using the numerical techniques described in Appendix~\ref{app:marginal}. In particular, for each value of $\alpha$, we calculated numerically $S_R^{(1)}$ for several values of the subsystem size $\ell$ up to 300; the  results were fitted with $a\ln \ell +b$, and we plot here $c_{\rm eff}^{(1)}$ extracted from the
 best fit of $a$ as a function of $\alpha$.  These results imply that teleportation protocols corrupted by marginal imperfection preserve long-range entanglement of Alice's original critical state, but with a prefactor that decays smoothly to zero with the degree of imperfection.}
    \label{fig:X_centralcharge}
\end{figure}
 
Our approach views Bob's final state in Eq.~\eqref{eq:X} as the result of evolving the critical wavefunction $\ket{\psi_c}$ with a non-Hermitian Hamiltonian $i \alpha M_X^+$, similar to what has been done in Ref.~\onlinecite{xhek2}. As we show in Appendix~\ref{app:marginal}, this time evolution can be exactly computed upon mapping the model to free fermions through the Jordan-Wigner transformation in Eq.~\eqref{eq:JW}, followed by a Fourier transform to momentum space. Both the von Neumann and R\'enyi entanglement entropies as well as all correlation functions of Bob's final state $\ket{\psi^{\mathrm{tele}}_{\tilde{a}}}$ can be expressed as determinants of matrices with a special structure, known as Toeplitz matrices---just as for the ground state of the critical Ising model. The asymptotic behavior of block-Toeplitz determinants can be studied using a generalization of the Fisher-Hartwig conjecture~\cite{ares2018}. We refer the interested reader to  Appendix~\ref{app:marginal} for all technical details; here we simply report the main results.

\begin{figure*}[ht]
    \centering
    \includegraphics[width=\linewidth]{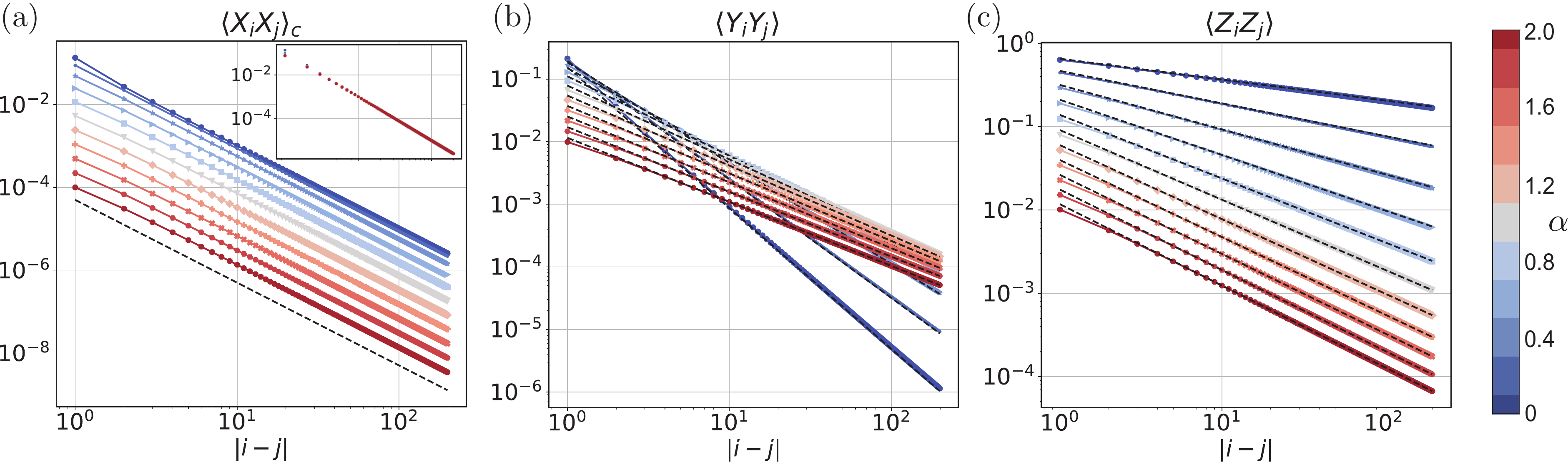} 
    \caption{\textbf{Correlations with marginal imperfection.} Correlations in the teleported state received by Bob [Eq.~\eqref{eq:X}] following an imperfect protocol with $\bm{n} = \hat{\bm{x}}$ and with Alice's most probable measurement outcome.  Data points were obtained using infinite DMRG for $\alpha \in [0,2]$ in steps of $0.2$.  (a) Two-point $X$ correlators exhibit unmodified power-law exponents (dashed line shows the slope 
arising with decay exponent 2), though imperfection suppresses the amplitude.   \emph{Inset:} Same correlations scaled by $\sech(2\alpha)^{-2}$; the collapse indicates that Eq.~\eqref{eq:correlators_critical2} indeed captures the amplitude's $\alpha$ dependence.   (b,c) Two-point $Y$ and $Z$ correlators exhibit modified power-law exponents dependent on the imperfection strength.  Dashed lines correspond to the analytical predictions in Eqs.~\eqref{eq:correlators_critical2}, after fitting the overall scale from numerical data. } 
    \label{fig:X_correlations}
\end{figure*}

In the thermodynamic limit $L\to\infty$, the R\'enyi entropies for a large interval of size $\ell \gg 1$ read 
\begin{equation}\label{Z_kappa_n_0_1}
S_R^{(n)}=\frac{c^{(n)}_{\mathrm{eff}}}{6}\left(\frac{n+1}{n}\right)\ln \ell+\mathcal{O}(1),
\end{equation}
where $c^{(n)}_{\mathrm{eff}}$ is given by the real integral
\begin{equation}\label{eq:c_eff}
 c^{(n)}_{\mathrm{eff}}=\frac{12}{\pi^2}\hspace{-1.3em}\int\limits_{\tanh(2\alpha)}^1\hspace{-1.3em}
 g_n(\lambda )\ln\frac{\sqrt{1-\lambda^2}}
 {\sqrt{\lambda^2-\tanh^2(2\alpha)}+\mathrm{sech}(2\alpha)} \,d\lambda
\end{equation}
with 
\begin{equation}
 g_n(\lambda)=\frac{n^2}{1-n^2 }\frac{(\lambda +1)  (1-\lambda )^n+(\lambda -1)  (\lambda +1)^n}{\left(\lambda ^2-1\right) \left[(1-\lambda
   )^n+(\lambda +1)^n\right]}.
\end{equation}
At $\alpha = 0$ one obtains $c_{\rm eff}^{(n)} = 1/2$, i.e., the coefficients reduce to the central charge of the Ising CFT for any $n$ in agreement with Eq.~\eqref{eq:pris_ent}.  In the limit $n \rightarrow 1$, where $S_R^{(n)}$ reduces to the von Neumann entanglement entropy, $c_{\rm eff}^{(1)}$ defines an effective central charge.  For this case the integral in Eq.~\eqref{eq:c_eff} explicitly evaluates to
\begin{equation}\label{eq:c_eff1}
\begin{split}
  c^{(1)}_{\mathrm{eff}}= & -\frac{3}{\pi^2}\left[(x+1) \text{Li}_2(-x)+(1-x) \text{Li}_2(x)+\right.\\&\left.\ln (x) ((1-x) \ln (1-x)+(x+1) \ln (x+1))\right],
  \end{split}
\end{equation}
where $x=\cosh^{-1}(2\alpha)$ and $\text{Li}_2$ denotes the dilogarithm function~\cite{abramowitz}. Figure~\ref{fig:X_centralcharge} shows the $\alpha$-dependence of $c^{(1)}_{\mathrm{eff}}$---which decays from the ideal value of 1/2 to zero as $\alpha$ increases (see caption for details).  Consequently, while for any finite $\alpha$ the von Neumann entanglement entropy still follows a logarithmic-law, its coefficient---the effective central charge---is diminished by protocol imperfection relative to Alice's original critical state.
This result agrees with that of Ref.~\onlinecite{JianMeasurementIsing}, where the mapping to a single-bond-defect Ising model was used. To our knowledge, the general expression in Eq.~\eqref{eq:c_eff} for $c_{\textrm{eff}}^{(n)}$ characterizing logarithmic growth of R\'enyi entropies with $n\neq 1$ for the state~\eqref{eq:X} has not been reported previously.  These coefficients similarly decay smoothly to zero as the protocol imperfection increases in a manner that depends (somewhat weakly) on $n$ and  yields a finite $\alpha$-dependent value at $n \rightarrow \infty$.  The situation should be contrasted to Eq.~\eqref{eq:pris_ent} for the pristine Ising theory, where a single parameter $c$ characterizes the logarithmic entanglement growth for any $n$.  We will return to these general R\'enyi entropies shortly from the perspective of the $\alpha \gg 1$ limit.

We now report the results of two-point correlation functions arising from marginally imperfect teleportation.   
In the limit $|i-j|\gg 1$ they read 
\begin{equation}\label{eq:correlators_critical2}
\begin{split}
&\langle X_iX_j \rangle_c = \frac{\text{sech}^2(2 \alpha )}{\pi ^2 |i-j|^2},\\
 &\langle Y_iY_j \rangle\propto \frac{1}{|i-j|^{2\Delta_Y(\alpha)}}, \,\,\,\,\Delta_Y(\alpha)=\frac{2[\arctan(e^{2\alpha})-\pi]^2}{\pi^2}\\
&\langle Z_iZ_j \rangle \propto  \frac{1}{|i-j|^{2\Delta_Z(\alpha)}},\quad  \Delta_Z(\alpha)=\frac{2\arctan^2(e^{2\alpha})}{\pi^2}.
\end{split}
\end{equation}
Compared to the correlations in Eq.~\eqref{eq:correlators_critical} operative for perfect teleportation, the following salient features arise: The power-law exponent for the connected $\langle X_iX_j \rangle_c$ correlator does not vary upon turning on non-zero $\alpha$, although the amplitude is suppressed and vanishes smoothly as $\alpha\to  \infty$.  Power-law exponents for two-point $Y$ and $Z$ correlators, by contrast, do depend on the degree of protocol imperfection: as $\alpha$ increases from $0$ to $\infty$, $\Delta_Y$ decreases from $9/8$ to $1/2$ while $\Delta_Z$ increases from $1/8$ to $1/2$.  DMRG simulations presented in Fig.~\ref{fig:X_correlations} confirm the scaling behavior predicted by Eq.~\eqref{eq:correlators_critical2}.  As an aside, although we focus on $\alpha\geq 0$, our analytical results for both the $c_{\rm eff}^{(n)}$ entanglement coefficients and the correlators in Eq.~\eqref{eq:correlators_critical2} apply for either sign of $\alpha$; the $c_{\rm eff}^{(n)}$ coefficients are invariant under $\alpha\rightarrow-\alpha$ while the scaling dimensions $\Delta_Z$ and $\Delta_Y$ are not.

Using the formalism from Sec.~\ref{sec:largealpha}, we can approximate Bob's penultimate wavefunction (again prior to applying the final outcome-dependent unitary) as
\begin{equation}
\label{eq:ZZ_measXbasis}
\ket{\psi_{\tilde{a}}}\approx\frac{1}{\sqrt{\mathcal{N}}} e^{-\frac{u^2}{2}\sum_{j\neq k}V_{jk}Z_jZ_k}\ket{\psi_B}.
\end{equation}
The unitary operator in Eq.~\eqref{psi_m_goal} is absent here since $q(j)$ vanishes by symmetry. 
In the non-unitary operator, the strange correlator $V_{jk}$ was evaluated in Ref.~\onlinecite{usmeasurementaltered} and shown to scale as $V_{jk} \sim 1/|j-k|$.  In Sec.~\ref{sec:largealpha} we argued that the perturbative expansion leading to Eq.~\eqref{eq:XX_measZbasis}---and the associated Hermitian parent Hamiltonian from Eq.~\eqref{Hau}---was expected to be controlled provided $V_{jk}$ decays \emph{faster} than $1/|j-k|$.  Nevertheless, we can recover asymptotic features present in our earlier non-perturbative computations by boldly calculating correlations and entanglement using the above large-$\alpha$ form of Bob's penultimate state.  Methods similar to Sec.~\ref{sec:relevant} and detailed further in Appendix~\ref{app:perturbative} show that $\langle Y_i Y_j \rangle$ and $\langle Z_i Z_j\rangle$ both scale as $u^2 V_{ij}$ whereas $\langle X_i X_j\rangle_c$ scale as $u^4 V_{ij}^2$.  Power-law exponents for the $Y$ and $Z$ correlators agree with the asymptotic $\alpha \rightarrow \infty$ limit of Eq.~\eqref{eq:correlators_critical2}; the $X$ correlator recovers the exact exponent as well as the leading large-$\alpha$ dependence of the prefactor in Eq.~\eqref{eq:correlators_critical2} upon using $u \approx e^{-\alpha}$.  We can, moreover, analytically calculate the $n = 2$ R\'enyi entropy upon Taylor expanding the exponential in Eq.~\eqref{eq:ZZ_measXbasis} (see Appendix~\ref{app:EE}).  To leading nontrivial order we obtain 
\begin{equation}\label{eq:pertS2}
S_R^{(2)}=\frac{4u^2}{\pi^2}\ln \ell+\mathcal{O}(1),
\end{equation}
which agrees with Eq.~\eqref{Z_kappa_n_0_1} in the $\alpha \gg 1$ regime.

To summarize, compared with relevant imperfection treated in Sec.~\ref{sec:relevant}, in the imperfect protocol studied here Bob inherits a far more faithful version of Alice's critical wavefunction: power-law correlations and long-range entanglement both persist, albeit with modifications that vary smoothly with $\alpha$.  In the next subsection we will see how resilience to imperfection can be further improved by following a protocol that employs yet a different quantization axis $\bm{n}$---though, as we will see, the improvement is not quite to a degree that a naive analysis would initially suggest.

\subsection{Disguised marginal imperfection}\label{sec:irrelevant}

\begin{figure*}[ht]
    \centering
    \includegraphics[width=\linewidth]{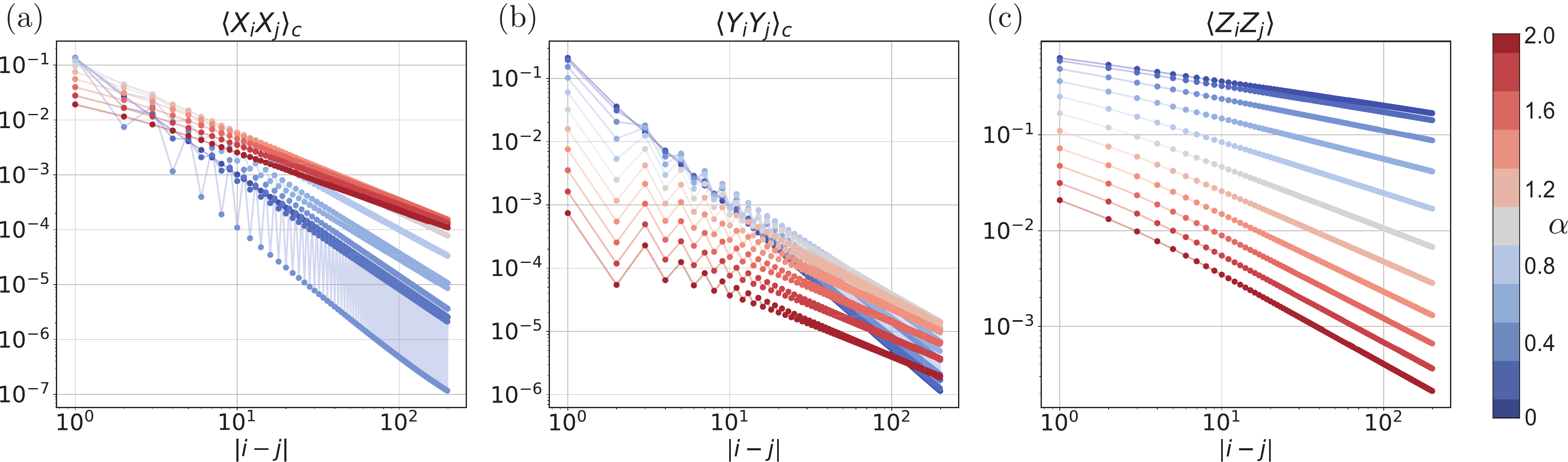} 
    \caption{\textbf{Correlations with disguised marginal imperfection.}  Correlations in the teleported state received by Bob [Eq.~\eqref{eq:teley}] following an imperfect protocol with $\bm{n} = \hat{\bm{y}}$ and with Alice's most probable measurement outcome.  Data points were obtained using infinite DMRG with for $\alpha \in [0,2]$ in steps of $0.2$. 
    Setup with Bob's initial wavefunction as a product state in the $\bm{\hat{y}}$ basis. All correlators exhibit continuously modified power-law exponents as the imperfection strength increases---in agreement with naively irrelevant imperfections generating marginal corrections under renormalization.  }
    \label{fig:Y_correlations}
\end{figure*}

Consider now a teleportation protocol with $\bm{n} = \hat{\bm{y}}$.  In this case we find that whenever the entangling gate $U=\prod_j e^{iuZ^A_jZ^B_j}$ is imperfect, Alice's most probable measurement outcomes are the two symmetry-related N\'eel states $|\tilde a,\hat{\bm{y}}\rangle$ with $a_j = \pm (-1)^j$ (at least for an even number of sites with periodic boundary conditions \footnote{With an odd number of sites, there are $\mathcal{O}(L)$ degenerate states corresponding to N\'eel states with a single local ferromagnetic bond.  We only consider even system sizes here.}).  For these outcomes Bob's final wavefunction is
\begin{equation}\label{eq:teley}
\ket{\psi^{\mathrm{tele}}_{\tilde{a}}}= \frac{1}{\sqrt{\mathcal{N}}}e^{\alpha M_Y^s}\ket{{\psi}_c},
\end{equation}
where we defined the staggered $Y$ magnetization $M_Y^s = \sum_j (-1)^j Y_j/2$.  What is the appropriate defect-line perturbation $\delta S$ generated by the non-unitary operator above at small $\alpha$?  Given that the low-energy expansion $Y_j \sim i \partial_\tau \sigma$ contains only smoothly varying field components, the $(-1)^j$ in the staggered magnetization $M_Y^s$ appears to imply that only highly oscillatory---and thus strongly irrelevant---terms appear in $\delta S$.  Actually, had we post-selected for (lower-probability) uniform outcomes rather than staggered measurement outcomes, the $(-1)^j$ oscillatory factor would disappear, yielding $\delta S \sim \alpha \int_x i \partial_\tau \sigma(x,\tau = 0)$---which is still irrelevant.  One might be tempted to conclude then that, for either the uniform or staggered measurement outcomes, Bob faithfully inherits the universal features of Alice's critical correlations and entanglement over at least a finite window of protocol imperfection.  

Such a conclusion would be incorrect, however.  Under renormalization, additional symmetry-allowed terms will be generated that may be more relevant than those in the `bare' theory (see, e.g., Ref.~\onlinecite{Starykh2007}).  In our context, with either the uniform or staggered measurement outcomes, symmetry allows generation of the \emph{marginal} term
\begin{equation}
  \delta S' \sim \alpha^2 \int_x \varepsilon(x,\tau = 0)
  \label{deltaSp}
\end{equation}
at $\mathcal{O}(\alpha^2)$.  For instance, one can view $\alpha$ in Eq.~\eqref{eq:teley} as odd under time reversal $\mathcal{T}$, $\mathbb{Z}_2$ spin-flip symmetry, and single-site translations; $\alpha^2$ is then invariant under all symmetries, so that $\delta S'$ is indeed symmetry-allowed.    
For another perspective on Eq.~\eqref{deltaSp} let us examine further the uniform measurement outcome. Using Lorentz invariance of the unperturbed continuum theory, we can Wick rotate to interchange imaginary time and space to obtain a defect-line action that acts at one spatial position $x = 0$ but all $\tau$.  This Wick-rotated defect describes a conventional static quantum impurity that microscopically descends from a critical Hamiltonian with a $Y$ perturbation of strength $\alpha J$ at a single site:
\begin{equation}
    H_{\textrm{imp}} = -J\bigg(\sum_j Z_jZ_{j+1}+\sum_{j\neq 0}X_j\bigg) 
    - J(X_0 + \alpha Y_0). 
\end{equation}
Note that we grouped the transverse field and the $Y$ perturbation at the impurity site, $j = 0$.  Locally rotating the $j = 0$ spin about $Z$ by an angle $\theta$ satisfying $\tan(2\theta) = \alpha$ yields the unitarily equivalent form
\begin{align}
    e^{i\theta Z_0}H_{\textrm{imp}}e^{-i\theta Z_0} &= -J\bigg(\sum_j Z_jZ_{j+1}+\sum_{j\neq 0}X_j\bigg)
    \nonumber \\
    &-J \sqrt{1+\alpha^2}X_0
    \approx H_c - \frac{J}{2} \alpha^2 X_0.
\end{align}
On the bottom right we assumed $\alpha \ll 1$ to arrive at the uniform critical Ising Hamiltonian $H_c$ from Eq.~\eqref{eq:Hamiltonian_Ising} perturbed solely by a local $\mathcal{O}(\alpha^2)$ shift in the transverse field.  In this form it is clear that the original $\mathcal{O}(\alpha)$ $Y$ perturbation indeed germinates an $\mathcal{O}(\alpha^2)$ $\varepsilon$ perturbation in the low-energy theory, as previously deduced on symmetry grounds.  

Because a marginal perturbation appears only upon a careful inspection that accommodates higher-order terms, we refer to this situation as `disguised marginal imperfection'.  Section~\ref{sec:marginal} already analyzed the impact of a marginal defect-line action on the teleported state.  Here we similarly expect that imperfection smoothly modifies long-distance entanglement and correlations away from their ideal universal forms, but to a still less dramatic degree at $\alpha \ll 1$ since corruption kicks in only at $\mathcal{O}(\alpha^2)$.  
Figure~\ref{fig:Y_correlations} presents two-point correlators obtained with DMRG assuming the most-likely staggered measurement outcome and confirms this expectation.  Indeed, in the small-$\alpha$ regime, the correlations in Fig.~\ref{fig:Y_correlations} behave qualitatively similarly to those for the $\bm{n} = \hat{\bm{x}}$ protocol in Fig.~\ref{fig:X_correlations}.  [The main difference arises for two-point $X$ correlators, whose power-law exponent does now evolve nontrivially with $\alpha$. This feature can be understood from the large-$\alpha$ limit.  We observe that in the present case $q(j)$ is non-zero in Eq.~\eqref{psi_m_goal}, which produces different behavior with respect to measuring in the $\hat{\bm{x}}$-basis.] 

\begin{figure}[ht]
    \centering
    \includegraphics[width=\linewidth]{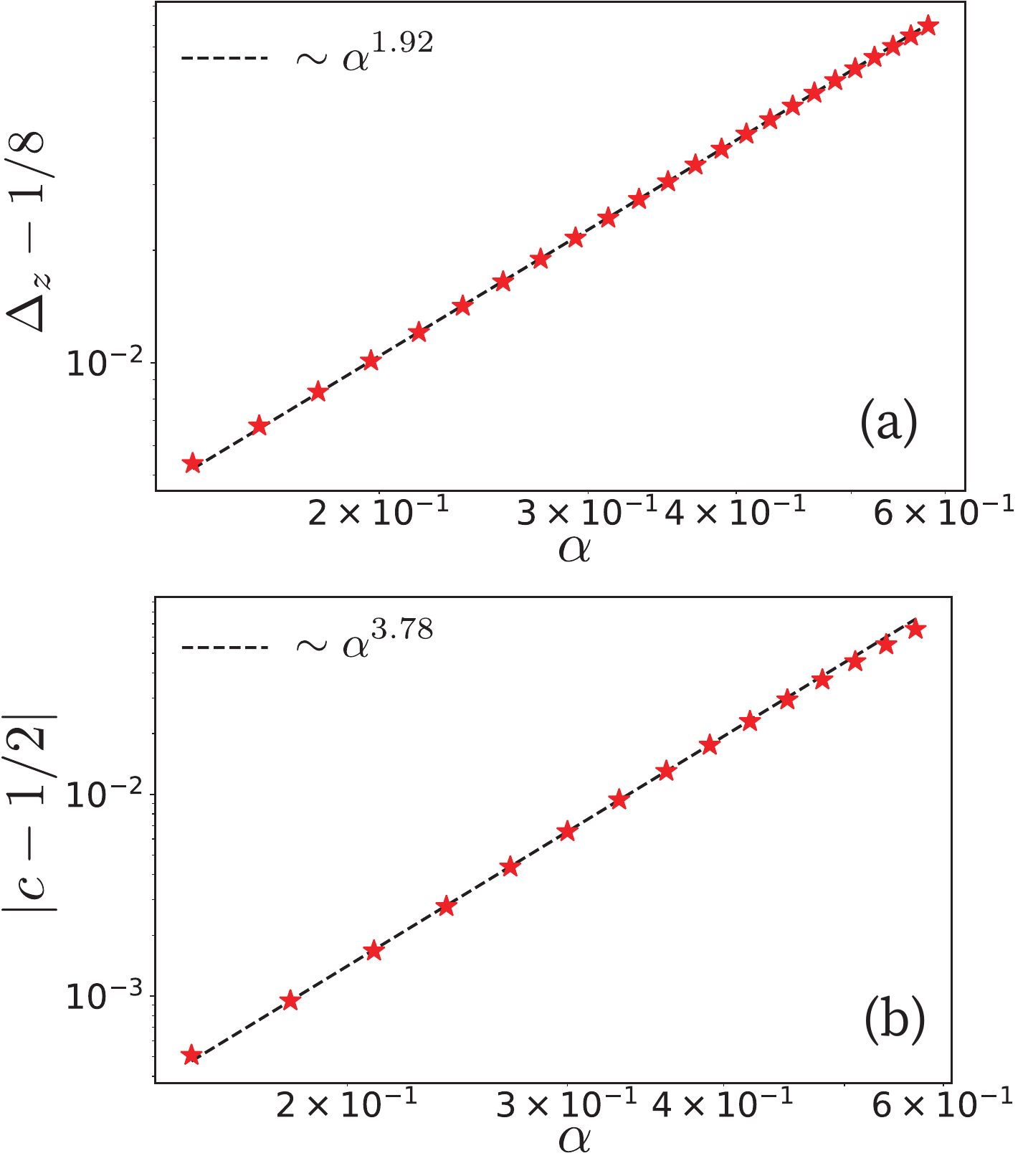}
    \caption{\textbf{Scaling dimension and effective central charge with disguised marginal imperfection.} (a) Power-law dependence of the scaling dimension of $Z$ as a function of $\alpha$. Red stars represent data obtained from infinite DMRG [Fig.~\ref{fig:Y_correlations}(c)], while the dashed line is a power-law fit $\Delta_z (\alpha)-1/8 \sim \alpha^{1.92}$.  The fit agrees well with the $\alpha^2$ scaling expected from the generation of a marginal term at second order in the imperfection strength $\alpha$; see Eq.~\eqref{deltaSp}.  (b) Dependence of the effective central charge on $\alpha$. Red stars represent the data obtained using finite DMRG from fitting $S_{L/2}$ vs $\ln L$ for $L = [20, 40, 60, 80, 100, 150, 200]$. Dashed line represents a power-law fit $|c-1/2|\sim \alpha^{3.78}$---which agrees reasonably well with $\alpha^4$ scaling expected from an $\mathcal{O}(\alpha^2)$ marginal imperfection term.} 
    \label{fig:Y_Ent_ceff}
\end{figure}

To further bolster our predictions, Fig.~\ref{fig:Y_Ent_ceff}(a) presents the imperfection-induced change in the $\langle Z_i Z_j\rangle \sim 1/|i-j|^{2\Delta_Z}$ scaling dimension relative to the ideal value of $\Delta_Z = 1/8$.  We numerically find the relation $\Delta_Z(\alpha)-1/8\sim \alpha^{1.92}$ at small $\alpha$, in good agreement with the scaling $\Delta_Z(\alpha)-1/8\sim \alpha^2$ that follows from Eq.~\eqref{deltaSp}.  One can also recover that scaling directly from Eq.~\eqref{eq:correlators_critical2} upon sending $\alpha \rightarrow \alpha^2$.
Figure~\ref{fig:Y_Ent_ceff}(b)  shows that the change in the effective central charge characterizing logarithmic entanglement growth scales approximately as $|c^{(1)}_{\textrm{eff}}-1/2|\sim \alpha^4$---as expected from Eq.~\eqref{eq:c_eff1} upon similarly replacing $\alpha \rightarrow \alpha^2$. 
Therefore, protocols with $\bm{n} = \hat{\bm{y}}$ yield additional resilience to weak imperfection when post-selecting for the highest-probability measurement outcomes, despite universal features still undergoing modifications.  

\begin{figure}[ht]
    \centering
    \includegraphics[width=\linewidth]{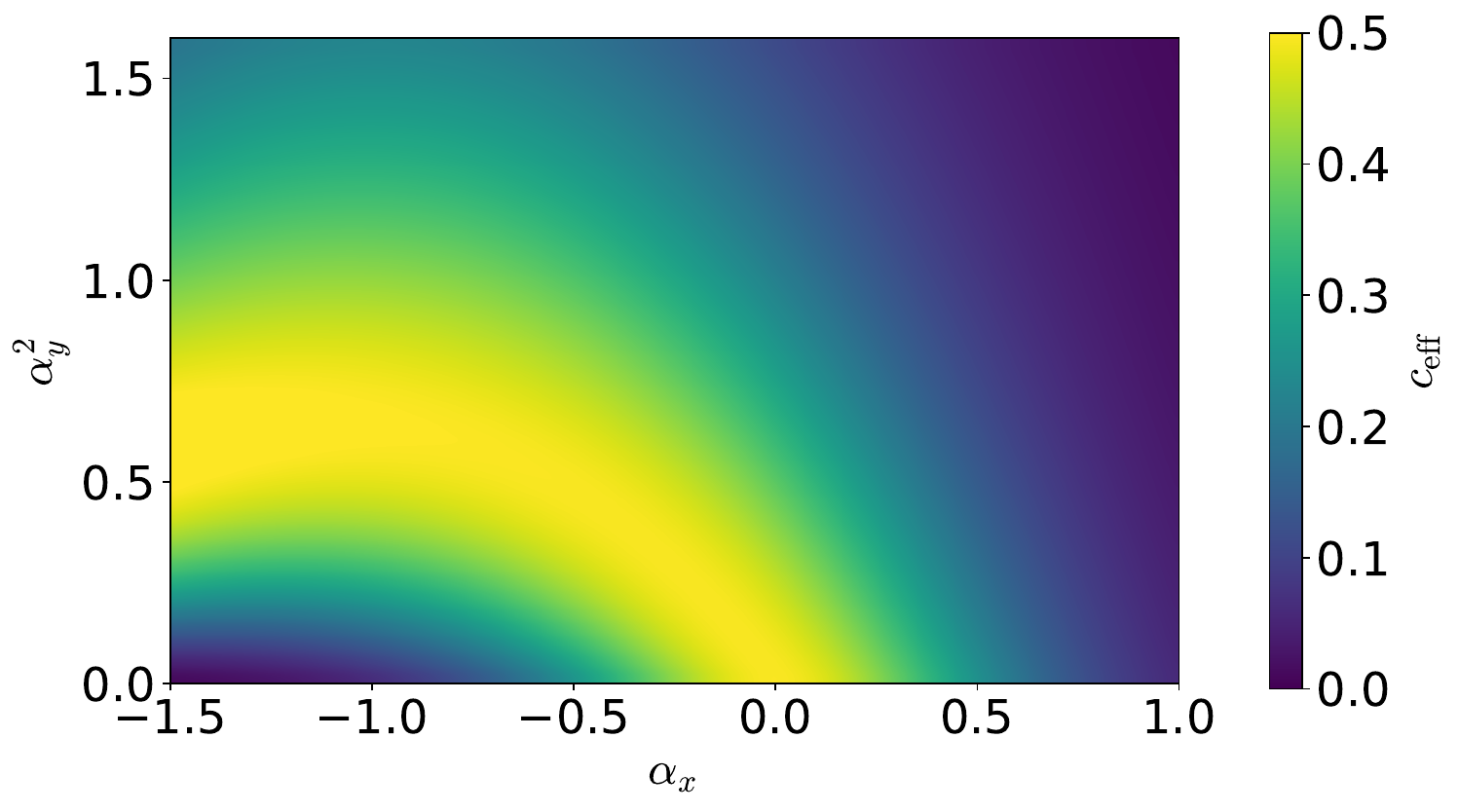}
    \caption{\textbf{Cancellation of marginal imperfections.} 
    Effective central charge for the generalized teleported state in Eq.~\eqref{eq:teley2} as a function of $\alpha_x$ and $\alpha_y^2$. Data were obtained using infinite DMRG (see Appendix \ref{app:dmrg} for details).  The bright yellow arc highlights fine-tuned parameters for which imperfection-induced marginal terms cancel---in turn producing a teleported state with pristine central charge $c = 1/2$.  Note the persistence of this fine-tuned arc even for sizable imperfection strengths $\alpha_{x,y}$. 
 }
    \label{fig:optimal}
\end{figure}

Further improvement is technically possible even when post-selecting for similar measurement outcomes.  Suppose that we generalize Eq.~\eqref{eq:teley} to 
\begin{equation}\label{eq:teley2}
\ket{\psi^{\mathrm{tele}}_{\tilde{a}}}= \frac{1}{\sqrt{\mathcal{N}}}e^{\alpha_x M_X^+ + \alpha_y M_Y^s}\ket{{\psi}_c}.
\end{equation}
The above form of Bob's final state arises upon replacing $\bm{n} = \hat{\bm{y}}$ with a position-dependent quantization axis $\bm{n}_j =(-1)^j \sin\varphi \hat{\bm{x}} + \cos\varphi \hat{\bm{y}}$ and again post-selecting for a staggered outcome with $a_j = \pm(-1)^j$. In the weak-imperfection limit, we have seen that both $M_X^+$ and $M_Y^s$ generate marginal $\varepsilon$ defect-line perturbations---the former at $\mathcal{O}(\alpha_x)$ and the latter at $\mathcal{O}(\alpha_y^2)$. 
When the two $\varepsilon$ contributions come with opposite sign, one can cancel off the marginal term by fine-tuning $\alpha_{x,y}$ to values satisfying $|\alpha_x| \sim \alpha_y^2$.  Here the imperfectly teleported state retains the quantized central charge $c = 1/2$ and power-law correlations characteristic of the pristine Ising theory.  By continuity this fine-tuned limit must persist also in the limit where $\alpha_{x,y}$ are not necessarily weak.  We support this scenario in Fig.~\ref{fig:optimal}, which shows the $\alpha_{x,y}$ dependence of the effective central charge of Eq.~\eqref{eq:teley2} obtained using DMRG.  An extended region with $c = 1/2$ is clearly visible in agreement with our predictions.  Although this limit is doubly fine-tuned in the sense of requiring both post-selection and tuning of $\alpha_x$ given some $\alpha_y$, we learn two lessons: First, it serves as a nontrivial check on our theory of disguised marginal imperfection.  And second, it shows that in principle one can teleport the ideal long-distance universal Ising criticality properties even with simple, spatially periodic measurement outcomes and with superficially `large' protocol imperfections (the $c = 1/2$ line in Fig.~\ref{fig:optimal} indeed persists to sizable $|\alpha_x|$ values where one might naively anticipate dramatically altered critical properties).

\subsection{Imperfect teleportation with typical measurement outcomes} 
\label{sec:typical}

\begin{figure*}[ht]
    \centering
    \includegraphics[width=\linewidth]{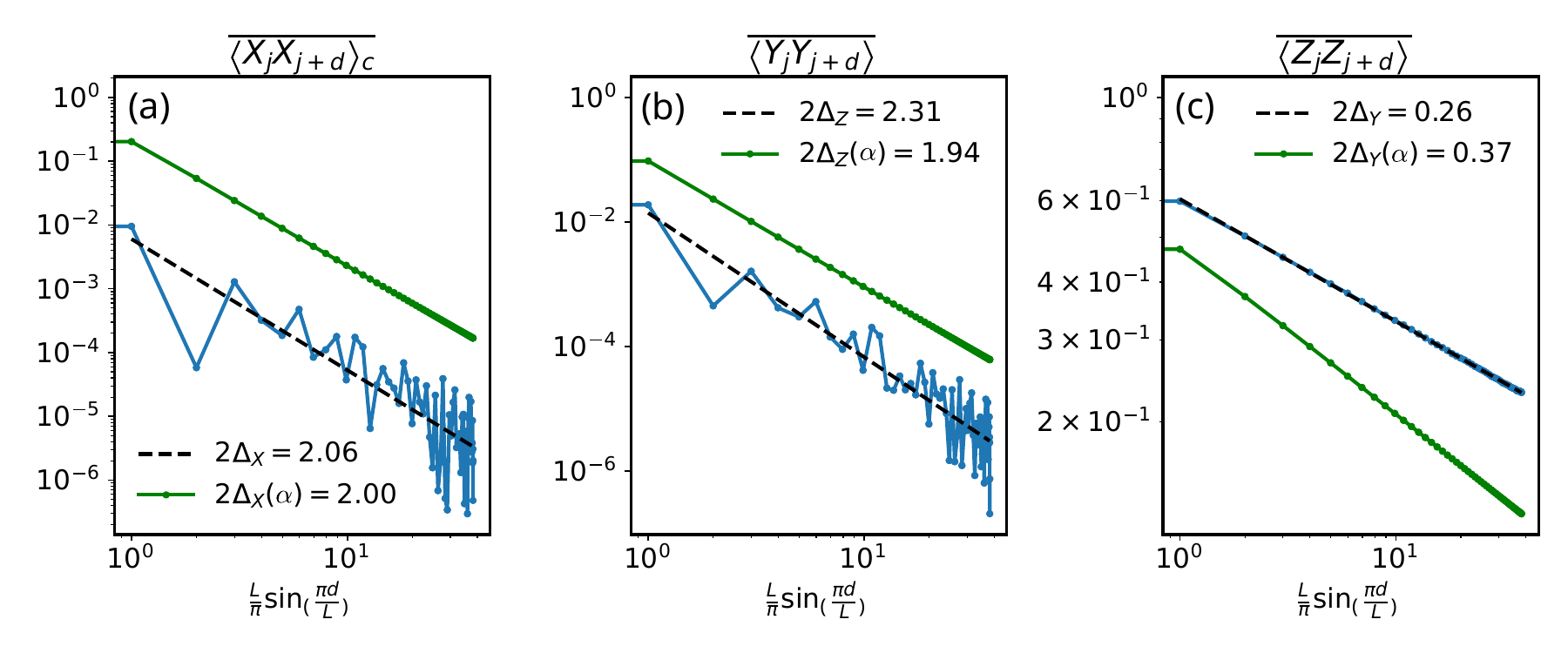}
    \caption{\textbf{Correlations with typical measurement outcomes.} Correlations in the teleported state received by Bob following an imperfect protocol with $\bm{n} = \hat{\bm{x}}$, now with a typical measurement outcome randomly sampled from Born's probability distribution as prescribed in Ref.~\onlinecite{JianMeasurementIsing}.  Data were obtained using finite DMRG with bond dimension $\chi=800$ and system size $L=120$ with periodic boundary conditions.  
    Results shown in blue correspond to $u=0.7$ ($\alpha\approx 0.17$) after averaging the correlation functions over the center of mass coordinate, i.e., $\overline{\langle \hat{O}_j  \hat{O}_{j+d}\rangle}=\sum_{j=1}^L \langle \hat{O}_{j}\hat{O}_{j+d}\rangle/L$. For comparison, the green points show correlations obtained with Alice's most probable measurement outcome for the same value of $u$.  Contrary to the latter case, with typical outcomes imperfections are irrelevant, and hence the averaged correlators are consistent with power laws expected for pristine Ising criticality.}
    \label{fig:typical}
\end{figure*}

Above we focused on the effects of an imperfect teleportation protocol when Alice obtains her most likely measurement outcomes.  However, the probabilities for such outcomes, while maximal, are still exponentially small in the number of teleported qubits.  Hence, it is important to investigate the quality of teleportation in the case of \emph{typical} measurement outcomes $\tilde a$---i.e., those appearing when sampling according to the ($\alpha$-dependent) Born probability distribution $p_{\tilde a}$ corresponding to the state $\ket{\psi_U} = U \ket{\psi_c}|\tilde{b},\bm{n}\rangle$ obtained after Alice and Bob entangle their qubits.  Reference~\onlinecite{AltmanMeasurementLL} already addressed a related question when probing measurement-altered criticality in a Luttinger liquid.  We will repeat and adapt their findings to our teleportation protocols with quantization axis/measurement basis $\bm{n}$.

Let $\tilde a$ now denote a typical outcome consisting of bits $a_j = \pm1$ associated with the Pauli operators $O^A_j({\bm n})$ that Alice measures.  We would like to use a coarse-graining procedure to take the continuum limit of the non-unitary operator $e^{-\frac{\alpha}{2} H_{\tilde a}(\bm{n})}$ encoding teleportation-protocol imperfection, obtain a corresponding defect-line action in the $\alpha \ll 1$ regime, and analyze its effect on the long-distance properties of Bob's final wavefunction.   For a perfect teleportation protocol with $\alpha = 0$, all outcomes are equally likely, and hence the $a_j$ bits are completely uncorrelated.  An imperfect protocol with $\alpha \neq 0$---which we assume hereafter---biases the measurement-outcome distribution, yielding correlations among the bits that are inherited from structure in the pre-measurement entangled state $\ket{\psi_U}$.  For instance, averaging a given $a_j$ over measurement outcomes gives
\begin{equation}
     \overline{a}_j = \sum_{\tilde a} p_{\tilde a} a_j = \cos(2u) \bra{\psi_c}\hat{O}_j^A({\bm n})\ket{\psi_c},
\end{equation}
while similarly averaging $a_j a_k$ for $j \neq k$ yields the connected correlation
\begin{align}
    \overline{a_j a_k}- \overline{a}_j\overline{a}_k = \cos^2(2u) \bra{\psi_c}\hat{O}_j^A({\bm n})\hat{O}_k^A({\bm n})\ket{\psi_c}_c.
    \label{eq:average_random}
\end{align}
(Both quantities above vanish in the perfect protocol limit $u = \pi/4$ as required.)  Equation~\eqref{eq:average_random} in particular limits the spatial extent of correlations among bit strings in a typical measurement outcome: on average, they decay at long distances according to the leading power-law contribution $\sim 1/|j-k|^{2\Delta_{\rm sl}}$ arising from the critical correlation of Alice's initial state on the right-hand side.

Following Kadanoff's decimation argument~\cite{Kadanoff}, next we  analyze the block-spin variable 
$a(x;b)=1/b\sum_{j=x-b/2}^{x+b/2}a_j$, which reduces the number of degrees of freedom by grouping the $\tilde{a}$ variables within a block of size $b\gg1$.  Intuitively, the more rapidly correlations among different $a_j$'s decay with their separation, the faster the block-spin variable averages to zero as $b$ increases. One can estimate the decay rate with $b$ for a typical measurement outcome by examining the variance of $a(x;b)$: 
\begin{align}
    \overline{[a(x;b)]^2} - [\overline{a}(x;b)]^2 &= \frac{1}{b^2} \sum_{j,k = x-b/2}^{x+b/2} (\overline{a_j a_k}  - \overline{a}_j \overline{a}_k)
    \nonumber \\
    & \!\!\!\!\!\!\!\!\!\!\!\!\!\!\!\!\!\!\!\!\!\!\!\!\! \propto \frac{1}{b^2} \sum_{j,k = x-b/2}^{x+b/2}\bra{\psi_c}\hat{O}_j^A({\bm n})\hat{O}_k^A({\bm n})\ket{\psi_c}_c.
\end{align}
For large $b$, in the regime $\Delta_{\rm sl} < 1/2$ the second line scales as $\frac{1}{b^2} \times b^{2(1-\Delta_{\rm sl})} = 1/b^{2\Delta_{\rm sl}}$ due to the power-law contribution from the long-distance $|j-k| \gg 1$ part of the sums.  With $\Delta_{\rm sl}> 1/2$, however, the long-distance part vanishes sufficiently quickly that the short-distance part dominates, yielding the faster scaling $\frac{1}{b^2} \times b = 1/b$.  Thus we obtain
\begin{equation}
   \overline{[a(x;b)]^2} - [\overline{a}(x;b)]^2 \sim \begin{cases}
        b^{-2\Delta_{\mathrm{sl}}},\qquad &\Delta_{\mathrm{sl}}<1/2 \\
        b^{-1},\qquad &\Delta_{\mathrm{sl}}>1/2
    \end{cases}.
    \label{eq:var}
\end{equation}
Notice that the scaling with $b$ in the lower line is the same as one would obtain at $\alpha = 0$, where all measurement outcomes are equally likely and the $a_j$'s at different sites are completely uncorrelated.  The key upshot is that $a(x;b)$ realizes a smoothly varying function whose scaling dimension follows from the square root of Eq.~\eqref{eq:var}; i.e., the dimension is either $\Delta_{a} = \Delta_{\rm sl}$ (for $\Delta_{\rm sl} < 1/2$) or $\Delta_{a} = 1/2$ (for $\Delta_{\rm sl} > 1/2$).

We can now take the continuum limit of $e^{-\frac{\alpha}{2} H_{\tilde a}(\bm{n})}$.  Recalling the definition of $H_{\tilde a}(\bm{n})$ in Eq.~\eqref{Han_def}, the coarse-grained function $a(x;b)$ couples to the low-energy expansion of $\hat{O}_j(\bm{n})$ given by the dictionary in Eq.~\eqref{eq:dictionary}.  Hence, if $\hat{O}_j(\bm{n}) \sim O_{\rm sl}$ is the leading contribution (which has the same scaling dimension $\Delta_{\rm sl}$ from earlier), then the corresponding defect-line action reads 
\begin{equation}
  \delta S \sim \alpha \int_x a(x;b) O_{\rm sl}(x,\tau = 0).
  \label{deltaS_typ}
\end{equation}
For $\Delta_{\rm sl} < 1/2$, the integrand has scaling dimension $\Delta_{\rm sl} + \Delta_{\rm sl} < 1$---implying that $\delta S$ constitutes a relevant perturbation.  For $\Delta_{\rm sl} > 1/2$, in contrast, the dimension becomes $1/2 + \Delta_{\rm sl} > 1$ so that $\delta S$ is then irrelevant.  

Teleportation protocols with $\bm{n} = \hat{\bm{x}}$ yield $O_{\rm sl} = \varepsilon$ and $\Delta_{\rm sl} = 1$.  Post-selecting for typical measurement outcomes thereby demotes the marginal defect-line action studied in Sec.~\ref{sec:marginal} to an irrelevant perturbation.  At least on sufficiently long length scales, Bob inherits Alice's universal correlations and entanglement despite imperfection in the protocol.  We numerically tested this prediction by computing two-point $X$, $Y$, and $Z$ correlators in Bob's final state obtained for an imperfect entangling gate with $u = 0.7$ and when Alice measures a typical outcome randomly sampled according to Born's probability distribution.  Specifically, following Ref.~\onlinecite{JianMeasurementIsing} we measured one of Alice's qubit (sampled according to Born's rule), obtained the resulting post-measurement state, then moved on to measure her next qubit, and so on. Our results for the spatially averaged quantities $\sum_{j=1}^L\langle O_j O_{j+d} \rangle/L$ with $O_j=X,Y$ and $Z$ are shown in the blue curves from Fig.~\ref{fig:typical}; for reference, the green curves correspond to correlators obtained with Alice's most probable outcome (Sec.~\ref{sec:marginal}).  
The blue curves in the $X$ and $Y$ panels appear noisier than for $Z$.  We attribute this behavior to their faster power-law decay, which in turn suppresses averaging out of the randomness from the measurement outcome.  
Nevertheless, in all three panels, power-law behavior for the typical outcome is indeed compatible with the results of Eq.~\eqref{eq:correlators_critical} obtained for the pure Ising CFT.

Using similar logic, $\delta S$ is also irrelevant for protocols with $\bm{n} = \hat{\bm{y}}$.  In Sec.~\ref{sec:irrelevant} we saw for the highest-probability outcome that an irrelevant term appearing at leading order in $\alpha$ generated a marginal $\varepsilon$ term [Eq.~\eqref{deltaSp}] at $\mathcal{O}(\alpha^2)$.  With typical outcomes, we also expect an $\varepsilon$ term to be generated---but now with a prefactor $[\alpha a(x;b)]^2$.  The appearance of $a(x;b)$ in particular renders that second-order term irrelevant as well.  Thus protocols with $\bm{n} = \hat{\bm{y}}$ also enjoy resilience against imperfection in the case of typical measurement outcomes.  

When $\bm{n} = \hat{\bm{z}}$, however, imperfection continues to yield dramatic consequences even for typical outcomes.  Here $O_{\rm sl} = \sigma$ and $\Delta_{\rm sl} = 1/8$, and consequently $\delta S$ is relevant (albeit less so compared to the case of uniform measurement outcomes).  Here we expect $\langle Z_j\rangle$ to take on a non-zero, position-dependent expectation value that roughly follows the sign of the imperfection-induced longitudinal-field-like perturbation encoded in $\delta S$.  Two-point correlators and entanglement can be attacked from the large-$\alpha$ limit using the parent Hamiltonian formalism from Sec.~\ref{sec:largealpha}.  We saw in Sec.~\ref{sec:relevant} that this formalism captures both power-law correlations and area-law entanglement when post-selecting for the highest-probability measurement outcome; recall Eq.~\eqref{eq:parent_Z} and the accompanying discussion.  With typical outcomes, the main difference stems from the form of the strange correlator $V_{jk}$ that governs `weak' long-range interactions in the parent Hamiltonian.  At large separation $|j-k|$, the mean of $V_{jk}$ is exactly $0$, while the variance of $V_{jk}$ vanishes as $\sim 1/|j-k|^4$ (see Ref.~\onlinecite{usmeasurementaltered} for the computation).  This relatively fast power-law decay suggests that $(i)$ the perturbative treatment underlying the large-$\alpha$ analysis is valid, $(ii)$ the entanglement entropy exhibits area-law behavior also for typical outcomes, and $(iii)$ spatially averaged spin-spin correlations decay as the square root of the variance of $V_{jk}$, i.e., as $\sim 1/|j-k|^2$.  (Verifying this power-law decay numerically, however, is challenging.)   We conclude that regardless of whether Alice obtains the most probable measurement outcome or a typical outcome, even weak imperfection in protocols using $\bm{n} = \hat{\bm{z}}$ qualitatively alters the universal correlations and entanglement of the Ising critical state undergoing teleportation.

\section{Generalized protocol} \label{sec:general_main}

In our imperfect teleportation protocol treated so far we restricted to entangling on-site unitaries $U=\prod_j U_j$ with $U_j$ generated by the tensor product of two Pauli operators $\hat{O}^A_j(\bm{n}^{\perp})\hat{O}^B_j(\bm{n}^{\perp})$.  Recall that $\bm{n}^\perp$ is orthogonal to the vector $\bm{n}$ specifying the basis for Alice's measurement and Bob's initialization.  We will now generalize the protocol to incorporate an additional possible imperfection by assuming that Alice and Bob entangle their qubits via 
\begin{equation}
U_j=e^{iu\hat{O}^A_j(\bm{m})\hat{O}^B_j(\bm{m})},
 \label{Ugeneral}
\end{equation} 
where $\bm{m}$ is now a general unit vector not necessarily orthogonal to $\bm{n}$.  As in the previous, more restricted protocol, Bob is allowed to perform a unitary dependent on Alice's measurement outcome $\tilde a$ (as well as $\tilde b, \bm{n}$, and $\bm{m}$) to obtain the final teleported state.  
In what follows we discuss how the quality of teleportation is altered by imperfections in both the entangling gate strength (parametrized by $u$) and `orientation' (parametrized by $\bm{m}$). Other error sources may of course also arise and can be studied similarly. 

We first examine the structure of Bob's penultimate wavefunction $\ket{\psi_{\tilde{a}}}$, prior to applying the final outcome-dependent unitary. 
With details consigned to Appendix~\ref{sec:general}, the expression---valid for an arbitrary initial state $\ket{\psi_A}$ for Alice---can be written as 
\begin{equation}
    \begin{aligned} \label{eq:gen_psis_main}
        \ket{\psi_{\tilde{a}}}&=\frac{1}{\sqrt{\mathcal{N}}}e^{i\frac{\pi}{4}H_{\tilde{b}}(\bm{n})} e^{-\frac{\alpha}{2}H_{\tilde{b}}(\bm{n})}\\ & \times  \mathcal{U}_{\tilde{b}}(\bm{m}\times\bm{n}^\perp)U_{\tilde{b}\leftarrow \tilde{a}}\mathcal{U}_{\tilde{a}}^\dagger(\bm{m}\times\bm{n}^\perp)\ket{\psi_A}.
    \end{aligned}
\end{equation}
The first two exponentials involve the Hermitian operator $H_{\tilde{b}}(\bm{n})=-\sum_j b_j \hat{O}_j(\bm{n})$, while   
$\mathcal{U}_{\tilde{d}}(\bm{m}\times\bm{n}^\perp)$ (for $d = a,b$) are on-site transformations that are in general non-unitary.  They take the explicit form
$\mathcal{U}_{\tilde{d}}(\bm{m}\times\bm{n}^\perp)=\prod_j[e^{i\theta \frac{(\bm{m}\times\bm{n}^\perp)}{||\bm{m}\times\bm{n}^\perp||}\cdot \bm{\sigma}_j}P^d_j+\mathds{1}(1-P^d_j)]$, where $\theta$ is the angle between $\bm{m}$ and $\bm{n}^\perp$, and $P^d_j$ are projectors defined as $P^d_j = \frac{1}{2}[1-\tilde{d}_j \hat{O}_j(\bm{n})]$.
Thus the $\mathcal{U}$ factors in Eq.~\eqref{eq:gen_psis_main} either act as the identity, or rotate the spins about the axis $\bm{m}\times\bm{n}^\perp$.  Notice that these rotations become trivial if $\bm{m}$ was chosen parallel to $\bm{n}^\perp$ (as arose in our previously treated protocols). When they act nontrivially, the projectors $P_j^d$ correct the `mismatch' in Eq.~\eqref{Ugeneral} that prevents $U_j$ from performing a full flip from $a_j\to -a_j$ and/or $b_j\to -b_j$.  
With these definitions, we can view the first three operators acting on $\ket{\psi_A}$ in Eq.~\eqref{eq:gen_psis_main} as follows:  First, $\mathcal{U}^\dagger_{\tilde{a}}(\bm{m}\times\bm{n}^\perp)$ `readjusts' the $\bm{m}$ direction such that it lies along $\bm{n}^\perp$.  Next, $U_{\tilde{b}\leftarrow \tilde{a}}=e^{i\frac{\pi}{4}H_{\tilde{b} \leftarrow \tilde{a}}(\bm{n}^\perp)}$ is a unitary that flips the measurement outcome $\tilde{a}$ to $\tilde b$ as in Eq.~\eqref{eq:compact}. 
Finally, $\mathcal{U}_{\tilde{b}}(\bm{m}\times\bm{n}^\perp)$ readjusts $\bm{m}$ by rotating it into the $\bm{n}^\perp$ direction.

Deducing the fate of the imperfectly teleported wavefunction   
as written in Eq.~\eqref{eq:gen_psis_main} is nontrivial since a series of interspersed unitary and non-unitary operators acts on $\ket{\psi_A}$.  A more useful form arises upon re-organizing the expression such that unitary factors act to the left of manifestly Hermitian, non-unitary factors.  We will not attempt a completely general rewriting of this type, but rather examine the situation where $\bm{m}$ is weakly misaligned but close to $\bm{n}^\perp$. In this limit one can write, after dropping an unimportant overall phase, 
\begin{multline}
    \mathcal{U}_{\tilde d}(\bm{m}\times \bm{n}^{\perp}) \approx  e^{\frac{i}{2} \sum_j \hat{O}_j(\bm{m} \times \bm{n}^\perp)} \\ \times e^{\frac{(\bm{m} \cdot\bm{n})}{2}\sum_jd_j \hat{O}_j(\bm{n}^\perp)}.
\end{multline}
Using this approximation and allowing Bob to undo an overall unitary operator, we obtain the final (unnormalized) teleported state
\begin{multline}\label{eq:gen2}
    \ket{{\psi}^{\rm tele}_{\tilde{a}}}= e^{-\frac{\alpha}{2}H_{\tilde{a}}(\bm{n})+\frac{\alpha}{2}\sum_j\epsilon_{-,j}\hat{O}_j(\bm{n}^\perp)}\\
    \times e^{\frac{1}{2}\sum_j\epsilon_{+,j}\hat{O}_j(\bm{n}^\perp)}\ket{\psi_A},
\end{multline}
where $\epsilon_{\pm,j}=(b_j\pm a_j)(\bm{m}\cdot\bm{n})$ are small factors originating from the new source of imperfection that we have allowed.  

Let us now specialize to the case $\ket{\psi_A} = \ket{\psi_c}$ so that $\ket{\psi_{\tilde a}^{\rm tele}}$ forms an imperfectly teleported Ising critical wavefunction. 
Section~\ref{sec:ising_teleported} showed that, for our previously studied protocol, orienting $\bm{n}$ in the $(x,y)$ plane optimized resilience to errors in the sense that imperfections were generically marginal (for the highest probability outcomes) or irrelevant (for typical outcomes).  Equation~\eqref{eq:gen2} reveals that misaligning $\bm{m}$ away from $\bm{n}^\perp$---which importantly would orient along $\hat{\bm{z}}$ for such $\bm{n}$'s---supplements the non-unitary operator with a $Z$ component that comprises a \emph{relevant} imperfection.  The heart of the issue is that misalignment of $\bm{m}$ with this otherwise optimal choice of $\bm{n}$ causes the entangling unitary to violate $\mathcal{T}\times \mathbb{Z}_2$ symmetry (contrary to the case where the unitary only involves $Z$ operators).  This symmetry reduction trickles down to the final teleported state via generation of a $\sigma$ perturbation to the defect-line action that is no longer symmetry-forbidden.  Zooming out, we anticipate that the effects of a still broader class of possible protocol imperfections can be analogously classified on symmetry grounds.

\section{Average teleported mixed-state}\label{sec:decoding}

The results of Sec.~\ref{sec:ising_teleported} quantify the character of an imperfectly teleported quantum critical state following a \emph{single} protocol implementation.  In particular, we saw that the role of imperfections depends on the quantization axis $\bm{n}$ used for initialization and measurement, as well as Alice's particular measurement outcome.  Suppose now that Alice and Bob repeat an (identically) imperfect teleportation protocol many times, and that after each iteration, Bob measures observables with respect to the final 
teleported state $\ket{\psi_{\tilde a}^{\rm tele}}$ in Eq.~\eqref{eq:compact_2}.  Upon averaging over Alice's measurement outcomes $\tilde a$, 
 are the nontrivial effects that we captured for a single implementation still observable?  Or do imperfections average out, allowing Bob to  sample Alice's pristine quantum critical state even if each implementation is flawed? 

To address these questions, we examine the mixed-state density matrix
\begin{equation}
  \rho^{\rm tele} = \sum_{\tilde{a}} p_{\tilde{a}}\ket{\psi_{\tilde{a}}^{\rm tele}}\bra{\psi_{\tilde{a}}^{\rm tele}}
  \label{rhotele}
\end{equation}
describing an ensemble of imperfectly teleported states, where again $p_{\tilde a}$ is the probability of Alice obtaining outcome $\tilde a$.  Using Eq.~\eqref{Nprelation} and inserting resolutions of the identity allows us to write
\begin{align}\label{eq:rhotele}
  \rho^{\rm tele} &= \sum_{\tilde{a},\tilde{a}'}\left\{\prod_j [\sin(2u)]^{\frac{1-a_ja_j'}{2}}\right\} \langle\psi_{c}|{\tilde a}^{\prime},\bm{n}\rangle \langle {\tilde a},\bm{n}|\psi_{c} \rangle
  \nonumber \\
& \times \ket{{\tilde a},\bm{n}}\bra{{\tilde a}^{\prime},\bm{n}}.
\end{align}
[Equation~\eqref{eq:rhotele} actually holds for an arbitrary initial state $\ket{\psi_A}$ for Alice upon simply replacing $\psi_c \rightarrow \psi_A$ in the corresponding bra and ket.]
In the perfect-protocol limit with $u = \pi/4$, the factor in braces becomes unity, and hence the density matrix reduces to the pure-state form $\rho^{\rm tele} = \ket{\psi_{c}}\bra{\psi_{c}}$.  
Here any correlations and entanglement probed by Bob exactly mirror the properties in Alice's original state.  

Away from this limit $\rho^{\rm tele}$ describes a mixed state, and the extent to which Bob's correlations match those in Alice's original critical state depends on the operators under consideration.  Defining a string of Bob's Pauli operators $\hat{O}_{\mathcal{K}}(\bm{d})=\prod_{a=1}^{|\mathcal{K}|}\hat{O}_{j_a}(\bm{d})$ acting along quantization axis $\bm{d}$, we in particular find
\begin{equation}  \begin{split}\label{eq:correlators_dec}
       & \mathrm{Tr}[\rho^{\rm tele}\hat{O}_{\mathcal{K}}(\bm{n})]= \bra{\psi_c}\hat{O}_{\mathcal{K}}(\bm{n})\ket{\psi_c},\\
        & \mathrm{Tr}[\rho^{\rm tele}\hat{O}_{\mathcal{K}}(\bm{n}^\perp)]= [\sin(2u)]^{|\mathcal{K}|}\bra{\psi_c}\hat{O}_{\mathcal{K}}(\bm{n}^\perp)\ket{\psi_c},\\
         & \mathrm{Tr}[\rho^{\rm tele}\hat{O}_{\mathcal{K}}(\bm{n}\times\bm{n}^\perp)]=[\sin(2u)]^{|\mathcal{K}|}\bra{\psi_c}\hat{O}_{\mathcal{K}}(\bm{n}\times\bm{n}^\perp)\ket{\psi_c}.
    \end{split}
\end{equation}
The first line indicates that Bob exactly recovers critical correlations of operators oriented along $\bm{n}$ \emph{for any choice of $u$}.  At $u = \pi/4$ this result is clear since the teleportation is perfect in every iteration as already highlighted above.  For the opposite extreme, at $u = 0$---where Alice and Bob altogether fail to entangle---the same conclusion persists for the following reason: Upon receiving Alice's measurement outcome, the unitary that Bob implements simply converts his initial product state into Alice's post-measurement product state, i.e., he probes correlations in the wavefunction $\ket{\psi^{\rm tele}} = \ket{\tilde a,\bm{n}}$. In effect, Alice has simply farmed out the measurement of correlations in her state to Bob, and hence averaging over outcomes reproduces her critical correlations as captured by the first line of Eq.~\eqref{eq:correlators_dec}.
The bottom two lines indicate that Bob also recovers critical correlations of Alice's state for Pauli operators oriented orthogonal to $\bm{n}$, but with a suppressed amplitude that eventually vanishes as $u$ decreases from $\pi/4$ towards zero.  

Consequently, averaging over sequential imperfect teleportation runs indeed allows Bob to measure Alice's pristine universal power-law exponents---even for $\bm{n} = \hat{\bm{z}}$ protocols where imperfection constitutes a relevant perturbation with both uniform and typical measurement outcomes.  This conclusion relates deeply to the `post-selection problem' commonly plaguing measurement-induced phenomena that manifest only when following specific measurement outcomes whose probability decays exponentially with system size.  From the quantum teleportation perspective, however, the preceding results comprise a feature rather than a bug in that averaging over sequential imperfectly teleported states suppresses errors in correlations that appear in any particular implementation.

In this context, the Rényi entropies we  discussed earlier cannot discern between classical and quantum correlations, and hence are not adequate measures of mixed-state entanglement. Instead we consider the entanglement negativity---which does provides a good metric~\cite{peres1996separability,vw-02}. 
We refer the interested reader to Appendix~\ref{app:negativity} for more details about the behavior of entanglement negativity for the mixed-state density matrix $\rho^{\rm tele}$. The main lesson we learn is that long-range entanglement encoded in the negativity in the pure-state $u = \pi/4$ limit persists with marginal or irrelevant imperfections but is spoiled by relevant imperfections---which is natural given that, in the latter case, individual protocols runs yield area-law entanglement with both uniform and typical measurement outcomes.  Thus entanglement in the mixed state is less resilient to imperfections compared to correlations. 
These results are compatible with those in Ref.~\onlinecite{Zou2023} regarding the effect of local quantum channels on quantum critical states. Indeed, by analyzing a proxy for the entanglement negativity, the authors find that when their channel corresponds to a relevant perturbation, the negativity obeys an area law.

It is instructive to consider an alternative averaging scheme wherein Bob does not implement the final outcome-dependent unitary in each protocol iteration, and instead directly probes the state $\ket{\psi_{\tilde a}}$ given in Eq.~\eqref{eq:compact}.  In this scenario the relevant mixed-state density matrix reads
\begin{equation}\label{eq:rhonoprime}
\rho=\sum_{\tilde{a}} p_{\tilde{a}}\ket{\psi_{\tilde{a}}}\bra{\psi_{\tilde{a}}},
\end{equation}
which is simply the average of Bob's penultimate wavefunction.
For an arbitrary observable $\hat{O}$ of Bob's qubits, we have 
$\mathrm{Tr}[\hat{O}\rho]=\langle \tilde b,\bm{n}|\bra{\psi_c}U^\dagger \hat{O}U\ket{\psi_c}|\tilde b,\bm{n}\rangle$, where $U$ is the  entangling gate employed in the protocol [Eq.~\eqref{eq:unitary}].  That is, this alternative averaging scheme returns the expectation value of observables evaluated in Alice and Bob's entangled state before she measures her qubits.  
The analogue of Eq.~\eqref{eq:correlators_dec} accordingly becomes
\begin{equation}  \begin{split}
       & \mathrm{Tr}[\rho\hat{O}_{\mathcal{K}}(\bm{n})]= [\cos(2u)]^{|\mathcal{K}|}  \prod_{a=1}^{|\mathcal{K}|} {b}_{j_a}, \\
        & \mathrm{Tr}[\rho\hat{O}_{\mathcal{K}}(\bm{n}^\perp)]= \bra{\tilde{b},\bm{n}}\hat{O}_{\mathcal{K}}(\bm{n}^\perp)\ket{\tilde{b},\bm{n}}=0,\\
         & \mathrm{Tr}[\rho\hat{O}_{\mathcal{K}}(\bm{n}\times\bm{n}^\perp)]=[\sin(2u)]^{|\mathcal{K}|}\prod_{a=1}^{|\mathcal{K}|} {b}_{j_a} \\ & ~~~~~~~~~~~~~~~~~~~~~~~~\times \bra{\psi_c}\hat{O}_\mathcal{K}(\bm{n}^\perp)\ket{\psi_c}.
    \end{split}
    \label{rhoaverages}
\end{equation}
At $u = \pi/4$, only the bottom correlation returns a nonzero result, which we can understand as follows.   In any given perfect-protocol implementation, Bob's state $\ket{\psi_{\tilde a}}$ corresponds to the pristine quantum critical wavefunction $\ket{\psi_c}$ modified by an outcome-dependent unitary.  For the case of $\bra{\psi_{\tilde a}}\hat{O}_{\mathcal{K}}(\bm{n}\times\bm{n}^\perp)\ket{\psi_{\tilde a}}$ correlations, the outcome-dependent unitary factors rotate $\hat{O}_{\mathcal{K}}(\bm{n}\times\bm{n}^\perp)$ in a manner that depends on the bits $\tilde b$ for Bob's initialization but \emph{not} on Alice's outcome $\tilde a$.  Hence averaging over measurement outcomes for such correlators does not `scramble' the signal obtained from individual protocol implementations.  In contrast, for $\bra{\psi_{\tilde a}}\hat{O}_{\mathcal{K}}(\bm{n})\ket{\psi_{\tilde a}}$ or $\bra{\psi_{\tilde a}}\hat{O}_{\mathcal{K}}(\bm{n}^\perp)\ket{\psi_{\tilde a}}$ correlators, the rotation depends on $\tilde a$, and averaging over outcomes therefore returns zero.  The amplitude for the third correlation in Eq.~\eqref{rhoaverages} becomes suppressed for an imperfect protocol, while the first correlation actually gets revived (but in a way that reveals no information about Alice's critical correlations).  Indeed in the trivial $u = 0$ protocol limit the upper line simply reveals a non-zero expectation value from Bob's initial product state.

Comparing Eqs.~\eqref{eq:correlators_dec} and~\eqref{rhoaverages}, we see that averaging observables measured after each full teleportation protocol run captures Alice's critical correlations far more completely compared to the scenario where Bob skips applying the final outcome-dependent unitary.  One can view the former averaging scheme as amending the latter with a decoding protocol very similar to that studied recently in Ref.~\onlinecite{lu2023} (related frameworks are also common in adaptive circuits; see, e.g., Refs.~\onlinecite{piroli2021,nat22,nat22_2,Lu22}).  Interestingly, in the present context such decoding algorithms are automatically incorporated into the quantum teleportation protocols.

\section{Conclusions and outlook}
\label{sec:outlook}

Quantum teleportation refers to a protocol that transmits an arbitrary, and in-principle unknown, quantum state from Alice to Bob over potentially long distances, enabled by entanglement, measurements, and classical communication.  When the entangling and measurement stages are accomplished perfectly, Bob recovers Alice’s precise original state up to a unitary transformation that he can decode (i.e., eliminate) upon receiving classical information corresponding to her measurement outcome.  In retrospect, it is remarkable that combining entangling unitaries and projective measurements acting on just a subsystem does not yield a general transformation containing both unitary \emph{and} non-unitary operators acting on Alice’s original state.  Quantum teleportation utilizes a set of fine-tuned operations that renders both the non-unitary and unitary (once the classical channel has been exploited) pieces trivial.  From this perspective, it becomes natural that when these operations are subject to imperfections, Bob generically receives a non-unitarily corrupted cousin of Alice's original state---as we showed explicitly in many instances throughout this paper.  

Understanding how general many-body wavefunctions behave under  imperfect teleportation protocols defines a very broad and rich problem.  Relative to the single-qubit case, many-body states exhibit vastly more structure characterized, e.g., by local observables, multi-body correlations, entanglement, symmetry quantum numbers,  topological properties, etc. Our study establishes two broadly applicable principles in this realm: 
First, imperfect teleportation deeply relates to the survival of correlations and entanglement in many-body states subjected to weak measurements.  The fact that an imperfect teleportation translates into an imaginary-time evolution holds in general, and applies to any many-body wavefunction, including those showcasing topological order. In fact, those types of imaginary-time evolutions have been previously studied in the literature (see e.g., Refs.~\onlinecite{Haegeman_15,GuoYi_19}), leading to conclusions regarding their robustness under teleportation protocols.  The link between imperfect teleportation and the effects of weak measurements on many-body states identifies a potential quantum networking application (see below) for the growing body of literature on the latter topic. 

Second, fidelity defined by the overlap between Alice's original wavefunction and the state Bob receives after decoding, while a natural metric for the single-qubit case, need not provide a useful measure for many-body wavefunctions.  
Fidelity based on wavefunction overlap blurs the distinction between unitary and non-unitary errors, which can be understood as a consequence of an Anderson orthogonality catastrophe: Consider a many-body wavefunction $\ket{\psi}$, and apply to each qubit $j$ an on-site linear transformation $M_j(\epsilon)$ that reduces to the identity as $\epsilon \rightarrow 0$.  Then the overlap $|\langle \psi|\prod_j M_j(\epsilon)|\psi\rangle|$ is expected to decay exponentially as the number of qubits $L$ increases for any small but finite $\epsilon$---regardless of whether $M_j(\epsilon)$ is unitary or non-unitary. Alternatively, one could consider the `local fidelity', defined as the $L$-th root of the global fidelity; this quantity, however, still does not cleanly distinguish between unitary and non-unitary errors that can yield drastically different consequences on many-body states.  It is crucial to instead focus on physically meaningful diagnostics of imperfectly teleported many-body wavefunctions that emphasize salient universal features over fine details.  Ultimately, Bob may be content receiving an imperfectly teleported wavefunction that overlaps negligibly with Alice's original state, provided that his state belongs to the same phase of matter, exhibits the same symmetries, displays the same universal correlations and entanglement characteristics, and so on.  This perspective becomes particularly essential in a scenario where Alice has limited information about her original wavefunction; for instance, she may know only that it is the ground state of some local Hamiltonian whose parameters are uncertain yet realizes a known phase or critical point.  The performance of different teleportation protocols could be naturally assessed by exploring how universal features of the most sensitive regions in the phase diagram respond to imperfections.

With this last message in mind, rather than analyzing the detailed character of arbitrary many-body wavefunctions under imperfect teleportation, we specialized to Ising quantum critical states as an illuminating test case.  Here universal power-law correlations and long-range entanglement encoded in Alice's original state provide well-motivated quantities for benchmarking the character of an imperfectly teleported wavefunction received by Bob.  Moreover, as shown in a collection of recent works~\cite{AltmanMeasurementLL,Lee2023,JianMeasurementIsing,EhudMeasurementIsing,usmeasurementaltered,Lee2023}, quantum critical points are expected to sensitively respond to even weak imperfections by virtue of their gaplessness.  The dichotomy between unitary and non-unitary errors becomes especially sharp here: On-site unitary errors preserve both power-law exponents and long-range entanglement, albeit with locally scrambled manifestations.  Non-unitary contributions, on the other hand, can qualitatively modify both properties in ways that we elucidated in detail for protocols employing different measurement bases and in different measurement outcome sectors.  Specifically, we classified protocol imperfections as relevant, marginal, disguised marginal, or irrelevant depending on their impact on universal correlations and entanglement---though in all of these cases the wavefunction-overlap fidelity decreases exponentially with system size.  In a nutshell, fidelity does not distinguish between corruption of short- versus long-distance physics arising from local errors. 
 
En route to establishing these results, we nontrivially extended the theory of measurement-altered criticality by, e.g., relating measurement effects to modifications of full counting statistics; uncovering the phenomenon of disguised marginal perturbations generated under renormalization by naively irrelevant measurement-induced terms; adding a new perspective on decoding protocols put forward in Ref.~\onlinecite{lu2023}, which quantum teleportation protocols automatically incorporate; and  deriving parent Hamiltonians for non-unitarily corrupted states.  The last item allowed us to make contact with the physics of long-range interacting chains---whose ground states resemble weakly measured/imperfectly teleported critical wavefunctions---as well as with non-Hermitian systems that even at criticality can display area-law entanglement and an emerging finite correlation length~\cite{Dora_2022}.  We hope that many of these tools can be directly exported to understand measurement effects on other systems such as those realizing strongly interacting CFTs and deconfined critical points.

We close with some questions related to future directions.  Are there practical applications of teleporting \emph{particular} many-body wavefunctions, e.g., Ising critical states?  As one possibility, quantum critical wavefunctions, by virtue of their long-range entanglement, are nontrivial to prepare on quantum hardware.  If Alice owns a special-purpose device that allows her to efficiently accomplish the task, she may (for a fee) quantum teleport critical wavefunctions to remote interested parties over quantum networks.  To this end, it would be highly desirable to adopt a protocol that optimizes against imperfections following insights from this work.  Among the infinite variety of possible many-body wavefunctions, which other classes are particularly interesting to examine through the lens of imperfect teleportation \footnote{We thank Guo-Yi Zhu for sharing a preprint on imperfectly teleporting surface codes [F.~Eckstein, B.~Han, S.~Trebst, and G.-Y.~Zhu, Robust teleportation of a surface code and cascade of topological quantum phase transitions].}?   Is it possible to teleport only a subset of qubits from a many-body wavefunction while still allowing the receiver to back out universal features of the complete state?  Finally, can one devise near-term experiments for testing the resilience of various many-body teleportation protocols against errors?

\begin{acknowledgments}

It is a pleasure to acknowledge enlightening conversations with  Ehud Altman, Filiberto Ares, Hannes Bernien, Pasquale Calabrese, Luca Capizzi, Mario Collura, Bal\'azs D\'ora, Jerome Dubail, Sam Garratt, Giuseppe Di Giulio, Chao-Ming Jian, Dan Mao, David Mross, Thomas Schuster, David Stephen, Nat Tantivasadakarn, Ruben Verresen, and Guo-Yi Zhu.  Tensor network calculations were performed using the TeNPy Library~\cite{tenpy}. This work was primarily led and supported by the U.S. Department of Energy, Office of Science, National Quantum Information Science Research Centers, Quantum Science Center (numerical and analytical explorations of the teleportation protocols, YL and JA).  Additional support was provided by the Caltech Institute for Quantum
Information and Matter, an NSF Physics Frontiers Center with support of the Gordon and Betty Moore Foundation through Grant GBMF1250, and the Walter Burke
Institute for Theoretical Physics at Caltech.
\end{acknowledgments}

\appendix

\onecolumngrid
\pagebreak[4]

\section{Derivation of imperfectly teleported many-body state}
\label{sec:general}

In this appendix we derive the form of the imperfectly teleported wavefunction received by Bob, prior to application of the final outcome-dependent unitary.  We will consider the generalized protocol from Sec.~\ref{sec:general}, recovering the main result of Sec.~\ref{sec:wavefunction} as a limiting case.
For clarity, let us recall the general setup. Alice and Bob prepare the initial wavefunction $\ket{\psi}_A\ket{\psi}_B$, where $\ket{\psi}_B=\ket{\tilde{b},\bm{n}}$ is a product state along the common quantization axis $\bm{n}$. In most of the text we considered Alice's wavefunction is given by the critical state $\ket{\psi_c}$ although this is not necessary. Then, they couple their degrees of freedom via the on-site unitary $U=\prod_j U_j$ with $U_j=\exp(iu\hat{O}^A_j(\bm{m})\hat{O}^B_j(\bm{m}))$ with $\bm{m}$ a unit vector. Finally, Alice measures her qubits along  $\bm{n}$ obtaining the outcome $\tilde{a}=\{a_j\}$ with probability $p_{\tilde{a}}$.  All together, a generic setup is parametrized by the tuple $\{\tilde{a}, \tilde{b},\bm{n}; \bm{m}\}$. Similar computations as those leading to Eq.~\eqref{eq:compact} in the main text, allow us to write 
\begin{equation} \label{eq:app1A}
    \begin{aligned}
        \ket{\psi_{\tilde{a}}}&=\frac{1}{\sqrt{p_{\tilde{a}}}}\sum_{N_f=0}^N\sum_{i_1<i_2<\dots < i_N}i^{N_f}\cos^{N-N_f}(u)\sin^{N_f}(u)\prod_{j=1}^{N_f}\hat{O}^B_{i_j}(\bm{m})\ket{\tilde{b},\bm{n}}\bra{\tilde{a},\bm{n}}\prod_{j=1}^{N_f}\hat{O}^A_{i_j}(\bm{m})\ket{\psi_c}.
    \end{aligned}
\end{equation}
Next, we explicitly include the flipping Pauli operators $\hat{O}^A_j(\bm{n}^\perp),\hat{O}^B_j(\bm{n}^\perp)$ mapping $a_j\to -a_j$, $b_j\to -b_j$ respectively by writing
\begin{equation}
    \hat{O}^D_j(\bm{m})=\hat{O}^D_j(\bm{m})\hat{O}^D_j(\bm{n}^\perp)^2=\underbrace{\left(\bm{m}\cdot\bm{n}^\perp+i\bm{m}\times \bm{n}^\perp\cdot \bm{\sigma} \right)}_{\equiv U^D_{j}(\bm{m}\times \bm{n}^\perp)}\hat{O}^D_j(\bm{n}^\perp)
\end{equation}
with $D=A,B$. Notice that $U^D_{j}(\bm{m}\times \bm{n}^\perp)=\exp(i\theta \hat{O}_j(\bm{m}\times \bm{n}^\perp/||\bm{m}\times \bm{n}^\perp ||))$ with $\cos(\theta)=\bm{m}\cdot\bm{n}^\perp$,  $\sin(\theta)=||\bm{m}\times \bm{n}^\perp||$. We also need to relate the product wavefunctions $\ket{\tilde{b},\bm{n}}, \ket{\tilde{a},\bm{n}}$ by applying a unitary that connects the string of numbers $\tilde{a}$ and $\tilde{b}$. This is achieved via
\begin{equation}
    \ket{\tilde{b},\bm{n}}=\underbrace{\left(\prod_{j:a_j\neq b_j}\hat{O}^B_j(\bm{n}^\perp)\right)}_{\equiv U^B_{\tilde{b}\leftarrow \tilde{a}}}\ket{\tilde{a},\bm{n}}=U^B_{\tilde{b}\leftarrow \tilde{a}}\ket{\tilde{a},\bm{n}},
\end{equation}
where we notice that this untary satisfies $(U^B_{\tilde{b}\leftarrow \tilde{a}})^2=\mathds{1}$.
Bringing all together and up to an overall factor we find
\begin{equation} 
    \begin{aligned}
        \ket{\psi_{\tilde{a}}}&=\frac{1}{\sqrt{\mathcal{N}}}\sum_{N_f=0}^N\sum_{i_1<\dots < i_N}i^{N_f}\tan^{N_f}(u)\prod_{j=1}^{N_f}U^B_{i_j}(\bm{m}\times \bm{n}^\perp)U^B_{\tilde{b}\leftarrow \tilde{a}}\\ 
        &  \times \left(\prod_{j=1}^{N_f}\hat{O}^B_{i_j}(\bm{n}^\perp)\ket{\tilde{a},\bm{n}}\bra{\tilde{a},\bm{n}}\prod_{j=1}^{N_f}\hat{O}^A_{i_j}(\bm{n}^\perp)\right) \prod_{j=1}^{N_f} U^{A}_{i_j}(-\bm{m}\times \bm{n}^\perp)\ket{\psi_c}.
    \end{aligned}
\end{equation}
where we have used that $[\hat{O}^B_{i_j}(\bm{n}^\perp),U^B_{\tilde{b}\leftarrow \tilde{a}}]=0$. To move further we recall that Alice's and Bob's Hilbert spaces are isomorphic and as such, we can evaluate 
\begin{equation}
  \bra{\tilde{a},\bm{n}}\prod_{j=1}^{N_f}\hat{O}^A_{i_j}(\bm{n}^\perp)U^{A}_{i_j}(-\bm{m}\times \bm{n}^\perp)\ket{\psi_c}=\bra{\tilde{a},\bm{n}}\prod_{j=1}^{N_f}\hat{O}^B_{i_j}(\bm{n}^\perp)U^{B}_{i_j}(-\bm{m}\times \bm{n}^\perp)\ket{\psi_c}
\end{equation}
on Bob's degrees of freedom. Finally, when summing over all possible values of $N_f$ for any set of locations $i_1<i_2<\dots i_N$, the state $\ket{a',\bm{n}}\equiv \prod_{j=1}^{N_f}\hat{O}^B_{i_j}(\bm{n}^\perp)\ket{\tilde{a},\bm{n}}\in \mathcal{H}_B$ runs over all possible product states in the local $\bm{n}$ basis. This allows us to express the Bob's wavefunction as 
\begin{equation} 
    \begin{aligned}
        \ket{\psi_{\tilde{a}}}=\frac{1}{\sqrt{\mathcal{N}}}\sum_{a'}(i\tan(u))^{N_f(a',\tilde{b})}\prod_{j: a^\prime_j\neq \tilde{b}_j}U^B_{j}(\bm{m}\times \bm{n}^\perp)U^B_{\tilde{b}\leftarrow \tilde{a}}\ket{a^\prime,\bm{n}}\bra{a^\prime,\bm{n}} \prod_{j: a^\prime_j\neq \tilde{a}_j}U^{B}_{j}(-\bm{m}\times \bm{n}^\perp)\ket{\psi_c},
    \end{aligned}
\end{equation}
where the sum is over binary chains $a^\prime=\{\pm 1\}$ of size $N$, and $N_f(a^\prime, \tilde{b})=\sum_{j=1}^N \frac{1}{2}(1-a^\prime_j \tilde{b}_j)$ counts the number of entrees where $a^\prime_j\neq \tilde{a}_j$ for a reference string $\tilde{a}$. This expression can be recast as
\begin{equation}
    \begin{aligned}
        \ket{\psi_{\tilde{a}}}&=\frac{1}{\sqrt{\mathcal{N}}}e^{i\frac{\pi}{4}H_{\tilde{b}}(\bm{n})}e^{-\frac{\alpha}{2}H_{\tilde{b}}(\bm{n})}\mathcal{U}_{\tilde{b}}(\bm{m}\times \bm{n}^\perp) U_{\tilde{b}\leftarrow \tilde{a}}\mathcal{U}_{\tilde{a}}^\dagger(\bm{m}\times\bm{n}^\perp)\ket{\psi_c},
    \end{aligned}
\end{equation}
where the transformation $\mathcal{U}_{\tilde{b}}(\bm{m}\times \bm{n}^\perp), \mathcal{U}_{\tilde{a}}(\bm{m}\times\bm{n}^\perp)$ are defined via
\begin{align}
   & \mathcal{U}_{\tilde{b}}(\bm{m}\times \bm{n}^\perp)=\prod_{j=1}^N\left[U^B_{j}(\bm{m}\times \bm{n}^\perp)\frac{1}{2}(1-\tilde{b}_j\hat{O}^B_j(\bm{n}))+\mathds{1}\frac{1}{2}(1+\tilde{b}_j\hat{O}^B_j(\bm{n}))\right],\\
   & \mathcal{U}_{\tilde{a}}(\bm{m}\times \bm{n}^\perp)=\prod_{j=1}^N\left[U^B_{j}(\bm{m}\times \bm{n}^\perp)\frac{1}{2}(1-\tilde{a}_j\hat{O}^B_j(\bm{n}))+\mathds{1}\frac{1}{2}(1+\tilde{a}_j\hat{O}^B_j(\bm{n}))\right],
\end{align}
while $H_{\tilde{b}}(\bm{n})= -\sum_j \tilde{b}_j \hat{O}_j^B(\bm{n})$ as defined in the main text.
Finally, we can combine $\mathcal{U}_{\tilde{b}}(\bm{m}\times \bm{n}^\perp)$ and $\mathcal{U}_{\tilde{a}}^\dagger(\bm{m}\times\bm{n}^\perp)$ using that $\mathcal{U}_{\tilde{b}}(\bm{m}\times \bm{n}^\perp) U_{\tilde{b}\leftarrow \tilde{a}}\mathcal{U}_{\tilde{a}}^\dagger(\bm{m}\times\bm{n}^\perp) = U_{\tilde{b}\leftarrow \tilde{a}} \mathcal{U}_{\tilde{a}}^\dagger(\bm{m}\times \bm{n}^\perp) \mathcal{U}_{\tilde{a}}^\dagger(\bm{m}\times\bm{n}^\perp)$, leading to the expression
\begin{equation}
    \begin{aligned}
        \ket{\psi_{\tilde{a}}}&=\frac{1}{\sqrt{\mathcal{N}}}U_{\tilde{b}\leftarrow \tilde{a}}e^{i\frac{\pi}{4}H_{\tilde{a}}(\bm{n})}e^{-\frac{\alpha}{2}H_{\tilde{a}}(\bm{n})}\left( \mathcal{U}_{\tilde{a}}^\dagger(\bm{m}\times \bm{n}^\perp)\right)^2\ket{\psi_c}.
    \end{aligned}
\end{equation}
If we restrict to the case $\bm{m}=\bm{n}^\perp$, then $\mathcal{U}_{\tilde{a}}^\dagger(\bm{m}\times \bm{n}^\perp)$ simply reduces to the identity and we recover Eq.~\eqref{eq:compact} in the main text.

\section{Probability distribution of the longitudinal field}\label{app:longitudinal}

\begin{figure}[t]
    \centering
    \includegraphics[width=0.5\textwidth]{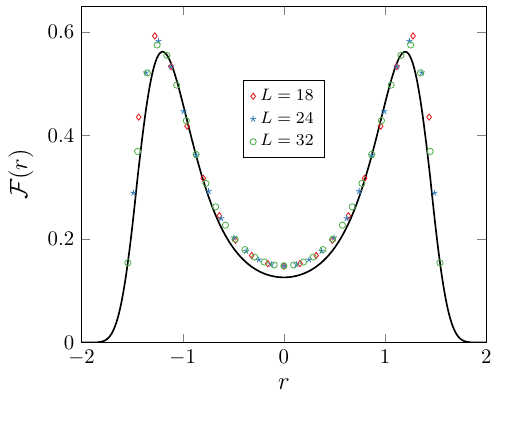}
    \caption{{\bf Universal scaling function $\mathcal{F}(r)$.} The solid line in Eq.~\eqref{eq:Fr}, while the symbols are the numerical data obtained using exact diagonalization for a system size $L=18,24,32$. 
   }\label{fig:Fx}
\end{figure}

The goal of this appendix is to analyze the probability distribution of the longitudinal field $M_Z=\sum_j Z_j$ in the ground state of the critical Ising model, $P(f,L)$. This amounts to compute the probability density for finding Alice’s initial state in a given magnetization sector in the $Z$-basis in a system of size $L$. Taking advantage of the results found in~\cite{Lamacraft08}, we can introduce the scaling variable $r=m/s$, where $s^2=\langle(M^+_Z)^2\rangle$ is the variance of the order parameter $M^+_Z$, and using $f=m/L+1/2=sr/L+1/2$, we are able to compute the asymptotic expression for the universal scaling function $\mathcal{F}(r)=sP(sr/L+1/2,L)$. We remark that, in order to have an analytic prediction for $\mathcal{F}(r)$ which captures the main features of the probability distribution, we use the asymptotic result for the generating function of the moments of the distribution, and therefore our result is not very accurate for $r$ close to $0$. In particular, we found that
\begin{equation}\label{eq:Fr}
\mathcal{F}(r)=
\begin{cases}\frac{1.05 e^{0.05 r^6-0.07 r^8}}{| r| ^{15/16}}, \qquad|r|\geq 1.44, \\
e^{-0.29 r^6+0.11 r^4+1.48 r^2-2.07} \qquad \text{otherwise}.
\end{cases}
\end{equation}
In order to benchmark out this prediction, we have tested it by exact diagonalization of the lattice Hamiltonian~\eqref{eq:Hamiltonian_Ising} for system sizes $L=18,24,32$ in Fig.~\ref{fig:Fx}. As we expected, we observe that the agreement is not perfect close to $r=0$, but $\mathcal{F}(r)$ correctly captures the shape of the numerical data, which is enough for our purposes. 

\begin{figure}[ht]
\centering
\subfigure
{\includegraphics[width=0.32\textwidth]{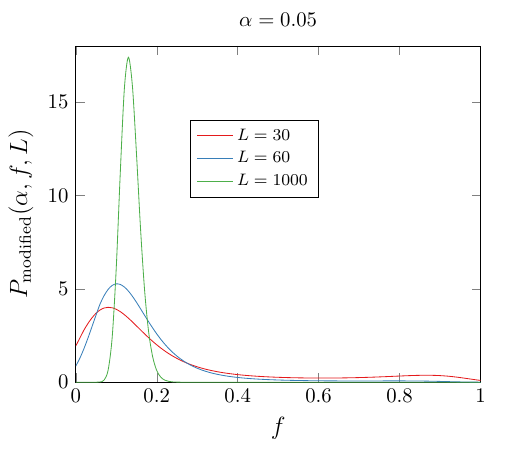}}
\subfigure
{\includegraphics[width=0.32\textwidth]{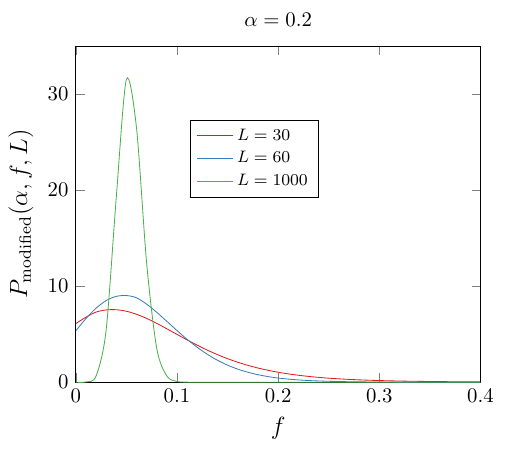}}
\subfigure
{\includegraphics[width=0.32\textwidth]{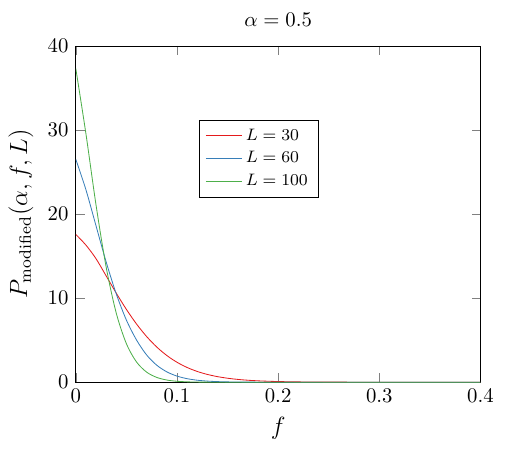}}
\caption{{\bf $P_{\rm modified}(\alpha,f,L)$ as a function of the intensive magnetization sector $f$ for different $\alpha$ and $L$.} The left panel shows how, for small $\alpha$ and $L$ (30,60) the left peak moves towards left and the double-peak structure is suppressed. As $L$ increases ($L=1000$), as well as $\alpha$ (middle panel), the peak we observe is due to the tail of the probability distribution $P(f,L)$. Indeed, for $\alpha=0.2$, its position is not shifted. As we increase $\alpha$ (right panel) $P_{\rm modified}(\alpha,f,L)$ becomes peaked around $f=0$, even though the exact details of the probability distribution require a more accurate analytical prediction in Eq.~\eqref{eq:Fr}. 
   }\label{fig:Pmodified}
\end{figure}

We numerically checked that $s=1/2\,L^{7/8}$ and, from the expression of $\mathcal{F}(r)$, we can compute $P(f,L)=\frac{1}{\mathcal{N} } \mathcal{F}(2(f-1/2)L^{1/8})$, where $\mathcal{N} $ is fixed such that the probability distribution is correctly normalized. Once we have the result for $P(f,L)$, we can compute $P_{\rm modified}(\alpha,f,L)=\frac{e^{-2\alpha fL}}{\mathcal{N'} } P(f,L)$ and also study its behavior as a function of $\alpha$ and $L$. We observe that, for $\alpha=0$, $P(f,L)$ shows a double peak structure (see Fig.~\ref{fig:Fx}). As we slightly increase $\alpha$, the right peak is suppressed, while the left peak drastically moves to the left until it falls outside the region $ |r|\leq 1.44$. In particular, in the large $L$ limit, the saddle point analysis reveals that we have a single saddle point at $f^*=1/2-\kappa'\alpha^{1/5}L^{1/20}$, which we can observe only until the condition $ |r|\leq 1.44$ is satisfied, i.e. for small values of $\alpha,L$ (see the left panel of Fig.~\ref{fig:Pmodified} for $L=30,60$). Here $\kappa'$ is only a numerical prefactor. If we focus on the behavior of the tail of the distribution, i.e. $r<-1.44$, we can observe a peak for $f^*=1/2-\kappa''\alpha$ (when $f$ satisfies the condition $r<-1.44$), where $\kappa''$ is a constant, which is pushed towards $f=0$ as we increase $\alpha$, while its position does not change by varying $L$. This can be clearly observed in the middle panel of Fig.~\ref{fig:Pmodified}. In order to have a complete analysis of the behavior for $f$ close to $0$, we need to know the analytic behavior of $\mathcal{F}(r)$ for any real value of $r$, which is not accessible from Ref.~\cite{Lamacraft08}. However, with the data at our disposal, we can already capture the main qualitative features of the distribution $P_{\rm modified}(\alpha,f,L)$, checking how the effect of a relevant perturbation can drastically spoil the double-peak structure, shifting the weight of the distribution towards $f^*=0$.

\section{Two-point correlators along the boundary}\label{app:boundary}
The goal of this appendix is to derive the results in Eq.~\eqref{eq:correlators_critical2_Z} through a field theoretical approach.
We can perform a spacetime rotation such that Eq.~\eqref{eq:Z} describes a different one-dimensional critical Ising spin chain everywhere in space except at a defect located in the middle point of the chain. Being the defect line relevant, we can think of it as a fixed boundary condition. In order to fix the ideas, we can focus on a system with spatial coordinate $x\in [-L/2,L/2]$ and we aim to study the behavior of the correlation functions of the critical Ising chain along the boundary denoted by $(x=0,\tau)$, where $\tau$ is the imaginary time.
We can take advantage of the results found in~\cite{Burkhardt91,Burkhardt93} to write down the magnetization profile with fixed boundary conditions (e.g. spin up ($+$)) at $x=0$ and at $x=L/2$, which reads
\begin{equation}
\langle\sigma(x)\rangle_{+}=2^{1/8}\left(\frac{L}{2\pi} \sin \frac{2\pi x }{L}\right)^{-1/8}.
\end{equation}
Notice that here $x$ denotes the distance from the boundary, while what we are really interested in are the correlation functions along the boundary. In the limit $x \to \epsilon$ (it plays the role of an ultra-violet cutoff), it turns out that, along the boundary the expectation value of the magnetic field is just a constant.
Similarly, we can exploit the known results for the two-point correlation functions on the half-plane, which read~\cite{Burkhardt91}
\begin{equation}\label{eq:2-point}
\langle\sigma(1)\sigma(2)\rangle_{+}=(4y_1 y_2)^{-1/8}(u+u^{-1})^{1/2},
\end{equation}
with 
$u=\left(1+4y_1y_2/r^2\right)^{1/4}$, $r=\sqrt{(v_1-v_2)^2+(y_1-y_2)^2}$. Since the half-space with fixed boundary conditions is transformed into a strip of width $L/2$ with boundary conditions on the edges under the conformal mapping $w=L/(2\pi)\ln z$, we find $z=v+i y=e^{2\pi/L (ix+\tau)}=e^{2\pi \tau/L}(\cos \frac{2\pi x}{L}+i\sin \frac{2\pi x}{L})$. Taking into account that under the conformal transformation $z \to w$, correlation functions of primary operators transform according to 
\begin{equation}
    \langle\prod_i \phi(z_i,\Bar{z}_i)\rangle=\prod_i w'(z_i)^{\Delta_i}\overline{w'(z_i)}^{\bar{\Delta}_i}\langle\prod_i \phi(w_i,\Bar{w}_i)\rangle,
\end{equation}
in this way, we can map the result in Eq.~\eqref{eq:2-point} for the half-space into the strip geometry. 
It reads
\begin{equation}\label{eq:2-point}
\begin{split}
\langle\sigma(1)\sigma(2)\rangle_{+}=\left(\frac{L}{\pi}\right)^{-1/4}\left( \sin \frac{2\pi x_1 }{L} \sin \frac{2\pi x_2 }{L}\right)^{-1/8}(u+u^{-1})^{1/2},\quad \\u=1+4\frac{e^{\frac{2\pi}{L}(\tau_1+\tau_2)}\sin \frac{2\pi x_1 }{L} \sin \frac{2\pi x_2 }{L}}{(e^{\frac{2\pi\tau_1}{L}}\sin(\frac{2\pi x_1}{L})-e^{\frac{2\pi\tau_2}{L}}\sin(\frac{2\pi x_2}{L}))^2+(e^{\frac{2\pi\tau_1}{L}}\cos(\frac{2\pi x_1}{L})-e^{\frac{2\pi\tau_2}{L}}\cos(\frac{2\pi x_2}{L}))^2}
\end{split}
\end{equation}
In this way, we get in the limit $x_1\to x_2 \to \epsilon$ and $L\to \infty$
\begin{equation}\label{eq:sigma_b}
   \langle\sigma(1)\sigma(2)\rangle_{+}-\langle\sigma(1)\rangle_{+}\langle\sigma(2)\rangle_{+} \simeq  \frac{\epsilon^{15/4}}{|\tau_1-\tau_2|^4}.
\end{equation}
Strictly speaking, the right-hand side vanishes when $\epsilon=0$, which here plays the role of a ultra-violet cutoff.
We can repeat the same analysis for the connected two-point correlator of the energy field $\varepsilon$. Its one-point correlation function in the presence of a boundary reads
\begin{equation}
  \langle\varepsilon(x)\rangle_+=  \left(\frac{L}{2\pi} \sin \frac{2\pi x }{L}\right)^{-1},
\end{equation}
while the 2-point correlation function on the complex plane is
\begin{equation}  
\begin{split}
&\langle \varepsilon(1)\varepsilon(2)\rangle_{+}=4\left(\frac{1}{4y_1y_2}+\frac{1}{r^2}-\frac{1}{r^2+4y_1y_2}\right),\\
&\langle\varepsilon(1)\varepsilon(2)\rangle_{+,c}\equiv \langle\varepsilon(1)\varepsilon(2)\rangle_{+}-\langle\varepsilon(1)\rangle_{+}\langle\varepsilon(2)\rangle_{+}=4\left(\frac{1}{r^2}-\frac{1}{r^2+4y_1y_2}\right).
\end{split}
\end{equation}
By applying the conformal transformation, we can evaluate the connected correlator in the original geometry, and in the limit $x_1\to x_2 \to \epsilon$ and $L\to \infty$ we get
\begin{equation}\label{eq:epsilon_b}
\langle\varepsilon(1)\varepsilon(2)\rangle_{+,c}\simeq \frac{\epsilon^2}{|\tau_1-\tau_2|^4}.  
\end{equation}
By doing a spacetime rotation back, both the results in Eqs.~\eqref{eq:sigma_b} and~\eqref{eq:epsilon_b} state that the correlation function for the spin and energy field decay as $1/|x|^4$ along the boundary, which is the content of Eq.~\eqref{eq:correlators_critical2_Z}.

\section{Correlators from the perturbative approach}\label{app:perturbative}
Let us start from computing the correlator $\langle X_iX_{i'} \rangle$ in Eq.~\eqref{eq:XX_measZbasis}. It can be written as
\begin{equation}\label{eq:XX_class}
\langle X_iX_{i'} \rangle_c \propto\mathrm{Tr}[e^{-u^2\sum_{j\neq k}V_{jk}X_jX_k}X_iX_{i'} ]\sim u^2V_{ii'}\sim\frac{u^2}{|i-i'|^4}.
\end{equation}
where the $\mathrm{Tr}$ denotes a sum over all possible spin configurations.
By expanding the exponential in the limit $u\to 0$, the first non-vanishing contribution is 
\begin{equation}
\langle X_iX_{i'} \rangle_c \sim u^2V_{ii'}\sim\frac{u^2}{|i-i'|^4}.
\end{equation}
If we now consider the correlator $\langle Y_iY_{i'} \rangle_c$, we get (we denote $\ket{Z}\equiv \ket{z}^{\otimes L},  q(j)\equiv q$) 
\begin{equation}
\begin{split}
\langle Y_iY_{i'} \rangle=\bra{Z}e^{-u^2\sum_{j\neq k, \neq i,i'}V_{jk}X_jX_k} e^{\frac{u^2}{2}X_i\sum_{k\neq i}V_{jk}X_k} e^{\frac{u^2}{2}X_{i'}\sum_{k\neq i'}V_{jk}X_k}e^{2iuq X_i+2iuqX_{i'}}Y_iY_{i'}\ket{Z}\\\propto \mathrm{Tr}[e^{-u^2\sum_{j\neq k, \neq i,i'}V_{jk}X_jX_k} e^{\frac{u^2}{2}X_i\sum_{k\neq i}V_{jk}X_k} e^{\frac{u^2}{2}X_{i'}\sum_{k\neq i'}V_{jk}X_k}e^{2iuq X_i+2iuqX_{i'}}X_i X_{i'}]\sim u^2 V_{ii'},
\end{split}
\end{equation}
where we have used that $Y=-iXZ$. Finally, using $Z_i Z_{i'}\ket{Z}=\ket{Z}$, we get 
\begin{equation}
\begin{split}
\langle Z_iZ_{i'} \rangle_c=\bra{Z}e^{-u^2\sum_{j\neq k, \neq i,i'}V_{jk}X_jX_k} e^{\frac{u^2}{2}X_i\sum_{k\neq i}V_{jk}X_k} e^{\frac{u^2}{2}X_{i'}\sum_{k\neq i'}V_{jk}X_k}e^{2iuq X_i+2iuqX_{i'}}Z_iZ_{i'}\ket{Z}\\\propto \mathrm{Tr}[e^{-u^2\sum_{j\neq k, \neq i,i'}V_{jk}X_jX_k} e^{\frac{u^2}{2}X_i\sum_{k\neq i}V_{jk}X_k} e^{\frac{u^2}{2}X_{i'}\sum_{k\neq i'}V_{jk}X_k}e^{2iuq X_i+2iuqX_{i'}}]\sim u^4 V_{ii'}.
\end{split}
\end{equation}

We can repeat the computations above starting from Eq.~\eqref{eq:ZZ_measXbasis}. The easiest correlator to compute is  
\begin{equation}\label{eq:ZZ_class}
\langle Z_iZ_{i'} \rangle=\mathrm{Tr}[e^{-u^2\sum_{j\neq k}V_{jk}Z_jZ_k}Z_iZ_{i'} ]\sim u^2V_{ii'}\sim\frac{u^2}{|i-i'|}.
\end{equation}
If we now consider the correlator $\langle X_iX_{i'} \rangle_c$, we get (we denote again $\ket{X}\equiv \ket{x}^{\otimes L}$)
\begin{equation}
\langle X_iX_{i'} \rangle_c=\bra{X}e^{-u^2/2\sum_{j\neq k}V_{jk}Z_jZ_k}X_iX_{i'} e^{-u^2/2\sum_{j\neq k}V_{jk}Z_jZ_k}\ket{X}.
\end{equation}
Observing that $X_je^{-\frac{u^2}{2}Z_j\sum_{k\neq j}V_{jk}Z_k}=e^{\frac{u^2}{2}Z_j\sum_{k\neq j}V_{jk}Z_k}X_j$, we get 
\begin{equation}\label{app:XX}
\begin{split}
\langle X_iX_{i'} \rangle_c=\bra{X}e^{-u^2\sum_{j\neq k, \neq i,i'}V_{jk}Z_jZ_k} e^{\frac{u^2}{2}Z_i\sum_{k\neq i}V_{jk}Z_k} e^{\frac{u^2}{2}Z_{i'}\sum_{k\neq i'}V_{jk}Z_k}X_iX_{i'}\ket{X}=\\\mathrm{Tr}[e^{-u^2\sum_{j\neq k, \neq i,i'}V_{jk}Z_jZ_k} e^{\frac{u^2}{2}Z_i\sum_{k\neq i}V_{jk}Z_k} e^{\frac{u^2}{2}Z_{i'}\sum_{k\neq i'}V_{jk}Z_k}]\sim u^4 V^2_{ii'},
\end{split}
\end{equation}
where we have used that $X_i X_{i'}\ket{X}=\ket{X}$. We remark that this result not only properly captures the algebraic behavior of $\langle X_iX_{i'} \rangle_c$ in Eq.~\eqref{eq:correlators_critical2}, but also the amplitude, since by using that $\alpha=-\ln[\tan(u)]$, $\alpha$ behaves at leading order in $u$ as $\sech(2\alpha)^2=u^4$. The result in Eq.~\ref{app:XX} refers to the connected correlator, since the disconnected part vanishes.
Finally, in order to compute $\langle Y_iY_{i'} \rangle$, we use that $Y_iY_{i'} \ket{X}=-Z_iZ_{i'}\ket{X}$ and therefore we can exploit the result in Eq.~\eqref{eq:ZZ_class}, $\langle Y_iY_{i'} \rangle\sim u^2 V_{ii'}$, in agreement with Eq.~\eqref{eq:correlators_critical2}.

\section{Entanglement entropy from the perturbative approach}\label{app:EE}

We consider the entanglement entropy of the teleported wavefunction, 
\begin{equation}\label{eq:f1}
    \ket{\psi} = \exp(\frac{\alpha}{2}\sum_j \hat{O}_j(\bm{n})) \ket{\psi_A} \quad\text{(up to normalization)},
\end{equation}
where $\hat{O}(\bm{n})$ is a Pauli operator along $\bm{n}$ axis and $\ket{\psi_A}$ is the initial state (which in this paper is the ground state of a critical Ising chain). 
In the $\alpha \to \infty$ limit, it becomes a product state for finite systems. However, the $\alpha < \infty$ case is non-trivial and we can compute the second R\'enyi entropy perturbatively.

Let us assume $\alpha>0$ and define ${\ket{0}\equiv}\ket{\tilde{b},\bm{n}}$ to be the product state of one-site eigenstates with $\hat{O}(\bm{n})$ eigenvalue $+1$. We denote by $\ket{j,k,l,\dots}{\equiv \hat{O}_j(\bm{n}^\perp)\hat{O}_k(\bm{n}^\perp)\hat{O}_l(\bm{n}^\perp)\dots \ket{0}}$ the state with spin flips on the $j$-, $k$-, $l$-$\dots$-th sites. The teleported state can be written as 
\begin{equation}
   |\psi\rangle= \exp(\frac{\alpha}{2} \sum_j \hat{O}_j(\bm{n}))|\psi_A\rangle = \exp(\frac{\alpha}{2} L)\left[ \langle 0 | \psi_A \rangle |0\rangle + e^{-\alpha} \sum_j \langle j | \psi_A \rangle |j\rangle + e^{-2\alpha} \sum_{j<k} \langle j, k | \psi_A \rangle |j, k\rangle + \cdots\right]
\end{equation}
where $L$ is the length of the system. In the nearly-projective measurement limit ($\alpha \gg 1$), one can neglect the terms in the ellipsis since they are exponentially suppressed by $\alpha$. After keeping only the first three terms on the right-hand-side (RHS) and defining
\begin{equation}
    q(j,k,l\dots)=\frac{\braket{j,k,l,...}{\psi_A}}{\braket{0}{\psi_A}},
\end{equation}
we simplify the above form as
\begin{equation}\label{eq:expansion_approach}
    \ket{\psi}=\ket{0} + u \sum_j q(j) \ket{j} + u^2 \sum_{j< k} q(j,k) \ket{j,k} + O(u^3) \quad\text{(up to normalization)},
\end{equation}
where $u = e^{-\alpha}$. One can check that this gives the same results as the correlation functions in Eqs.~\ref{eq:correlators_critical2} and~\ref{eq:correlators_critical2_Z}, as we have shown in Appendix~\ref{app:perturbative}. In the following, however, we focus on the second R\'enyi entropy from Eq.~\ref{eq:expansion_approach}. Note that if we cut off higher orders and keep only the first two terms, the approximate state can be written as,
\begin{equation}
    \ket{\psi}_{\rm approx} = H_{\rm eff} \ket{0}, \quad H_{\rm eff} = 1 + u \sum_j q(j) \hat{O}_j(\bm{n}^\perp) + u^2 \sum_{j<k}q(j,k) \hat{O}_j(\bm{n}^\perp) \hat{O}_k(\bm{n}^\perp),
\end{equation}
$H_{\rm eff}$ can be written as a matrix product operator (MPO)~\cite{Schollwock2011Jan} with bond dimension $L+1$, $H_{\rm eff} = W_1 W_2 \cdots W_L$, where $W$'s are operator-valued matrices:
\begin{equation}
    \label{eq:mpo}
    W_i = \begin{pmatrix}
        \mathbbm{1} & 0 & \cdots & 0 & \hat{O}_{i} (\bm{n}^\perp) & u q(i) \hat{O}_{i}(\bm{n}^\perp) \\
        0 & \mathbbm{1} & \cdots & 0 & 0 & u^2 q(1,i) \hat{O}_{i}(\bm{n}^\perp) \\
        \vdots & \vdots & \ddots &\vdots& 0 & u^2 q(2,i)\hat{O}_{i}(\bm{n}^\perp)\\
        0 & 0 & \cdots &\mathbbm{1}& 0 & u^2 q(i-1,i)\hat{O}_{i}(\bm{n}^\perp)\\
        0 & 0 & \cdots & 0 & 0 & \mathbbm{1}\\
    \end{pmatrix}
\end{equation}
for $1<i<L$ is an $(i+1)\times(i+2)$ matrix, and on the first and last sites
\begin{equation}
    W_1 = \begin{pmatrix}
        \mathbbm{1}, & \hat{O}_{1}(\bm{n}^\perp), & \mathbbm{1} +  {u}q(1)\hat{O}_{1}(\bm{n}^\perp) \\
    \end{pmatrix}, \quad
    W_L = \begin{pmatrix}
        u q(L) \hat{O}_{L}(\bm{n}^\perp)\\
        u^2 q(1,L) \hat{O}_{L}(\bm{n}^\perp)\\
        u^2 q(2,L) \hat{O}_{L}(\bm{n}^\perp)\\
        \vdots \\
        u^2 q(L-1,L) \hat{O}_{L}(\bm{n}^\perp)\\
        \mathbbm{1}\\
    \end{pmatrix}.
\end{equation}
Therefore, the entanglement entropy of $\ket{\psi}_{\rm approx}$ is upper bounded by $O(\ln(L))$; hence the following calculation of entanglement entropy only makes sense when $u$ is small enough such that $S \lesssim \ln(L)$.

For a system bipartitioned into subsystems $R$ and $\bar{R}$, the 2nd R\'enyi entanglement entropy is defined to be, 
\begin{equation}
    S^{(2)}_R \equiv -\ln\frac{\text{Tr}(\rho_R^2)}{\text{Tr}(\rho_R)^2}
\end{equation}
where $\rho_R$ is the reduced density matrix for subsystem $R$. From now on, for simplicity we assume all $q({j,k,l...})$ are real.
One can then write down
\begin{align}
     \rho_R &= \left[ 1 + u^2 \sum_{k\in \bar{R} }q(k)^2 + u^4\sum_{j<k\in \bar{R}}q(j,k)^2 \right]\ket{0}\bra{0} + \sum_{j\in R} \left[u q(j) + u^3 \sum_{k\in \bar{R}}q(k) q({j,k})\right](\ket{0}\bra{j} + \ket{j}\bra{0})\\ \nonumber
      & + \sum_{j\in R}\sum_{j'\in R} \left[u^2 q(j) q({j'})+ u^4 \sum_{k\in \bar{R}} q({j,k}) q({j',k})\right] \ket{j}\bra{j'} + u^2 \sum_{j<k\in R} q({j,k}) (\ket{0}\bra{j,k} + \ket{j,k}\bra{0})\\ \nonumber
      &  + u^3 \sum_{j'\in R}\sum_{j<k\in R}q({j'}) q({j,k}) \left(\ket{j'} \bra{j,k} + \ket{j,k}\bra{j'}\right) + u^4 \sum_{j<k\in R}\sum_{j'<k'\in R} q({j,k}) q({j',k'}) \ket{j,k}\bra{j',k'}.\\ \nonumber
\end{align}
The second R\'enyi entropy takes a surprisingly simple form,
\begin{equation}\label{eq:Renyi2}
     S^{(2)}_R = 2u^4 \sum_{j\in R} \sum_{k\in \bar{R}} \left[ q({j,k}) - q(j) q(k)\right]^2 + O(u^6).
\end{equation}

When the subsystem $R$ consists of $\ell$ consecutive sites and $a_{jk}-a_j a_k \sim |j-k|^{-\kappa}$ decays as a power law with exponent $\kappa$, one can show by replacing the summation by integrals (i.e. going to the continuous limit),
\begin{equation}
\begin{aligned}
    S^{(2)}_R \simeq 2u^4 \sum_{1\leq j\leq \ell}\sum_{(\ell+1)\leq k \leq L} \frac{1}{|j-k|^{2\kappa}} & \approx 2u^4 \int_0^{\ell}dx \int_{\ell+1}^L dy\,\frac{1}{|x-y|^{2\kappa}} = 2u^4\, F_\kappa(\ell,L)\\
\end{aligned}
\end{equation}
where we consider an open chain, $R = [0, \ell]$, $\bar{R} = [\ell+1, L]$ and
\begin{equation}
    F_\kappa (\ell,L) \equiv \frac{-1 + \ell^{2-2\kappa} - L^{2 - 2 \kappa} + (L-\ell+1)^{2 - 2 \kappa}}{2 (1 - 2 \kappa) (\kappa - 1)}.
\end{equation}
For a segment in an infinite chain, one has $S^{(2)}_R \simeq 4u^4 \sum_{1\leq j\leq \ell}\sum_{(\ell+1)\leq k < \infty} \frac{1}{|j-k|^{2\kappa}}$, which is the case of Eq.~\ref{eq:pertS2}.

We can then consider different cases:
\begin{enumerate}
    \item For $\kappa > 1$, the half-chain entropy $\mathrm{max}_{\ell} \, F_\kappa(\ell,L) = F_{\kappa}\left(\frac{L}{2},L\right) = \frac{L^{2-2 \kappa }-2 \left(\frac{L}{2}+\frac{1}{2}\right)^{2-2 \kappa }+1}{2 (2 \kappa -1) (\kappa -1)}$ converges to a finite value in the thermodynamic limit. Therefore the system satisfies the area law.
    \item For $\kappa = 1$, we find
    \begin{equation}
        F_1 (\ell,L) = \frac{\partial_\kappa[-1 + \ell^{2-2\kappa} - L^{2 - 2 \kappa} + (L-\ell+1)^{2 - 2 \kappa}]}{\partial_\kappa[2 (1 - 2 \kappa) (\kappa - 1)]}\bigg|_{\kappa=1} = \ln(\ell (L-\ell)/L),
    \end{equation}
    indicating that the entanglement entropy grows logarithmically in the subsystem size.
    \item For $0 < \kappa < 1$, we find $F_{\kappa}(L/2,L) = \frac{L^{2-2 \kappa }-2 \left(\frac{L}{2}+\frac{1}{2}\right)^{2-2 \kappa }+1}{2 (2 \kappa -1) (\kappa -1)}\propto L^{2-2\kappa}$, and for small $\ell$, $F_\kappa(\ell,L) \approx \ell^{2-2\kappa} + (2-2\kappa)L^{1-2\kappa}\ell + \cdots$. Note that this approximation is only valid when $S_2 \lesssim \ln(L)$, otherwise it would  be required to scale $u$ with system size as $u \lesssim L^\frac{\kappa - 1}{2} \ln^{\frac{1}{4}}(L)$.
\end{enumerate}

\begin{figure}
    \centering
    \includegraphics[width=0.8\linewidth]{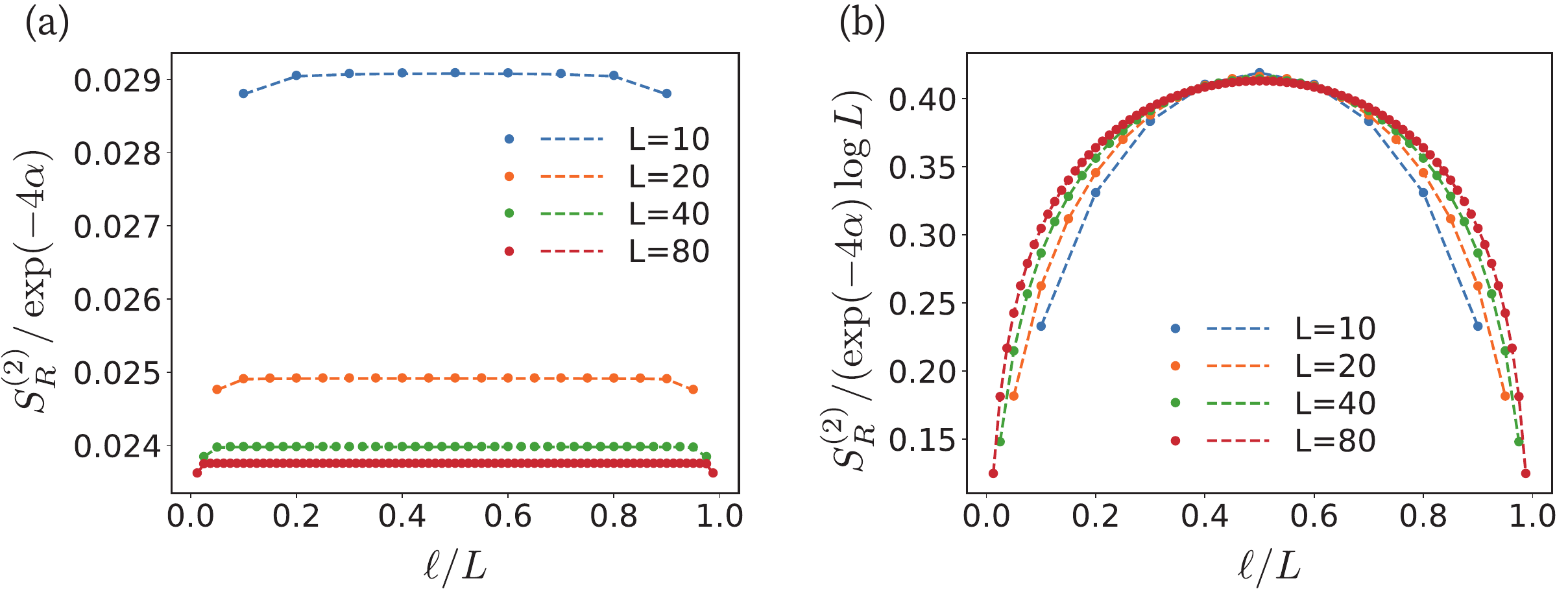}
    \caption{\textbf{Second R\'enyi entanglement entropy} for the state in Eq.~\ref{eq:f1} with $\alpha = 4$ and (a) $\bm{n}=\hat{\bm{z}}$ or (b) $\bm{n}=\hat{\bm{x}}$. The dots refer to the data obtained by DMRG on a periodic chain and dashed lines refer to the analytical prediction by Eq.~\ref{eq:Renyi2}.}
    \label{fig:app_expansion}
\end{figure}

When $\bm{n}=\hat{\bm{x}}$, we have $\kappa = 1$. Using that $V_{jk}=\frac{1}{\pi|j-k|}$, we find that the entanglement scaling remains logarithmic but $c^{(2)}_{\rm eff}$ for the 2nd R\'enyi entropy scales as $u^4$, which is consistent with the derivation in Appendix.~\ref{app:marginal} (see also Eq.~\eqref{eq:pertS2} for the result in the thermodynamic limit). For the $Z$ measurement, $\kappa = 4$. Therefore the teleported state satisfies entanglement area law at large $\alpha$, although concomitant with power-law decaying 2-point correlations. In Fig.~\ref{fig:app_expansion}, we numerically computed the entanglement entropies in Eq.~\ref{eq:f1} by DMRG for $\bm{n}=\hat{\bm{z}}$ and $\hat{\bm{x}}$, and check that they are close to the prediction from perturbation theory in Eq.~\ref{eq:Renyi2}.

\section{MPS representation for a relevant imperfection}\label{app:MPS_rel}
Given the atypical properties found in the presence of a relevant imperfection in Sec.~\ref{sec:relevant}, it is interesting to ask whether the imperfectly teleported state received by Bob can be efficiently expressed as an MPS.  An MPS with bond dimension $\chi$ on a Hilbert space $(\mathbb{C}^d)^{\otimes L}$ takes the form
\begin{equation}
    \ket{\psi_\chi}=\sum_{s_1,s_2\dots,s_L=1}^d \text{tr}\left(A_1^{[s_1]}A_2^{[s_2]}\cdots A_L^{[s_L]} \right)\ket{s_1,s_2\dots,s_L},
\end{equation}
where $\{A_j^{[s_j]}\}_{j,s_j}$ are $\chi\times \chi$ matrices. Hence, an MPS $\ket{\psi_\chi}$ has entanglement entropy less than $\ln\chi$ for any spatial cut.  While arbitrary states can be described by an MPS with exponentially large (in $L$) bond dimension, one can only efficiently calculate  expectation values of spatially local observables when such states can be approximated by an MPS with sufficiently small bond dimension. Hence, whether such an approximation exists poses a very relevant question. 

For 1D systems with strictly local interactions and a finite excitation gap, it has been rigorously proven that correlations decay exponentially and that the entanglement entropy satisfies an area law~\cite{Hastings_2007}---the latter justifying efficient description of ground states for such systems using MPS's. One then wonders to which extent relaxing strict locality can modify these conclusions, as relevant, e.g., for systems with van der Waals interactions or, as we have shown here, imperfectly teleported quantum critical states. Reference~\onlinecite{Saito_19} proved that the (non-degenerate) ground state $\ket{\psi}$ of generic 1D gapped systems with interactions decaying as a power law $\sim 1/|i-j|^{n}$ exhibits area-law entanglement for $n>2$. Evidently the gap condition imposes stronger constraints on the entanglement structure than on the decay of correlations. This important conclusion further implies that there exists an MPS $\ket{\psi_\chi}$ with bond dimension $\chi=\exp[c\ln^{5/2}(1/\delta)]$ and $c\sim \mathcal{O}(1)$ such that 
\begin{equation}
    ||\text{tr}_{X^c}\left(\ket{\psi_\chi}\bra{\psi_{\chi}} \right)-\text{tr}_{X^c}\left(\ket{\psi}\bra{\psi} \right)||_1\leq \delta |X|
\end{equation}
for an arbitrary concatenated subregion of size $|X|$, with $||\cdot ||_1$ the trace norm and $\delta$ the approximation error.  Applying these results to our problem, we conclude that $\ket{\psi_{\tilde a}}$---which is the non-degenerate ground state of the gapped Hamiltonian in Eq.~\eqref{eq:parent_Z}  with $\sim 1/|j-k|^4$ interactions---can be efficiently approximated by an MPS.  (And since $\ket{\psi_{\tilde a}}$ is the same as Bob's final wavefunction $\ket{\psi_{\tilde a}^{\rm tele}}$ up to a product of local unitary operators, the same conclusion holds for the latter.)

\section{Effective central charge calculation for \texorpdfstring{$x$}{x}-basis measurement}\label{app:marginal}
Let us consider the teleported wavefunction in Eq.~\eqref{eq:X}. We can study it as a quantum quench protocol in which the system at time $t = 0$ is initialized in the ground state $\ket{\psi_c}$ of the critical Ising chain of length $L$ with periodic boundary conditions and then is let evolve under the action of the non-Hermitian effective Hamiltonian 
\begin{equation}\label{eq:app1}    \ket{\psi^{\mathrm{tele}}_{\tilde{a}}}=\frac{e^{-it H_{+}(x)}\ket{\psi_c}}{||e^{-itH_{+}(x)}\ket{\psi_c}||}, \quad H_{+}(x)=i\sum_j X_j,
\end{equation}
with $t=\alpha/2$.
As already mentioned in the main text, this time-evolution can be exactly computed once the model is mapped to free fermions $(c,c^{\dagger})$ through Jordan-Wigner transformation, followed by Fourier transform into momentum space. Indeed, it can be written as
\begin{equation}
  H_{+}(x)=\sum_{k } \left[(c^{\dagger}_k \quad c_{-k})M \left(\begin{array}{c} c_{k} \\ c^{\dagger}_{-k} \end{array}\right)\right]  =\sum_k H_k
\end{equation}
where the momenta are semi-integer
\begin{equation}
    k = - \frac{L}{2} + \frac{1}{2}, \ldots, -\frac{1}{2}, \frac{1}{2}, \ldots, \frac{L}{2} - \frac{1}{2}\,, 
\end{equation}
(even parity sector) and 
\begin{equation}
    M=2\begin{pmatrix}
    i & 0 \\
    0 & -i
    \end{pmatrix}
\end{equation}
Since the state $\ket{\psi_c}$ is
translational invariant and with a well defined parity, then its dynamics can be decomposed in
\begin{equation}
\begin{split}   \ket{\psi_c}=\prod_k\ket{\psi_k(t=0)}, \quad \ket{\psi_{\tilde{a}}^{\rm tele}}=\prod_k\ket{\psi_k(t=\alpha/2)} \\
\ket{\psi_k(t=\alpha/2)}=\frac{\ket{\tilde{\psi}_k(t=\alpha/2)}}{||\ket{\tilde{\psi}_k(t=\alpha/2) }||}, \quad i\frac{d }{dt}\ket{\tilde{\psi}_k(t)}=H_k\ket{\tilde{\psi}_k}
    \end{split}
\end{equation}
where the $k$-momentum states are expressed as $\ket{\tilde{\psi}_k}=u_k(t)\ket{0}+v_k(t) c^{\dagger}_k c^{\dagger}_{-k}\ket{0}$. Thus, we have to solve a differential equation
\begin{equation}
  i\frac{d}{dt}  \begin{pmatrix} u_{k}(t) \\ v_{k}(t) \end{pmatrix}=-M\begin{pmatrix} u_{k}(t) \\ v_{k}(t) \end{pmatrix}
\end{equation}
with given initial conditions $u_k(0), v_k(0)$ fixed by $\ket{\psi_c}$. By solving it, we find
\begin{equation}\label{eq:ukvk}
\begin{split}
    u_k(t=\alpha/2)=&-\frac{1}{2} e^{-\alpha}\sqrt{2-2|\sin(\pi k/L)|}, \\
    v_k(t=\alpha/2)=& e^{\alpha}\frac{\sin(2\pi k/L)}{2\sqrt{(1-\cos(2\pi k/L))(1-|\sin(\pi k/L)|)}}.
    \end{split}
\end{equation}
Exploiting the Guassianity of the state~\eqref{eq:app1} and employing Wick's theorem, the correlation functions and the entanglement properties can be fully encoded in the two-point function of the Majorana fermions, which are given by the operators $\gamma_{A,j}=c_j+c_j^{\dagger}$, $\gamma_{B,j}=i(c_j^{\dagger}-c_j)$. Then, we find 
\begin{equation}
    \begin{split}
        &\langle \gamma_{A,n} \gamma_{A,m}\rangle=\langle \gamma_{B,n} \gamma_{B,m}\rangle=0\\
        &\langle \gamma_{A,n} \gamma_{B,m}\rangle=\langle \gamma_{B,m} \gamma_{A,n}\rangle^*=\frac{1}{L}\sum_ke^{2\pi i k/L(n-m)}\frac{u_k^2-v_k^2-2iu_kv_k}{u_k^2+v_k^2}
    \end{split}
\end{equation}
Taking the thermodynamic limit $L\to \infty$, the above sum is transformed into an integral, and we have
\begin{equation}\label{eq:correl_V}
\begin{split}
V\equiv\left\langle\left(\begin{array}{c}
		\gamma_{A,n}\\
		\gamma_{B,n}
\end{array}\right)\left(\begin{array}{cc}
		\gamma_{A,m}& \gamma_{B,m}
\end{array}\right)\right\rangle=
    &\displaystyle \int_{-\pi}^{\pi}\frac{d\theta}{2\pi}e^{i\theta(n-m)}\begin{pmatrix}
    &0 \,&\frac{u_{\theta}^2-v_{\theta}^2-2iu_{\theta}v_{\theta}}{u_{\theta}^2+v_{\theta}^2}\\
    &\frac{u_{\theta}^2-v_{\theta}^2+2iu_{\theta}v_{\theta}}{u_{\theta}^2+v_{\theta}^2}\, &0\end{pmatrix}\\
    &\displaystyle \int_{-\pi}^{\pi}\frac{d\theta}{2\pi}e^{i\theta(n-m)}\mathcal{G}(\theta),
    \end{split}
\end{equation}
where we took $2\pi k/L\to \theta$ in Eq.~\eqref{eq:ukvk}. 
The quantity $\mathrm{Tr}(\rho_A^n)$ can be computed from the two-point correlation matrix $V$ introduced in Eq.~\eqref{eq:correl_V}~\cite{Peschel_2009}. More specifically,
\begin{equation}\label{det_Z_kappa}
 \mathrm{Tr}\rho_A^n=\det\left[\left(\frac{I+V_A}{2}\right)^n
 +\left(\frac{I-V_A}{2}\right)^n\right],
\end{equation}
where $V_A$ denotes the restriction of $V$ to the subsystem $A$. If we take into account that the eigenvalues of $V_A$ lie on the real interval
$[-1, 1]$ and we use the residue theorem, then the previous expression can be rewritten as
the contour integral
\begin{equation}\label{contour_Z_kappa}
 \frac{1}{1-n}\ln \mathrm{Tr}_{A}^{(n)}=\frac{1}{2\pi i}\lim_{\varepsilon\to 1^+}
 \oint_{\mathcal{C}}f_n(\lambda/\varepsilon)\frac{d}{d\lambda}\ln D_{A}(\lambda)d\lambda,
\end{equation}
where the integration contour $\mathcal{C}$ encloses the interval $[-1, 1]$, 
\begin{equation}
 f_n(\lambda)= \frac{1}{1-n}\ln\Big[\left(\frac{1+\lambda}{2}\right)^n 
 +\left(\frac{1-\lambda}{2}\right)^n\Big],
\end{equation}
and $D_f(\lambda)$ denotes the characteristic polynomial of $V_A$, i.e. 
$D_f(\lambda)=\det(\lambda I-V_A)$.

Here we will focus 
on a single interval of $\ell$ contiguous sites.
In this case, the restriction $V_A$ is a $2\ell\times 2\ell$ block Toeplitz matrix
with symbol the $2\times 2$ matrix $\mathcal{G}(\theta)$ of Eq.~\eqref{eq:correl_V}. 
To deduce the large $\ell$ behaviour of $D_f(\lambda)$ and, therefore,
of $\mathrm{Tr}\rho_{A}^{n}$, we will apply the 
results on the asymptotic behaviour of block Toeplitz determinants. In particular, if the symbol $\mathcal{G}_\lambda(\theta)
=\lambda I-\mathcal{G}(\theta)$ satisfies $\det\mathcal{G}_\lambda(\theta)\neq 0$
and is a piecewise continuous function in $\theta$ with jump discontinuities 
at $\theta=\theta_1,\dots,\theta_R$, then 
\begin{equation}\label{asymp_log_D_A}
 \ln D_f(\lambda)= \frac{\ell}{4\pi}\int_{-\pi}^{\pi}\ln\det\mathcal{G}_\lambda(\theta)d\theta 
 +\frac{\ln \ell}{4\pi^2}\sum_{r=1}^R \Tr[\ln \mathcal{G}_{\lambda,r}^-(\mathcal{G}_{\lambda,r}^+)^{-1}]^2+\mathcal{O}(1),
\end{equation}
where $\mathcal{G}_{\lambda,r}^\pm$ are the lateral limits of $\mathcal{G}_\lambda(\theta)$ in 
the jump discontinuity at $\theta=\theta_r$,
\begin{equation}
 \mathcal{G}_{\lambda,r}^\pm=\lim_{\theta\to\theta_r^\pm}\mathcal{G}_\lambda(\theta).
\end{equation}
Eq.~\eqref{asymp_log_D_A} is a generalisation of the Fisher-Hartwig conjecture for block Toeplitz determinants~\cite{Basor1979,BASOR1994}.
If the symbol $\mathcal{G}_\lambda(\theta)$ has continuous entries in $\theta$, then there is no logarithmic term in the asymptotic 
expansion of $\ln D_f(\lambda)$, and Eq.~\eqref{asymp_log_D_A} simplifies to
\begin{equation}
 \ln D_f(\lambda)= \frac{\ell}{4\pi}\int_{-\pi}^{\pi}\ln\det\mathcal{G}_\lambda(\theta)d\theta+\mathcal{O}(1).
\end{equation}
This is the Szeg\H{o}-Widom theorem~\cite{WIDOM}. 
Now the 
source of discontinuities in the symbol $\mathcal{G}_\lambda(\theta)$ 
is around $\theta=0$, with lateral limits
\begin{equation}\label{lat_lim_G}
 \mathcal{G}_{\lambda,0}^\pm=\lambda I+\tanh(2\alpha) X\pm \mathrm{sech}(2\alpha) Y,
\end{equation}
where $X$ and $Y$ are the Pauli matrices.
According to Eq.~\eqref{asymp_log_D_A}, this discontinuity gives rise to a logarithmic 
term in $\ln D_f(\lambda)$. If we plug Eq.~\eqref{lat_lim_G} into Eq.~\eqref{asymp_log_D_A}, 
we find that 
\begin{equation}\label{asymp_log_D_A_nu_0_1}
 \ln D_f(\lambda)= \ln\left[\lambda^2-1\right]+b_0(\lambda) \ln \ell + \mathcal{O}(1),
\end{equation}
with 
\begin{equation}
 b_0(\lambda)=\frac{2}{\pi^2}\left(\ln\frac{\sqrt{\lambda^2-\tanh^2(2\alpha)}+\mathrm{sech}(2\alpha)}{\sqrt{\lambda^2-1}}\right)^2.
\end{equation}
The linear term is obtained from the integral
\begin{equation}
    \lim_{\varepsilon\to 1^+} 
 \oint_\mathcal{C} f_n(\lambda/\varepsilon)\frac{\lambda}{\lambda^2-1}d\lambda,
\end{equation}
which is zero by applying the Cauchy theorem, since $f_n(\pm 1 )=0$.
Therefore, 
we obtain the following asymptotic behaviour for the R\'enyi entropies
\begin{equation}
S^{(n)}_R=\frac{c^{(n)}_{\mathrm{eff}}}{6}\left(\frac{n+1}{n}\right)\ln \ell+\mathcal{O}(1),
\end{equation}
where the coefficient $c^{(n)}_{\mathrm{eff}}$ 
is given by the contour integral 
\begin{equation}\label{eq:c_effn}
 c^{(n)}_{\mathrm{eff}}=\frac{6n}{4\pi i(1+n)}\lim_{\varepsilon\to 1^+} 
 \oint_\mathcal{C} f_n(\lambda/\varepsilon)\frac{db_0(\lambda)}{d\lambda}d\lambda,
\end{equation}
which, following similar steps as in Ref.\cite{ares2018}, can be reduced to the real integral in Eq.~\eqref{eq:c_eff}.

Beyond the computation of the entanglement entorpy, we are also interested in computing the spin-spin correlation functions, for which we have to take into account the non-local effects of the Jordan-Wigner transformation. Using them and Wick's Theorem, we can expand the spin expectation values in terms of two-point correlation. In particular, for the $X$ correlators, we find
\begin{equation}
\langle X_0X_j\rangle=H(0)^2-H(j)H(-j), \quad H(j)=\langle\gamma_{B,n} \gamma_{A,m}\rangle.
\end{equation}
Therefore the connected two-point function reads
\begin{equation}
\begin{split}
&\langle X_0X_j\rangle_c=\int_{-\pi}^{\pi}\frac{d\theta}{2\pi}e^{i j \theta}\mathcal{M}(\theta,\alpha),\\
 &  \mathcal{M}(\theta,\alpha)=\frac{\left| \sin \left(\frac{k}{2}\right)\right|  \left(4+4 i e^{2 \alpha } \cot \left(\frac{k}{2}\right)\right)-2 i e^{2
   \alpha } \sin (k)+e^{4 \alpha } (\cos (k)+1)+\cos (k)-3}{4 \left| \sin \left(\frac{k}{2}\right)\right| -e^{4 \alpha }
   (\cos (k)+1)+\cos (k)-3}.
   \end{split}
\end{equation}
We can solve explicitly this integral and in the limit of large distances, $j$, we find 
\begin{equation}
    \langle X_iX_j\rangle_c=\frac{\text{sech}^2(2 \alpha )}{\pi ^2 |i-j|^2}.
\end{equation}
The other two correlators involve a non local string of operators and their Wick’s
expansion can be expressed as the determinant of a $L\times L$ matrix with elements given by all non-trivial
contractions
\begin{equation}
\langle Z_lZ_m\rangle=\det|H(i-j)|^{j=l+1\ldots m}_{i=l\ldots m-1}, \quad \langle Y_lY_m\rangle=\det|H(i-j)|^{j=l\ldots m-1}_{i=l+1\ldots m}.
\end{equation}
In order to derive the scaling behavior of these correlators, we can apply again the Fisher-Hartwig conjecture in Eq.~\eqref{asymp_log_D_A} but now our symbol $\mathcal{M}(k,\alpha)$ is not a matrix, but it is simply a function. In particular, for our purpose we want to compute the determinant of the following matrices:
\begin{equation}
    M^Z_{kj}=\displaystyle \int_{-\pi}^{\pi}\frac{dk}{2\pi}e^{i (j-k) \theta} \mathcal{M}(\theta,\alpha), \qquad
   M^Y_{kj}=\displaystyle \int_{-\pi}^{\pi}\frac{d\theta}{2\pi}e^{i (j-k) \theta}\widetilde{\mathcal{M}}(\theta,\alpha),\qquad k=l\ldots m-1,j=l+1\ldots m 
\end{equation}
with $\widetilde{\mathcal{M}}(\theta,\alpha)=\mathcal{M}(\theta,\alpha)e^{-2i\theta}$, such that 
\begin{equation}
    \langle Z_lZ_m\rangle=\det[M^Z], \qquad  \langle Y_lY_m\rangle=\det[M^Y], 
\end{equation}
 We observe that the imaginary part of $\mathcal{M}(\theta,\alpha)$ has a discontinuity for $\theta\to 0$, and 
 \begin{equation}
 \begin{split}
    & \mathcal{M}_{\pm}(\alpha)=\lim_{\theta\to 0^{\pm}}\mathcal{M}(\theta,\alpha)=\frac{e^{-2 \alpha } \mp i}{e^{-2 \alpha }\pm i}, \\
     & \widetilde{\mathcal{M}}_{+}(\alpha)=\lim_{\theta\to 0^{+}}\widetilde{\mathcal{M}}(\theta,\alpha)=\frac{e^{-2 \alpha } - i}{e^{-2 \alpha }+ i},\quad  \widetilde{\mathcal{M}}_{-}(\alpha)=\lim_{\theta\to 0^{-}}\widetilde{\mathcal{M}}(\theta,\alpha)=e^{-4i\pi}\frac{e^{-2 \alpha } + i}{e^{-2 \alpha }- i},
     \end{split}
 \end{equation}
The non-vanishing terms in the Fisher-Hartwig conjecture are now given by 
\begin{equation}
\begin{split}\label{eq:fisher-corr}
    &\ln \det[M^{Z}]=\frac{1}{4\pi^2}\left(\ln[\mathcal{M}_{-}\mathcal{M}^{-1}_{+}]\right)^2\ln |l-m|=\frac{1}{\pi^2}\left(\ln\left[\frac{e^{-2\alpha}-i}{e^{-2\alpha}+i}\right]\right)^2\ln |l-m|, 
    \\
    &\ln \det[M^{Y}]=\frac{1}{4\pi^2}\left(\ln[\widetilde{\mathcal{M}}_{-}\widetilde{\mathcal{M}}^{-1}_{+}]\right)^2\ln |l-m|=\frac{1}{\pi^2}\left(\ln\left[\frac{e^{-2\alpha}-i}{e^{-2\alpha}+i}\right]+2i\pi\right)^2\ln |l-m|.
    \end{split}
\end{equation}
We can now compare these results with the ones in Eq.~\eqref{eq:correlators_critical2}. By using that 
\begin{equation}
    \begin{split}
        \left(\ln\left[\frac{e^{-2\alpha}-i}{e^{-2\alpha}+i}\right]\right)^2&=-4\arctan^2(e^{2\alpha}),\\
        \left(\ln\left[\frac{e^{-2\alpha}-i}{e^{-2\alpha}+i}\right]+2i\pi\right)^2&=-4[\arctan(e^{2\alpha})-\pi]^2,
    \end{split}
\end{equation}
and Eq.~\eqref{eq:fisher-corr}, we can derive the correct power-law behavior for the correlators $\langle Z_lZ_m\rangle$ and $\langle Y_lY_m\rangle$ in Eq.~\eqref{eq:correlators_critical2}.

\section{Extracting the correlations and effective central charge from DMRG}\label{app:dmrg}

\begin{figure}[h]
    \centering
    \includegraphics[width=\linewidth]{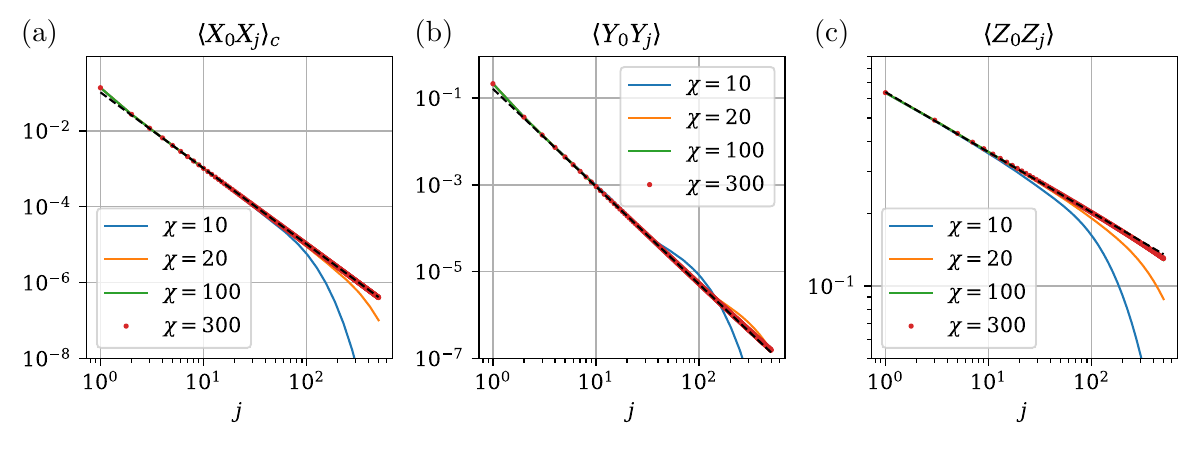}
    \caption{\textbf{Connected $XX$, $YY$ and $ZZ$ correlations of the pristine critical Ising chain.} Data obtained using infinite DMRG with different bond dimensions $\chi=10,20,200,300$. The black dashed lines represent the power-law decay with theoretical exponents, i.e., $-2$, $-9/4$, $-1/4$ in panels $(a)$, (b) and (c) respectively.}
    \label{fig:convergenceofDMRG}
\end{figure}

In this appendix, we provide details of DMRG simulations of correlation functions and how the effective central charges are extracted in Fig.~\ref{fig:Y_Ent_ceff}(b).  The critical Ising state (used for computing correlations in Figs.~\ref{fig:Z_correlations},~\ref{fig:X_correlations} and~\ref{fig:Y_correlations} is obtained by infinite DMRG with bond dimension $\chi = 300$. In Fig.~\ref{fig:convergenceofDMRG} we show that this bond dimension is enough to correctly extract the expected power-law exponent for distances up to $|j|\approx 500$ sites. We also used a bond dimension $\chi=300$ in Figs.~\ref{fig:Z_expansion} and \ref{fig:Y_Ent_ceff}.

\begin{figure}[h]
    \centering
    \includegraphics[width=0.7\linewidth]{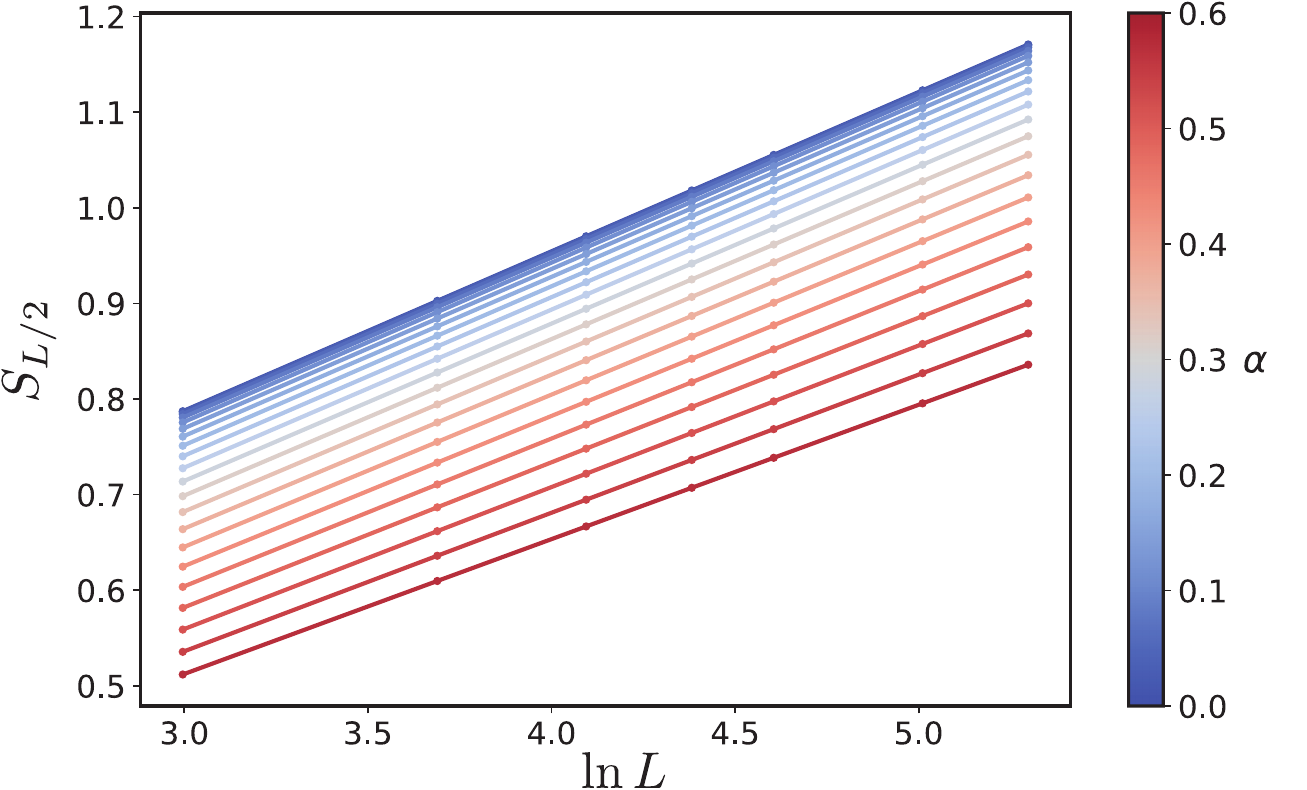}
    \caption{\textbf{Half-chain entanglement entropy for different system sizes and imperfection strength}. The effective central charges extracted from the slope of the lines are shown in Fig.~\ref{fig:Y_Ent_ceff}(b)}
    \label{fig:Ymeas_data}
\end{figure}

The effective central charge can be obtained using finite DMRG with different system sizes. In Fig.~\ref{fig:Y_Ent_ceff}(b), we extract $c_{\rm eff}$ from
\begin{equation}
    S_{L/2} \sim \frac{c_{\rm eff}}{3} \ln(L),
\end{equation}
where $S_{L/2}$ is the half-chain von Neumann entanglement entropy of the teleported wavefunction with periodic boundary and length $L$. In practice we take $L = [20, 40, 60, 80, 100, 150, 200]$ and fit $S(L/2)$ v.s. $\ln(L)$ for different imperfection strength $\alpha$ (see Fig.~\ref{fig:Ymeas_data}).

\begin{figure}[h]
    \centering
    \includegraphics[width=0.6\linewidth]{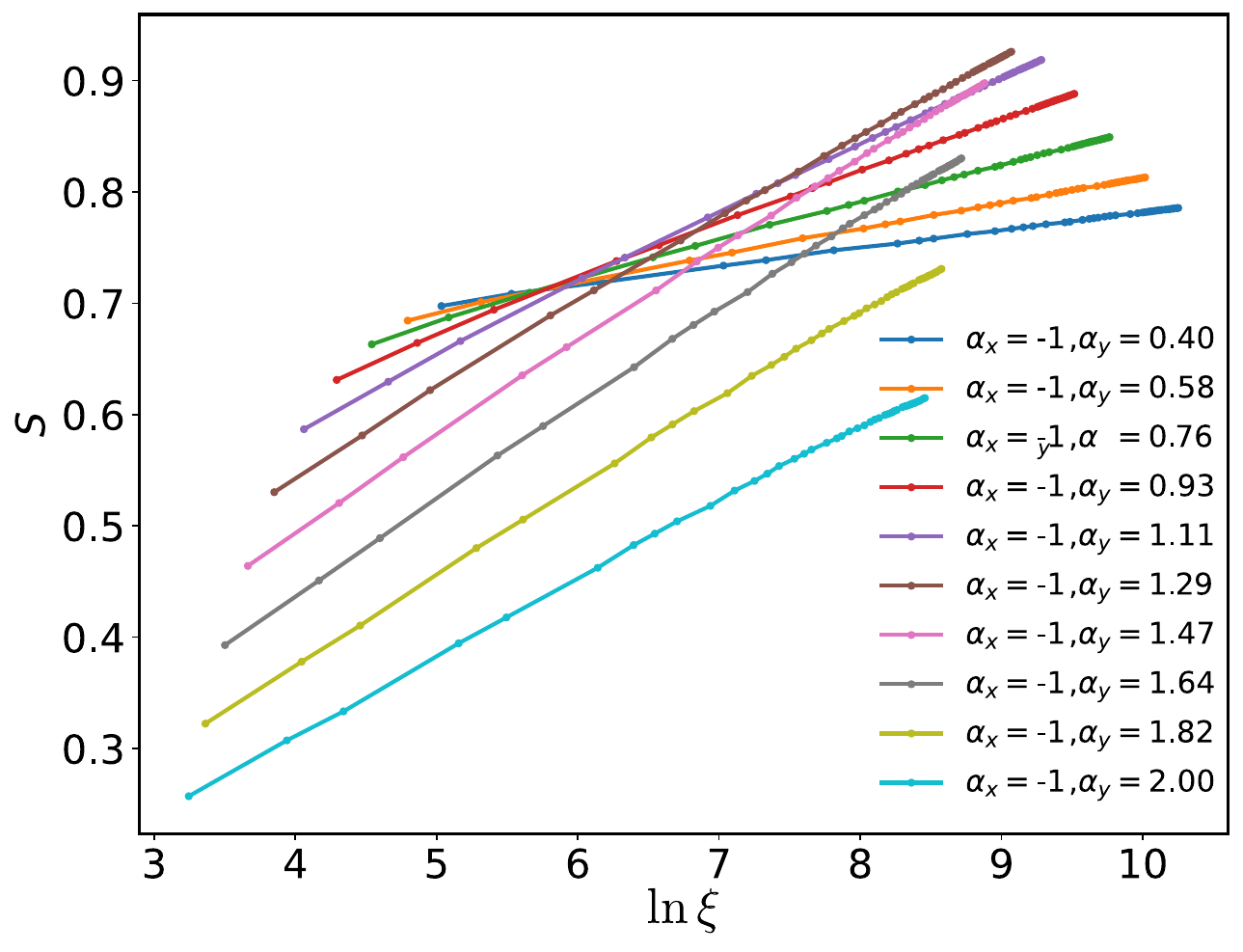}
    \caption{$S$-$\ln(\xi)$ curves for different $(\alpha_x, \alpha_y)$ in Fig.~\ref{fig:optimal}.}
    \label{fig:app_optimal}
\end{figure}

For the mixed $X$- and $Y$-basis measurement in Fig.~\ref{fig:optimal}, we make use of the relation between entanglement entropy and correlation length (as induced by a finite bond dimension) in infinite DMRG~\cite{Pollmann_2009},
\begin{equation}
    S \sim \frac{c_{\rm eff}}{3} \ln(\xi)
\end{equation}
where $\xi$ is the correlation length computed from the transfer matrix of the infinite MPS. Practically, for a given set of $(\alpha_x, \alpha_y)$ in Fig.~\ref{fig:optimal}, we use different bond dimensions $\chi \in [5,100]$ and compute the entanglement entropy $S$. The curve $S(\ln(\xi))$ is expected to be a straight line in the large $\xi$ limit for critical systems. We extract the effective central charge using 30 data points of each curve with the largest $\xi$. In Fig.~\ref{fig:app_optimal} we show $S$ vs $\ln{\xi}$ for some combinations of $(\alpha_x, \alpha_y)$.

 \section{Correlators of teleported mixed state}

In order to obtain the result in Eq.~\eqref{eq:correlators_dec}, let us start from $\hat{O}(\bm{n})$ (we suppress the subscript $\mathcal{K}$ to enlight the notation).
Using $\ket{\tilde{a},\bm{n}}=\hat{O}(\bm{n})\ket{\tilde{a},\bm{n}}$,
it is easy to check that
\begin{equation}
\begin{split}
 \mathrm{Tr}(\rho^{\rm tele}\hat{O}(\bm{n}))=&     
    \sum_{\tilde{a},\tilde{a}^\prime} \prod_j[\sin(2u)]^{\frac{1-a_ja'_j}{2}}\langle\psi_c|\tilde{a}^\prime,\bm{n}\rangle \langle \tilde{a},\bm{n}|\psi_c \rangle  \bra{\tilde{a}^\prime,\bm{n}}\hat{O}_{\mathcal{K}}(\bm{n})\ket{\tilde{a},\bm{n}}\\
    &=\bra{\psi_c}\hat{O}(\bm{n})\left( \sum_{\tilde{a} }\ket{\tilde{a},\bm{n}} \bra{\tilde{a},\bm{n}}\right)\ket{\psi_c}=\bra{\psi_c}\hat{O}(\bm{n})\ket{\psi_c}.
\end{split}
\end{equation}
On the other hand, if we consider $\hat{O}(\bm{n}^\perp)$, we need to take care of the fact that $\hat{O}(\bm{n}^\perp)$ acting on $\ket{\tilde{a},\bm{n}}$ sends $a_k\to -a_k$, which yields
\begin{equation}\label{eq:Operp}
\begin{split}
 \mathrm{Tr}(\rho^{\rm tele}\hat{O}_{\mathcal{K}}(\bm{n}^\perp))=&     
    \sum_{\tilde{a},\tilde{a}^\prime} \prod_j[\sin(2u)]^{\frac{1-a_ja'_j}{2}}\langle\psi_c|\tilde{a}^\prime,\bm{n}\rangle \langle \tilde{a},\bm{n}|\psi_c \rangle  \bra{\tilde{a}^\prime,\bm{n}}\hat{O}_{\mathcal{K}}(\bm{n}^{\perp}) \ket{\tilde{a},\bm{n}}\\
    &=\prod_{j\in \mathcal{K}}\sin(2u) \sum_{\tilde{a}}\bra{\psi_c}\hat{O}_{\mathcal{K}}(\bm{n}^{\perp})\ket{\tilde{a},\bm{n}} \bra{\tilde{a},\bm{n}}\ket{\psi_c}=\sin(2u)^{\mathcal{K}}\bra{\psi_c}\hat{O}_{\mathcal{K}}(\bm{n}^{\perp})\ket{\psi_c}.
\end{split}
\end{equation}
where we have used that $\bra{\tilde{a}^\prime,\bm{n}}\hat{O}(\bm{n}^{\perp}) \ket{\tilde{a},\bm{n}}=\prod_{j\notin \mathcal{K}}\delta_{a_j,a^\prime_j}\prod_{j\in \mathcal{K}}\delta_{a_j,-a^\prime_j}$ and $\ket{-\tilde{a},\bm{n}}=\hat{O}(\bm{n}^\perp)\ket{\tilde{a},\bm{n}}$. Finally, in order to get the result for $\hat{O}_{\mathcal{K}}(\bm{n}\times\bm{n}^\perp)$, we first notice that $\hat{O}(\bm{n}\times\bm{n}^\perp)=i\hat{O}(\bm{n}^\perp)\hat{O}(\bm{n})$ and then we apply the same strategy as in Eq.~\eqref{eq:Operp}.

\section{Entanglement negativity of teleported mixed state}\label{app:negativity}
\begin{figure*}[ht]
    \centering
    \includegraphics[width=\linewidth]{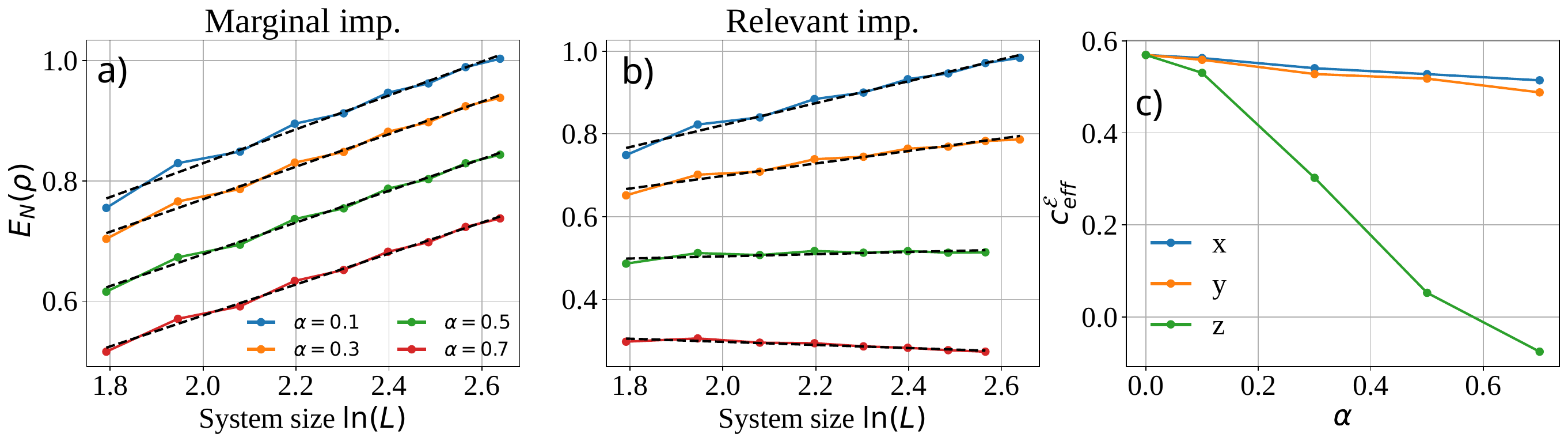}
    \caption{\textbf{Half-chain entanglement negativity as a function of system size for different deformations}. Scaling of the half-chain entanglement negativity of the teleported mixed state $\rho^{\rm tele}$ with system size in $L\in [6,14]$ as a function of the deformation strength $\alpha$ along $\bm{x}$ (panel (a)) and $\bm{z}$ (panel (b)) axes. (c) Scaling of the effective central charge $c_{\textrm{eff}}^{\mathcal{E}}$ as a function of $\alpha$ with imperfections in the $\bm{x}$, $\bm{y}$ and $\bm{z}$ quantization axes.}
    \label{fig:negativity}
\end{figure*}
Since $\rho^{\rm tele}$ from Eq.~\ref{rhotele} is a mixed state, the natural way to quantify its entanglement structure is using the entanglement negativity~\cite{peres1996separability,vw-02}, as we already pointed out in Sec.~\ref{sec:decoding}. Through a numerical study, the goal of this appendix is to show how the negativity is affected by the imperfections of our teleportation protocol.
Let us recall the definition of negativity. We consider a bi-partition of the system into a region $R$ and its complement $R\cup \bar{R}$, and we then perform a partial transposition of the density
matrix $\rho^{\rm tele}$
with respect to the system $R$, that we denote as $\rho^{\mathrm{tele},T_R}$. This operation is defined performing a standard transposition in the Hilbert space $\mathcal{H}_{R}$ associated to the region $R$, i.e. exchanging
the matrix elements in $R$, $\rho^{\mathrm{tele},T_R}=(T_R\otimes \mathbbm{1}_{\bar{R}})\rho^{\rm tele}$, leaving the rest untouched. The
presence of negative eigenvalues of $\rho^{\mathrm{tele},T_R}$ is a signature of
mixed state entanglement~\cite{peres1996separability}, which can be quantified
by the logarithmic negativity $\mathcal{E}=\ln \mathrm{Tr}||\rho^{\mathrm{tele},T_R}||_1$, where $||\cdot||_1$ denotes the trace norm. When $\alpha=0$, the negativity of a single interval of size $|R|=\ell$, embedded in an infinite system, behaves as $\mathcal{E}=c/2 \ln(\ell/\epsilon) $, where $c$ is the central charge of Alice's critical state $\ket{\psi_c}$~\cite{cct-13}. For $\alpha\neq 0$, we do not have any analytical prediction about the behavior of $\mathcal{E}$ and, for this reason, we focus on some specific examples. 
 
In the following, we numerically compute the negativity of $\rho^{\rm tele}$ with Alice in the critical state, initializing Bob's wavefunction on a uniform product state on either the $\hat{\bm{x}}$ basis (Fig.~\ref{fig:negativity}a)) or on the $\hat{\bm{z}}$ basis (Fig.~\ref{fig:negativity}B)), as in Secs.~\ref{sec:marginal} and~\ref{sec:relevant} respectively. We have used exact diagonalization (ED) for finite systems of length up to $L=14$ and computed the half-chain negativity of $\rho^{\rm tele}$. We numerically find that
$\mathcal{E}$ approximately obeys the scaling form 
\begin{equation}
    \mathcal{E}=\frac{c^{\mathcal{E}}_{\rm eff}}{2}\ln(L)+\Upsilon,
\end{equation}
where $\Upsilon$ is an $\alpha$-dependent term. From the plots in Fig.~\ref{fig:negativity}a)-c), we find that for $\alpha=0$, $c^{\mathcal{E}}_{\rm eff}\lesssim 0.6$, while we would expect $c^{\mathcal{E}}_{\rm eff}=1/2$~\cite{cct-13} and this is due to the small system sizes $L$ we can study with ED (similarly to Ref.~\cite{lu2023}). As $\alpha$ increases, we notice that $c^{\mathcal{E}}_{\rm eff}$ smoothly decreases when we measure in the $\hat{\bm{x}}$ basis (Fig.~\ref{fig:negativity}b)-c)), in agreement with the discussion of Sec.~\ref{sec:marginal} about a marginal imperfection, while it quickly drops to 0 in the $\hat{\bm{z}}$ basis (Fig.~\ref{fig:negativity}c)), in the same spirit of a relevant imperfection described in Sec.~\ref{sec:relevant}. This behavior suggests that the amount of long-range entanglement of Bob's wavefunction is maximal when Alice and Bob are maximally entangled (i.e. $u=\pi/4$), and it decreases otherwise. 
Finally, we comment that we do not report the plots of $\mathcal{E}$ as a function of $\ln(L)$ when measuring in the $\hat{\bm{y}}$ basis since its behavior is not different with respect to the one observed in Fig.~\ref{fig:negativity}a), consistently with the disguised marginal imperfection discussed in Sec.~\ref{sec:irrelevant}. This is confirmed by comparing the behavior of $c_{\rm eff}^{\mathcal{E}}$ in Fig.~\ref{fig:negativity}c) measuring in the $\hat{\bm{x}}$ (blue line) and $\hat{\bm{y}}$-basis (green line).

\twocolumngrid

%

\end{document}